\documentclass{article}

\usepackage[utf8]{inputenc}
\usepackage[usenames, dvipsnames]{color}
\usepackage{amsmath}
\usepackage{amssymb}
\usepackage{graphicx}
\usepackage{appendix}
\usepackage{geometry}
\usepackage{hyperref}
\usepackage{cite}

\hypersetup{
colorlinks=true,         
linkcolor=blue,          
citecolor=red,           
urlcolor=Violet          
}

\usepackage{color}
\usepackage[normalem]{ulem}
\usepackage{epsf,epsfig}
\usepackage{psfrag}
\usepackage{multirow}
\usepackage{diagbox}

\usepackage{mathtools}

\title{New class of plane waves for $\kappa$-noncommutative  Quantum Field Theory}
\date{}

\begin{document}

	\begin{center}
		\boldmath
		
		{\textbf{\LARGE New class of plane waves for $\kappa$-noncommutative  Quantum Field Theory}}
		\unboldmath
		
		\bigskip\bigskip
{Maria Grazia~Di Luca$^{1,2,3}$, Flavio Mercati $^3$\\}

\small{$^1$ Dipartimento di Fisica ``E. Pancini", Università di Napoli Federico II - INFN sezione di Napoli, Complesso Universitario di Monte S. Angelo Edificio 6, via Cintia, 80126 Napoli, Italy}\\
\small{$^2$ Scuola Superiore Meridionale, Largo S. Marcellino, 10, 80138 Napoli NA, Italy}\\
\small{$^3$ Departamento de F\'isica, Universidad de Burgos, 09001 Burgos, Spain}

\begin{abstract}
We discuss the construction of a free scalar quantum field theory on $\kappa$-Minkowski noncommutative spacetime. We do so in terms of $\kappa$-Poincaré-invariant $N$-point functions, \textit{i.e.} multilocal functions which respect the deformed symmetries of the spacetime. As shown in a previous paper by some of us, this is only possible for a lightlike version of the commutation relations, which allow the construction of a covariant algebra of $N$ points that generalizes the $\kappa$-Minkowski commutation relations. We solve the main shortcoming of our previous approach, which prevented the development of a fully covariant quantum field theory: the emergence of a non-Lorentz-invariant boundary of momentum space. To solve this issue, we propose to ``extend" momentum space by introducing a class of new Fourier modes and we prove that this approach leads to a consistent definition of the Pauli-Jordan function, which turns out to be undeformed with respect to the commutative case. We finally address the quantization of our scalar field and obtain a deformed, $\kappa$-Poincaré-invariant, version of the bosonic oscillator algebra.
\end{abstract}

\end{center}

\section{Introduction}

In the past few decades, quantum field theory (QFT) on noncommutative spacetimes has been extensively studied. An in-depth discussion of noncommutative spacetimes can be found in \cite{Szabo:2001kg,aschieri2011noncommutative,Wulkenhaar2019}, while here we just present a few main conceptual motivations. The first one can be traced back to the very early days of QFT: by introducing a noncommutative structure for spacetime coordinates at small length scales, the notion of infinitesimal point has to be relaxed into some sort of ``fuzzy"/non-local point.
One then hopes to get an effective ultraviolet cutoff and therefore to gain control on the divergences of the theory \cite{Szabo:2001kg}. On the other hand, noncommutative spacetimes are expected to play a crucial role in addressing the dichotomy between classical gravity and quantum physics. Already at an effective level of QFT coupled to classical general relativity, in fact, the support of a quantum field cannot be localized better than the Planck length $\ell_P=\sqrt{{G\hbar}/{c^3}}$ (see \cite{Wulkenhaar2019} for details). QFT could incorporate such a restriction on localisability via the introduction of uncertainty relations among noncommutative coordinate operators. A further strong indication that spacetime might be noncommutative at very small scales comes from the only model of quantum gravity that is well-understood as a QFT: 2+1-dimensional general relativity. This theory lacks local propagating degrees of freedom (gravitons) and can be therefore quantized with topological QFT methods. Coupling the theory to matter and integrating away the gravitational degrees of freedom gives rise to an effective theory of matter propagation on a noncommutative spacetime \cite{Matschull1997QuantumMO,PhysRevLett.96.221301}. The Planck scale in this model ends up playing the role of a scale of noncommutativity.

In this paper, we focus on a model of noncommutative spacetime that is valid in any dimension and is maximally symmetric under a deformation of Poincaré symmetry. This model has been widely studied in the literature \cite{Majid:1988we,Lukierski:1991ff,Lukierski:1991pn,Lukierski:1992dt,Majid:1994cy,Majid:1999td,Lukierski:2015zqa,Lizzi:2018qaf,Lizzi:2019wto,Lukierski:1993wxa,Kowalski-Glikman:2002eyl,Agostini:2002yd,Agostini:2002de,Kowalski-Glikman:2002oyi,Kowalski-Glikman:2003qjp,Agostini:2005mf,Amelino-Camelia:2007yca,Amelino-Camelia:2007rym,Carmona:2011wc,Carmona:2012un,Amelino-Camelia:2010yrn,Gubitosi:2011hgc,Mercati:2011aa,Meljanac:2016jwk,Loret:2016jrg,Lukierski:2016vah,Mercati:2018fba}, including many efforts to build a consistent QFT on it~\cite{Kosinski:1999ix,Kosinski:2001ii,Kosinski:2003xx,Arzano:2007ef,Daszkiewicz:2007az,Freidel:2007hk,Arzano:2009ci,Arzano:2017uuh,Juric:2015hda,Mathieu:2020ccc,Poulain:2018mcm,Poulain:2018two,Juric:2018qdi,Arzano:2018gii,Mercati:2018hlc,Mercati:2018ruw}. A lot of progress has been made on the problem, however a fully satisfactory formulation of QFT is still elusive. In this paper, we build upon the results obtained by some of us~\cite{PhysRevD.103.126009} and solve a technical issue that prevented the construction of a fully Lorentz-invariant QFT.
The paper is structured as follows: in the first section, we provide a detailed review of~\cite{PhysRevD.103.126009}, reporting all the main calculations and results. We end the section with a discussion of the problem of constructing the Pauli-Jordan function of our theory. This obstruction motivates us to ``extend" the Fourier theory of our model, introducing new Fourier modes. The second section is devoted to the characterization of such Fourier modes and their application to the construction of a classical (non-quantum but noncommutative) scalar field theory. Furthermore, we show how these modes allow us to define the Pauli-Jordan function, overcoming the problems that arose in~\cite{PhysRevD.103.126009}. In the third section, we apply the Pauli-Jordan quantization procedure to our scalar field and obtain a deformed bosonic oscillator algebra for the creation and annihilation operators, that is compatible with the deformation of Poincaré symmetry of the model. We conclude by presenting different possible directions in which this line of research might develop in the future.

\subsection{\texorpdfstring{$\kappa$}--Minkowski}

One of the most studied models of noncommutative spacetime is $\kappa$-Minkowski \cite{Lukierski:1991ff,Lukierski:1992dt,Majid:1994cy}. In this model the algebra of functions on Minkowski spacetime is deformed into a noncommutative $\ast$-algebra $\mathcal{A}$, generated by the coordinate operators ${x}^{\mu}$ such that

\begin{equation} \label{eq:kmink}
    [x^{\mu},x^{\nu}]=\frac{i}{\kappa}(v^{\mu}x^{\nu}-v^{\nu}x^{\mu}), \quad \mu=0,\dots,3, \quad (x^{\mu})^{\dagger}=x^{\mu}.
\end{equation}

Here $v^{\mu}$ is a set of four real numbers and $\kappa$ is an inverse-length scale, which one expects to be very small.

The symmetries of $\kappa$-Minkowski can be described within the formalism of Hopf algebras: let us consider the algebra of continuous functions on the Poincaré group $\mathbb{C}[ISO({\,3,1})]$, generated by the translation and Lorentz parameters, respectively $a^{\mu}$ and $\Lambda^{\mu}_{\!\ \!\ \nu}$. We can introduce a coproduct map $\Delta$, an antipode $S$ and a counit $\epsilon$ (see \cite{majid_2002})to codify the properties of group product, inverse and identity as follows:

\begin{equation} \label{eq:coalgebra}
    \begin{aligned}
    &\Delta[\Lambda^{\mu}_{\!\ \!\ \nu}]=\Lambda^{\mu}_{\!\ \!\ \alpha}\otimes \Lambda^{\alpha}_{\!\ \!\ \nu}, \quad \Delta[a_{\mu}]=\Lambda^{\mu}_{\!\ \!\ \nu} \otimes a^{\nu}+a^{\mu}\otimes 1 \\
    &S[\Lambda^{\mu}_{\!\ \!\ \nu}]=(\Lambda^{-1})^{\mu}_{\!\ \!\ \nu}, \quad S[a^{\mu}]=-(\Lambda^{-1})^{\mu}_{\!\ \!\ \nu}a^{\nu}, \quad \epsilon[{\Lambda^{\mu}_{\!\ \!\ \nu}}]=\delta^{\mu}_{\nu}, \quad \epsilon[a^{\mu}]=0.
    \end{aligned}
\end{equation}

Equations (\ref{eq:coalgebra}) encode everything about the Lie group structure of $ISO({\,3,1})$ in a perhaps unfamiliar, ``dual", form. (see \cite{Tjin:1991me} for an introduction).The Hopf algebra $\mathbb{C}[ISO({\,3,1})]$ is completed by trivial commutation relations of the form $[\Lambda^{\mu}_{\!\ \!\ \nu},\Lambda^{\rho}_{\!\ \!\ \sigma}]=[a^{\mu},a^{\nu}]=[a^{\gamma},\Lambda^{\mu}_{\!\ \!\ \nu}]=0$. The $\kappa$-Poincaré Hopf algebra $\mathbb{C}_{\kappa}[ISO({\,3,1})]$ is a noncommutative deformation of the algebraic sector of $\mathbb{C}[ISO({\,3,1})]$:

\begin{equation} \label{eq:algebra}
    \begin{aligned}
    &[\Lambda^{\mu}_{\!\ \!\ \nu},\Lambda^{\rho}_{\!\ \!\ \sigma}]=0, \quad [a^{\mu},a^{\nu}]=\frac{i}{\kappa}(v^{\mu}a^{\nu}-v^{\nu}a^{\mu})\\
    &[a^{\gamma},\Lambda^{\mu}_{\!\ \!\ \nu}]=\frac{i}{\kappa}[(\Lambda^{\mu}_{\!\ \!\ \alpha}v^{\alpha}-v^{\mu})\Lambda^{\gamma}_{\!\ \!\ \nu}+(\Lambda^{\alpha}_{\!\ \!\ \nu}g_{\alpha \beta}-g_{\nu \beta})v^{\beta}g^{\mu\gamma}]\\
    &\Lambda^{\mu}_{\!\ \!\ \alpha}\Lambda^{\nu}_{\!\ \!\ \beta}g^{\alpha \beta}=g^{\mu\nu}, \quad \Lambda^{\rho}_{\!\ \!\ \mu}\Lambda^{\sigma}_{\!\ \!\ \nu}g_{\rho \sigma}=g_{\mu\nu},
    \end{aligned}
\end{equation}

where $g_{\mu\nu}$ is assumed to be a symmetric invertible matrix with signature (+,-,-,-) and inverse $g^{\mu\nu}$. 

Consider the usual transformation of coordinates

\begin{equation} \label{eq:coaction}
    x'{}^{\mu}=\Lambda^{\mu}_{\!\ \!\ \nu}x^{\nu}+a^{\mu}.
\end{equation}

This can be understood as a map $\cdot \, {}' :\mathcal{A}\rightarrow \mathbb{C}_{\kappa}[ISO({\,3,1})] \otimes \mathcal{A}$, where we assume $[\Lambda^{\mu}_{\!\ \!\ \nu},x^{\rho}]=[a^{\mu},x^{\nu}]=0$.
Given the deformed algebra (\ref{eq:algebra}), it is easy to verify that the $\kappa$-Minkowski commutators in Eq. (\ref{eq:kmink}) are invariant, i.e. that the transformation is an algebra-homomorphism: observers connected by such transformations all agree on the commutation relations. As long as we deal with single-particle coordinates, then, we can get a fully relativistic theory out of three ingredients: a noncommutative algebra of coordinates (\ref{eq:kmink}), its deformed symmetries (\ref{eq:algebra}) and the usual transformation of coordinates under such symmetries (\ref{eq:coaction}).

\subsection{Braided tensor product algebra}
As pointed out in \cite{PhysRevD.103.126009}, generalizing this construction to more than one point (as required to discuss $N$-point functions) is not straightforward. In the commutative case, in fact, 2-point functions belong to the tensor product of two copies of the algebra of functions; however, the tensor product algebra $\mathcal{A}\otimes \mathcal{A}$ does not respect the deformed symmetries. Let us denote by 1 the identity of the algebra $\mathcal{A}$, $\mathcal{A}\otimes \mathcal{A}$ is generated by

\begin{equation}
    [x_1^{\mu},x_1^{\nu}]=\frac{i}{\kappa}(v^{\mu}x_1^{\nu}-v^{\nu}x_1^{\mu}), \quad     [x_2^{\mu},x_2^{\nu}]=\frac{i}{\kappa}(v^{\mu}x_2^{\nu}-v^{\nu}x_2^{\mu}), \quad [x_1^{\mu},x_2^{\nu}]=[x_1^{\mu},1^{\otimes^2}]=[x_2^{\mu},1^{\otimes^2}]=0,
\end{equation}

where $x_1^{\mu}$ and $x_2^{\mu}$ are the coordinates of the first and second point and $1^{\otimes^2}\equiv 1\otimes 1$ is the identity of $\mathcal{A}\otimes \mathcal{A}$. The coordinates $x_{1,2}^{\mu}$ each close a copy of $\kappa$-Minkowski, they commute with the identity and among each other. Moreover, the natural transformation rule for $x_{1,2}^{\mu}$ is

\begin{equation} \label{eq:transf}
\begin{aligned}
x'_{1}{}^{\mu}= \Lambda^{\mu}_{\!\ \!\ \nu}x_1^{\nu}+a^{\mu}, \quad x'_{2}{}^{\mu}= \Lambda^{\mu}_{\!\ \!\ \nu}x_2^{\nu}+a^{\mu}.
\end{aligned}
\end{equation}

We can immediately see that, because (\ref{eq:algebra}) is nonabelian, the commutators between the transformed coordinates of the first and second point is different from zero:

\begin{equation}
    [x'_1{}^{\mu},x'_2{}^{\nu}]=[\Lambda^{\mu}_{\!\ \!\ \rho},a^{\nu}](x_1^{\rho}-x_2^{\rho})+[a^{\mu},a^{\nu}]\neq 0.
\end{equation}
Therefore the tensor product algebra $\mathcal{A}\otimes \mathcal{A}$ is not invariant under the deformed symmetries.

A possible way out of this problem is to relax the commutativity between $x_1$ and $x_2$, requiring that the transformation in Equation (\ref{eq:transf}) leaves the commutator covariant: $[x'_1{}^{\mu},x'_2{}^{\nu}]=[x_1^{\mu},x_2^{\nu}]'$. A similar concept has been used in~\cite{Oeckl:2000eg,Wess:2003da,Chaichian:2004za,Koch:2004ud,Fiore:2007vg,Fiore:2007zz} to properly define QFT on the Moyal/canonical spacetime.
In \cite{PhysRevD.103.126009}, we proved that the only suitable form for the commutator, that is also linear in the coordinates and vanishes in the commutative limit ($v^{\mu}\longrightarrow 0$), is the following (from now on we choose units in which $\kappa=1$):

\begin{equation}
    [x_1^{\mu},x_2^{\nu}]=i[v^{\mu}x_1^{\nu}-v^{\nu}x_2^{\mu}-g^{\mu\nu}g_{\rho\sigma}v^{\rho}(x_1^{\sigma}-x_2^{\sigma})].
\end{equation}

This can be immediately generalized to the case of $N$ points as

\begin{equation} \label{eq:braiding}
    [x_a^{\mu},x_b^{\nu}]=i[v^{\mu}x_a^{\nu}-v^{\nu}x_b^{\mu}-g^{\mu\nu}g_{\rho\sigma}v^{\rho}(x_a^{\sigma}-x_b^{\sigma})], \quad a,b=1,...,{N},
\end{equation}
which for $a=b$ reduces to the usual $\kappa$-Minkowski commutator. We call this the ``braided-tensor product" algebra and we indicate it with the symbol $\mathcal{A}^{\otimes_{\kappa} ^N}$. If we now impose the Jacobi identities, we ensure the associativity of the algebra. These imply

\begin{equation}
    [x_a^{\mu},[x_b^{\nu},x_c^{\rho}]+\text{cyclic}=-g_{\alpha\beta}v^{\alpha}v^{\beta}[g^{\nu\rho}(x^{\mu}_c-x^{\mu}_b)+g^{\rho\mu}(x^{\nu}_a-x^{\nu}_c)+g^{\mu\nu}(x^{\rho}_b-x^{\rho}_a)]=0,
\end{equation}
which is only satisfied for $g_{\alpha\beta}v^{\alpha}v^{\beta}=0$, that is when $v$ is a lightlike vector with respect to the metric $g$. This case is called in the literature \textit{lightlike} $\kappa$-Minkowski~\cite{ballesteros1995new,1997PhLB..391...71B,Borowiec:2013lca,Borowiec:2014aqa,Juric:2014cla,Borowiec:2018rbr}.

\subsection{Representation of the braided algebra} \label{sec:repr}

In \cite{PhysRevD.103.126009}, we found a Hermitian infinite-dimensional representation for the braided algebra $\mathcal{A}^{\otimes_{\kappa} ^N}$; here we just want to sketch the main steps of the derivation. First of all, one should notice that coordinate differences, $\Delta x^{\mu}_{ab}\equiv x^{\mu}_{a}-x^{\mu}_{b}$, commute among each other:
\begin{equation}
    [\Delta x^{\mu}_{ab},\Delta x^{\nu}_{cd}]=0, \quad a,b,c,d=1,\dots, N \,\,\,\text{ and } \mu,\nu=0,\dots,3.
\end{equation}
Hence all the noncommutativity is collected by the center of mass degrees of freedom:
\begin{equation}
   x^{\mu}_{cm}=\frac{1}{N}\sum_{a=1}^N x^{\mu}_{a}.
\end{equation}
We can then define the relative positions, $y^{\mu}_{a}=x^{\mu}_{a}-x^{\mu}_{cm}$. These are 4$N$ commutative coordinates, of which only
 $4(N-1)$ are independent since we have the four constraints $\sum_{a=1}^N y^{\mu}_{a}=0$. In this basis, the braided algebra (\ref{eq:braiding}) takes the form of a semi-direct product algebra:
\begin{equation}\label{eq:cdmcommutators}
\begin{aligned}
    &[y^{\mu}_{a},y^{\nu}_{b}]=0,  \quad  [x^{\mu}_{cm},x^{\nu}_{cm}]={i}(v^{\mu}x^{\nu}_{cm}-v^{\nu}x^{\mu}_{cm}), \\ &[x^{\mu}_{cm},y^{\nu}_{a}]=i(\eta^{\mu\nu}\eta_{\rho\sigma}v^{\rho}y^{\sigma}_{a}-v^{\nu}y^{\mu}_{a}).
    \end{aligned}
\end{equation}
The maximal Abelian subalgebra is $4(N-3)$-dimensional and generated by the $y^{\mu}_{a}$ together with the projection $w$ of $x^{\mu}_{cm}$ along $v^{\mu}$:
\begin{equation}
    w=g_{\mu\nu}v^{\mu}x^{\nu}_{cm}, \quad [w,y^{\mu}_{a}]=0, \quad [x^{\mu}_{cm},w]=iv^{\mu}w.
\end{equation}
Assuming, without loss of generality, $g_{\mu\nu}=\eta_{\mu\nu}=diag\left\{1,-1,-1,-1\right\}$ and $v^{\mu}=(1,1,0,0)$, we find that $w=x^0_{cm}-x^1_{cm}=x^-_{cm}$ is the minus lightcone coordinate of the center of mass and, together with the $y_a^{\mu}$s, it can be represented as a multiplicative operator with real spectrum. The other components of $x^{\mu}_{cm}$, the ones perpendicular to $v^{\mu}$, $x^+_{cm}=x^0_{cm}+x^1_{cm}, \,x^2_{cm} $ and $x^3_{cm}$, are irreducibly noncommutative and can be represented as a sum of generators of Lorentz transformations of the $N-1$ coordinates $y_a^{\mu}$ and the generator of dilations of the real line $w$:
\begin{equation} \label{eq:repr}
 x^+_{cm}=2M^{10}+2iw\frac{\partial}{\partial w}+i, \quad x^2_{cm}=M^{12}-M^{02}, \quad x^3_{cm}=M^{13}-M^{03},
\end{equation}
with $M^{\mu\nu}=i\sum_{a=1}^{N-1}\bigg(y^{\mu}_a\eta^{\nu\rho}\frac{\partial}{\partial y_a^{\rho}}-y^{\nu}_a\eta^{\mu\rho}\frac{\partial}{\partial y_a^{\rho}}\bigg)$.

\bigskip

The braided algebra also admits finite-dimensional representations, but from the physical viewpoint it is not clear in which sense they could provide a localization procedure in the commutative limit $v^{\mu}\longrightarrow 0$. Moreover, no finite-dimensional representation of a noncompact Lie algebra like (\ref{eq:cdmcommutators}) is Hermitian, which challenges the interpretation of the $x^{\mu}_a$ operators as coordinates. For these reasons, we will work with the representation in Equation (\ref{eq:repr}).

\subsection{\texorpdfstring{$\kappa$}--Poincaré-invariant \texorpdfstring{$N$}--point functions} \label{sec:transinv}
The most relevant result of \cite{PhysRevD.103.126009} for QFT concerns the $\kappa$-Poincaré invariance of $N$-point functions: these can only depend upon coordinate differences and are therefore commutative. This statement hugely simplifies the interpretational framework of the theory, since, as in the commutative case, all physical information is encoded in a set of commutative $N$-point functions. Given the importance of this result, we reproduce here its proof. In the commutative case, an $N$-point function $f(x_a^{\mu})$ can be written as the N-dimensional Fourier transform of the function $\tilde{f}(k^a_{\mu})$ of momenta as

\begin{equation}
  f(x_a^{\mu})=\int d^4k^1\dots d^4k^N\tilde{f}(k^a_{\mu})e^{i\sum_a k^a_{\mu}x_a^{\mu}} .
\end{equation}
In our scenario, plane waves are replaced by elements of the (component connected to the identity of the) Lie group $\mathcal{G}_{\kappa}^N$ generated by the algebra $\mathcal{A}^{\otimes_{\kappa}^N}$ (in the case of a single coordinate, the group $\mathcal{G}_{\kappa}^1$ has been identified in the literature with the Lie group $AN(3)$ in the 3+1-dimensional case~\cite{Arzano:2010kz}, and with the affine group $Aff(1)$ in 1+1 dimensions). Of course, given the noncommutativity (\ref{eq:braiding}), group elements can be written uniquely once we choose an ordering prescription, i.e. a coordinate system on the group manifold (which we identify with momentum space). All physical predictions of the theory are independent of the ordering choice. The theory we are describing is invariant under general coordinate transformations of momentum space. This has been illustrated quite explicitly in \cite{PhysRevD.103.126009}, where all the main calculations are carried out with two different ordering prescriptions and lead to the same result.
Let us choose, without loss of generality, the following ordering:
\begin{equation}
    e^{ik_{\mu}^1x^{\mu}_1} \dots e^{ik_{\mu}^Nx^{\mu}_N},
\end{equation}
then, a Fourier transformable function can be written as
\begin{equation} \label{eq:four}
  f(x_a^{\mu})=\int d^4k^1\dots d^4k^N\tilde{f}(k^a_{\mu})e^{ik_{\mu}^1x^{\mu}_1} \dots e^{ik_{\mu}^Nx^{\mu}_N}.
\end{equation}
Notice that the form of $\tilde{f}(k^a_{\mu})$ is not unique and depends upon the ordering prescription. It is convenient to express our plane waves in terms of center of mass coordinates: $ e^{ik_{\mu}^ax^{\mu}_a}=e^{ik_{\mu}^a(y^{\mu}_a+x^{\mu}_{cm})}=e^{iq_{\mu}^a y^{\mu}_a}e^{ik_{\mu}^ax^{\mu}_{cm}}$. The last equality, with $q_{\mu}^a$ being a certain function of $k_{\nu}^b$, is a consequence of the Baker-Campbell-Hausdorff formula, given the form of the commutators (\ref{eq:cdmcommutators}). The $x_{cm}^{\mu}$ and $y_a^{\mu}$ generators close two Lie sub-algebras $\mathcal{A}_{cm},$ $\mathcal{B}^N\subset \mathcal{A}^{\otimes_{\kappa}^N}$. The commutation relations (\ref{eq:cdmcommutators}) give $\mathcal{A}^{\otimes_{\kappa}^N}$ the structure of a semidirect product $\mathcal{A}_{cm}\ltimes \mathcal{B}^N$. This, in turn, reflects into an action of the Lie group associated to $\mathcal{A}_{cm}$ on the one associated to $\mathcal{B}^N$: $e^{ik_{\mu}^ax^{\mu}_{cm}}e^{iq_{\mu}^by_b^{\mu}}=e^{i(k^a\vartriangleright q^b)_{\mu}y^{\mu}_b}e^{ik_{\mu}^ax^{\mu}_{cm}}$, where the map $\vartriangleright$ is the left action.
Lastly, because $\mathcal{A}_{cm}$ is a subalgebra, there exists an associative deformed sum of momenta $\boxplus:\mathbb{R}^4\times \mathbb{R}^4 \longrightarrow \mathbb{R}^4$, such that

\begin{equation}
    e^{ip_{\mu}x^{\mu}_{cm}}e^{iq_{\mu}x^{\mu}_{cm}}+e^{i(p\boxplus q)_{\mu}x^{\mu}_{cm}}.
\end{equation}
We can now use these identities to rearrange Equation (\ref{eq:four}):

\begin{equation} \label{eq:fourexp} \begin{aligned}
  &f(x_a^{\mu})=\int d^4k^1\dots d^4k^N\tilde{f}(k^a_{\mu})e^{ik_{\mu}^1x^{\mu}_1} \dots e^{ik_{\mu}^Nx^{\mu}_N}=\\
  &\int d^4k^1\dots d^4k^N\tilde{f}(k^a_{\mu})
  e^{iq_{\mu}^1y^{\mu}_1} e^{i(k^1\vartriangleright q^2)_{\mu}y_2^{\mu}} \dots e^{i(k^1\boxplus k^2 \boxplus \dots \boxplus k^{N-1})\vartriangleright q_{\mu}^Ny_N^{\mu}}e^{i(k^1 \boxplus \dots \boxplus k^{N})_{\mu}x^{\mu}_{cm}}.
\end{aligned}\end{equation}
If we transform our coordinates according to (\ref{eq:transf}), we obtain
\begin{equation}  \begin{aligned}
  &f(x'_a{}^{\mu})=\int d^4k^1\dots d^4k^N\tilde{f}(k^a_{\mu})
  e^{iq_{\mu}^1\Lambda^{\mu}_{\!\ \!\ \nu}y^{\nu}_1} e^{i(k^1\vartriangleright q^2)_{\mu} \Lambda^{\mu}_{\!\ \!\ \nu}y_2^{\nu}} \dots e^{i(k^1\boxplus k^2 \dots \boxplus k^{N-1})\vartriangleright q_{\mu}^N \Lambda^{\mu}_{\!\ \!\ \nu}y_N^{\nu}}e^{ip_{\mu}(\Lambda^{\mu}_{\!\ \!\ \nu}x^{\nu}_{cm}+a^{\mu})}, \\ &\text{where }p^{\mu}=(k^1 \boxplus \dots \boxplus k^{N})^{\mu}.
\end{aligned}\end{equation}
Let us focus on the last exponential, $e^{ip_{\mu}(\Lambda^{\mu}_{\!\ \!\ \nu}x^{\nu}_{cm}+a^{\mu})}$; from Equation (\ref{eq:algebra}) we see that the $\Lambda^{\mu}_{\!\ \!\ \nu}$ close an Abelian subalgebra and their commutators with $a^{\mu}$ give a linear combination of $a^{\mu}$. So, again, using Baker-Campbell-Hausdorff formula we can write $e^{i(k^1 \boxplus \dots \boxplus k^{N})_{\mu}(\Lambda^{\mu}_{\!\ \!\ \nu}x^{\nu}_{cm}+a^{\mu})}=e^{ic^{\nu}_{\!\ \!\ \mu}\Lambda^{\mu}_{\!\ \!\ \nu}}e^{i(k^1 \boxplus \dots \boxplus k^{N})_{\rho}a^{\rho}}$, where $c^{\nu}_{\!\ \!\ \mu}=c^{\nu}_{\!\ \!\ \mu}(x^{\rho}_{cm},p^{\sigma})$ are nonlinear functions of the center of mass coordinates and the momentum $p^{\sigma}$. The only way to have $f(x'_a{}^{\mu})$ independent of $a^{\mu}$ is then to set $\tilde{f}(k^a_{\mu})\propto \delta^4(p)=\delta^4(k^1 \boxplus \dots \boxplus k^{N})$. If we use this condition in Equation (\ref{eq:fourexp}), we see that there is no dependence on the center of mass coordinates.

\subsection{Plane waves}
From now on, we will focus on the 1+1-dimensional case. It is then convenient to write the braided algebra (\ref{eq:braiding}) in terms of lightcone coordinates $x^{\pm}_a=x^0_a\pm x^1_a$:
\begin{equation} \label{eq:1+1alg}
    \begin{aligned}
        &[x^+_a,x^+_b]=2i(x^+_a-x^+_b), \quad [x^+_a,x^-_b]=2ix^-_b, \quad [x^-_a,x^-_b]=0\\
        \Longrightarrow &[x^+_{cm},x^-_{cm}]=2ix^-_{cm}, \quad [x^+_{cm},y_a^{\pm}]=\mp 2i y_a^{\pm}, \quad [x^-_{cm},y_a^{\pm}]=0.
    \end{aligned}
\end{equation}
An ordering choice for single-coordinate plane waves is with $x_a^+$ to the right: $E_a[k]=e^{ik_-x^-_a}e^{ik_+x^+_a}$ (as we said no physical quantity will depend on the ordering choice);  such plane waves are closed under Hermitian conjugation:
\begin{equation}
   E_a^{\dagger}[k]= E_a[S(k)], \quad S(k)=(-e^{2k_+}k_-,-k_+),
\end{equation}
where the map $S: \mathbb{R}^2\longrightarrow \mathbb{R}^2$ is an involution called \textbf{antipode}. The product between two plane waves of the same point can be codified by a \textbf{coproduct} map $\oplus:\mathbb{R}^2\times \mathbb{R}^2 \longrightarrow \mathbb{R}^2$ as
\begin{equation}
    E_a[k]E_a[q]=E_a[k\oplus q], \quad k\oplus q=(k_-+e^{-2k_+}q_-,k_+=q_+).
\end{equation}
From the Lie group axioms it follows that
\begin{equation} \begin{aligned}
    &(k\oplus q)\oplus p=k\oplus(q\oplus p)\equiv k\oplus q\oplus p, \quad k\oplus S(k)=S(k)\oplus k= o, \\
    &S(k\oplus q)=S(k)\oplus S(q), \quad o\oplus k=k\oplus o=k.
\end{aligned} \end{equation}
Here $o=(0,0)$ are the coordinates of the origin of momentum-space, so that $o$ is the neutral element for the coproduct; moreover the identity of the algebra is the plane wave of momentum $o$, $E_a[o]=1$. 

As we saw, to ensure translational invariance, $N$-point functions can only depend upon coordinate differences. If we want to build $\kappa$-Poincaré invariant two-point plane waves, we need to consider products of different-point plane waves $E_1[k]$ and $E_2[q]$, that depend upon the coordinate differences $x_1^{\mu}-x_2^{\mu}$ only. In \cite{PhysRevD.103.126009}, we proved that this happens only for two possible products of plane waves, plus their Hermitian conjugates:
\begin{equation}
    E_1[k]E_2^{\dagger}[k], \quad  E_1^{\dagger}[k]E_2[k], \quad  E_2[k]E_1^{\dagger}[k], \quad E_2^{\dagger}[k]E_1[k].
\end{equation}

The explicit expressions for these two products are given by

\begin{equation} \label{eq:2pplwv} \begin{aligned}
       &E_1[k]E_2^{\dagger}[k]=e^{i\xi_-(x_1^--x^-_2)}e^{i\xi_+(x_1^+-x^+_2)}, \\
        &E_1^{\dagger}[k]E_2[k]=e^{i\chi_-(x_1^--x^-_2)}e^{i\chi_+(x_1^+-x^+_2)},
\end{aligned}
\end{equation}
where
\begin{equation} \label{eq:xidef}
    \xi_-=k_-,\quad \xi_+=\frac{e^{2k_+}-1}{2}, \quad \chi_-=e^{-2k_+}k_-,\quad \chi_+=\frac{e^{-2k_+}-1}{2}.
\end{equation}
The expression for the Hermitian conjugate products can be simply found by swapping the coordinates 1 and 2. 

We notice that the functions $\xi_{\pm}(k_{\pm})$ map $\mathbb{R}^2$ onto the half-plane $\xi_+>-\frac{1}{2}$ (and the same is true for $\chi_{\pm}(k_{\pm})$, with $\chi_+>-\frac{1}{2}$); as we will see, this condition has important consequences which are the main motivations for the present paper.

\subsection{\texorpdfstring{$\kappa$}  --Lorentz transformation of momenta- exiting momentum space}
If we want to construct a $\kappa$-Poincaré invariant field theory, we need to know how plane-wave momenta transform under the $\kappa$-Poincaré coaction on coordinates (\ref{eq:transf}). For this purpose, one evaluates the plane wave on the transformed coordinates, $E_a'[k]\equiv e^{ik_-x_a'^-}e^{ik_+x_a'^+}$. This is computed explicitly in \cite{PhysRevD.103.126009}. The result is
\begin{equation}
    E_a'[k]=E_a[\lambda(k,\Lambda)]\boldsymbol{a}[k],
\end{equation}
where $\boldsymbol{a}[k]=e^{ik_-a^-}e^{ik_+a^+}$ is an ordered plane wave of the translation parameters (with $a^{\pm}=a^0\pm a^1$) and $\lambda$ is a non-linear representation of the Lorentz group:
\begin{equation} \begin{aligned}
    &\lambda:\mathbb{R}^2 \times SO(1,1)\rightarrow \mathbb{R}^2 \, \text{ such that } \\ &\lambda(\lambda(k,\Lambda),\Lambda')=\lambda(k,\Lambda \cdot \Lambda'), \quad \lambda(k,\delta^{\mu}_{\,\,\,\nu})=k, \quad \lambda(o,\Lambda)=o.
\end{aligned} \end{equation}
In 1+1 dimensions, $\lambda$ depends only on the parameter $\omega$, \textit{i.e.} the boost rapidity ($\Lambda^{0}_{\!\ \!\ 0}=\Lambda^{1}_{\!\ \!\ 1}=\cosh(\omega), \, \, \, \Lambda^{1}_{\!\ \!\ 0}=\Lambda^{0}_{\!\ \!\ 1}=\sinh(\omega)$) and it takes the following form:
\begin{equation} \label{eq:kappatransf}
    \lambda(k,\omega)_-=e^{-\omega}k_-, \quad \lambda(k,\omega)_+=\frac{1}{2}\ln[1+e^{\omega}(e^{2k_+}-1)].
\end{equation}
Let us now recall the form of the momenta $\xi_{\pm}$ in terms of $k_{\pm}$ given in Equation (\ref{eq:xidef}); it is simple to verify that the non-linear transformations (\ref{eq:kappatransf}) act like usual linear Lorentz transformations on $\xi_{\pm}$:
\begin{equation}
    \xi_-\rightarrow e^{-\omega}\xi_-, \quad \xi_+\rightarrow e^{+\omega }\xi_+.
\end{equation}
This result could be obtained directly from the transformation of the product $E_1[k]E_2^{\dagger}[k]$:
\begin{equation}
  E'_1[k]E'_2{}^{\dagger}[k]=e^{i\xi_-(x'_1{}^--x'_2{}^-)}e^{i\xi_+(x'_1{}^+-x'_2{}^+)} = e^{i\xi_-e^{-\omega}(x_1^--x_2^-)}e^{i\xi_+e^{+\omega}(x_1^+-x_2^+)} ,
\end{equation}
because coordinate differences are insensitive to the translation part of the $\kappa$-Poincaré transformation.
The momenta $\xi_{\pm}$, then, transform linearly and for a large enough boost we can violate the boundary $\xi_+=-\frac{1}{2}$ and ``exit" momentum space. In Equation (\ref{eq:kappatransf}), this condition corresponds to the argument of the logarithm becoming negative. This happens when
\begin{equation}
    \text{For } k_+<0 \,\,\text{  and  }\,\, e^{\omega}>\frac{1}{1-e^{2k_+}}.
\end{equation}

Even if we know how to impose $\kappa$-Poincaré invariance under infinitesimal transformations, this result seems an obstruction to building a theory that is invariant under finite transformations. We will see that this fact is going to preclude the definition of a consistent Pauli-Jordan function.

\subsection{Mass-shell and \texorpdfstring{$\kappa$}--Klein-Gordon equation} \label{sec:massh}
As noticed in \cite{PhysRevD.103.126009}, since the $\xi_{\pm}$ coordinates transform linearly under $\kappa$-Lorentz, an invariant metric for momentum space is given by
\begin{equation}
    ds^2=d\xi_-d\xi_+=e^{2k_+}dk_-dk_+,
\end{equation}
\textit{i.e.} Minkowski's metric written in light-cone coordinates. Moreover any $\kappa$-Lorentz-invariant function can be expressed in terms of the geodesic distance between the origin $o=(0,0)$ and the point $(\xi_-,\xi_+)$ as
\begin{equation} \label{eq:casimir}
    \tilde C (k)=\xi_-\xi_+=\frac{1}{2}k_-(e^{2k_+}-1).
\end{equation}
This function generalized the notion of mass to the noncommutative case (and reduces to the usual special-relativistic mass at low energies $k_\pm<<1$).
On-shell plane waves are defined by the constraint
\begin{equation}
   \tilde C(k)=m^2 \Longrightarrow k_-=\omega_r(k_+)=\frac{2m^2}{e^{2k_+}-1};
\end{equation}
positve (negative)-frequency mass-shells correspond to $k_+>0$ ($k_+<0$).

We now have all the ingredients to construct the classical field theory of a Klein-Gordon complex scalar field $\phi(x_a)$. The ``$\kappa$-Klein-Gordon" equation is written as
\begin{equation} \label{eq:kleing}
    C \vartriangleright \phi (x_a)=m^2\phi(x_a),
\end{equation}
where we introduced the operator $C : \mathcal A \to \mathcal A$, which acts in the following way on  Fourier-transformable noncommutative functions:
\begin{equation} \label{eq:kkle}
    C \vartriangleright \phi (x_a)=\int d^2k\sqrt{-g(k)}\tilde{\phi}(k) \tilde C(k)E_a[k]=m^2\phi(x_a),
\end{equation}
where$\sqrt{-g(k)}=\frac{1}{2}e^{2k_+}$. The generic solution to (\ref{eq:kkle}) is
\begin{equation} \label{eq:soluzione}\begin{aligned}
    \phi(x_a)&=\int d^2k \sqrt{-g(k)}\delta(\tilde C(k)-m^2)\tilde{\phi}(k)E_a[k] \\
    &=\int d^2k\sqrt{-g(k)}\frac{\delta(k_--\omega_r(k_+))}{\frac{1}{2}|e^{2k_+}-1|}\tilde{\phi}(k)E_a[k].
\end{aligned}\end{equation}
If we split the function $\tilde{\phi}(k)$ into the two mass-shells with positive ($k_+>0$) and negative ($k_+<0$) frequencies,
\begin{equation}
    \tilde{\phi}(k)=a(k_+)\Theta(k_+)+\bar{b}(-k_+)\Theta(-k_+),
\end{equation}
Equation (\ref{eq:soluzione}) becomes
\begin{equation}
    \phi(x_a)=\int_0^{+\infty}dk_+\frac{e^{2k_+}}{|e^{2k_+}-1|}a(k_+)e_a(k_+)+\int_{-\infty}^0 dk_+ \frac{e^{2k_+}}{|e^{2k_+}-1|}\bar{b}(-k_+)e_a(k_+),
\end{equation}
with $e_a(k_+)=E_a[\omega_r(k_+),k_+]$ being on shell plane waves. One can easily prove that $e_a(-k_+)=e_a^{\dagger}(k_+)$ and so our scalar field is written as
\begin{equation}
    \phi(x_a)=\int_0^{+\infty}dk_+\frac{e^{2k_+}}{e^{2k_+}-1}(a(k_+)e_a(k_+) +e^{-2k_+}\bar{b}(k_+)e_a^{\dagger}(k_+)).
\end{equation}

\subsection{Two-point functions} \label{sec:twopoint}
Relying on all the previous results, we can write a two-point function that is a $\kappa$-Poincaré invariant element of the braided algebra $\mathcal{A}^{\otimes^2_{\kappa}}$ that solves the $\kappa$-Klein-Gordon equation. As explained in Section~\ref{sec:transinv}, when we write an $N$-point function in Fourier transform, we need to specify an ordering prescription for different-point plane waves, \textit{i.e.} $E_a(k)$ and $E_b(q)$ for $a \neq b$ (we stress again the no physical quantity will be affected by such prescription). We here choose to put the coordinates of the second point to the right.
In principle, there are two ways to do so as two are the translation-invariant two-point plane waves with this ordering: $E_1[k]E_2^{\dagger}[k]$ and  $E_1^{\dagger}[k]E_2[k]$. However, in \cite{PhysRevD.103.126009} we proved that the two expressions lead to identical results and we here only focus on the first one:
\begin{equation} \begin{aligned} \label{eq:twopoint}
    F(x_1^{\mu}-x_2^{\mu})&=\int d^2k \sqrt{-g(k)}E_1[k]E_2^{\dagger}[k]f(k)\delta(\tilde C(k)-m^2)\\&=\int d^2k \sqrt{-g(k)}E_1[k]E_2^{\dagger}[k]f(k)\frac{\delta(k_--\omega_r(k_+))}{\frac{1}{2}|e^{2k_+}-1|},
\end{aligned}\end{equation}
where $\sqrt{-g(k)}$ makes the integral measure $\kappa$-Lorentz invariant and $f(k)$ is supposed to be a Lorentz-invariant function of the momenta. As in the commutative case, we can assume $f(k)$ to be constant on the forward and backward light-cones in momentum space:
\begin{equation} \label{eq:formaf}
    f(k)=f_-\Theta(-k_+)+f_+\Theta(k_+),
\end{equation}
where of course $f_-$ and $f_+$ are constants. Such functions are $\kappa$-Lorentz invariant because, as we can see from Equation (\ref{eq:kappatransf}), $\kappa$-Lorentz transformations do not change the sign of $k_+$. We recall that our two-point function, however, cannot be invariant under finite boosts unless $f_-=0$, since the backwards light-cone is not closed under $\kappa$-Lorentz transformations. If, in fact, we plug the expression (\ref{eq:formaf}) into Equation (\ref{eq:twopoint}), we find, for the two point function (restoring $\kappa$ for the rest of the section)
\begin{equation} \label{eq:twopointfunction} \begin{aligned}
    F(x_1^{\mu}-x_2^{\mu})&=\frac{1}{2}f_+\int_{-\infty}^{+\infty} \frac{dp}{\sqrt{p^2+m^2}}e^{2i(\sqrt{p^2+m^2}(x_1^0-x_2^0)+p(x_1^1-x_2^1))}\\
    &+\frac{1}{2}f_-\int_{-\infty}^{m \, \sinh(\ln(\frac{\kappa}{2m}))} \frac{dp}{\sqrt{p^2+m^2}}e^{-2i(\sqrt{p^2+m^2}(x_1^0-x_2^0)+p(x_1^1-x_2^1))}.
\end{aligned} \end{equation}
The interested reader can find the explicit calculation in \cite{PhysRevD.103.126009}, or in Appendix~\ref{appendixD}. The first integral in (\ref{eq:twopointfunction}) is the undeformed , commutative, 2-point function, while the second one is clearly non-Lorentz-invariant, because of the integration boundary $\frac{\kappa }{4}-\frac{m^2}{\kappa } = m \, \sinh(\ln(\frac{\kappa}{2m}))$. This issue does not affect the definition of the Wightman function, which is the positive-frequency part of the 2-point function:
\begin{equation} \label{eq:wigh} \begin{aligned}
    \Delta_W(x_1^{\mu}-x_2^{\mu})&=\int d^2k \sqrt{-g(k)}E_1[k]E_2^{\dagger}[k]\Theta(k_+)\delta(\tilde C(k)-m^2)\\
    &=\int_{-\infty}^{+\infty} \frac{dp}{\sqrt{p^2+m^2}}e^{2i(\sqrt{p^2+m^2}(x_1^0-x_2^0)+p(x_1^1-x_2^1))}.
\end{aligned}\end{equation}
However, the non-Lorentz-invariant term in (\ref{eq:twopointfunction}) prevents the construction of the Pauli-Jordan function, \textit{i.e.} the anti-Hermitian part of $\Delta_W$. In fact, there is no way to write
\begin{equation}
    \Delta_W^{\dagger} (x_1^{\mu}-x_2^{\mu})  =\int_{-\infty}^{+\infty} \frac{dp}{\sqrt{p^2+m^2}}e^{-2i(\sqrt{p^2+m^2}(x_1^0-x_2^0)+p(x_1^1-x_2^1))}  \,,
\end{equation}
in terms of our plane waves when the $x_2^{\mu}$ coordinates are put to the right. One could be tempted to write as a workaround
\begin{equation} \label{eq:wigdagger}
\Delta_W^{\dagger}=\int d^2k \sqrt{-g(k)}E_2[k]E_1^{\dagger}[k]\Theta(k_+)\delta(\tilde C(k)-m^2)=\int_0^{+\infty}dk_+\frac{e^{\frac{2k_+}{\kappa}}}{e^{\frac{2k_+}{\kappa}}-1}e_2(k_+)e_1^{\dagger}(k_+);
\end{equation} 
however, as we will prove, we can re-order $x_1^{\mu}$ and $x_2^{\mu}$ in the product $e_2(k_+)e_1^{\dagger}(k_+)$ as
\begin{equation} \label{eq:anticipo}
   e_2(k_+)e_1^{\dagger}(k_+)=e_1\bigg(\frac{1}{2}\ln(2-e^{\frac{2k_+}{\kappa}})\bigg)e_2\bigg(-\frac{1}{2}\ln(2-e^{\frac{2k_+}{\kappa}})\bigg). 
\end{equation}
 Clearly this can only hold for $k_+<\frac{\kappa}{2}\ln(2)$. The integral in Eq.~(\ref{eq:wigdagger}) can thus be divided as follows:
 \begin{equation} \label{eq:nopauli}
\begin{aligned}
 \Delta_W^{\dagger} &= \int_0^{\frac{\kappa}{2} \ln (2)}dk_+\frac{e^\frac{2k_+}{\kappa}}{e^{\frac{2k_+}{\kappa}}-1}e_1\bigg(\frac{\ln(2-e^{\frac{2k_+}{\kappa}})}{2}\bigg)e_2\bigg(-\frac{\ln(2-e^{\frac{2k_+}{\kappa}})}{2}\bigg)
\\&+\int_{\frac{\kappa}{2} \ln (2)}^{+\infty}dk_+\frac{e^{\frac{2k_+}{\kappa}}}{e^{\frac{2k_+}{\kappa}}-1}e_2(k_+)e_1^{\dagger}(k_+)
\\
&= \frac{1}{2}\int_{-\infty}^{\frac{\kappa }{4}-\frac{m^2}{\kappa }} \frac{dp}{\sqrt{p^2+m^2}}e^{-2i(\sqrt{p^2+m^2}(x_1^0-x_2^0)+p(x_1^1-x_2^1))}
+\int_{\frac{\kappa}{2} \ln (2)}^{+\infty}dk_+\frac{e^{\frac{2k_+}{\kappa}}}{e^{\frac{2k_+}{\kappa}}-1}e_2(k_+)e_1^{\dagger}(k_+) \,.
\end{aligned}
\end{equation} 
In the second term above, the plane waves in the product $e_2(k_+)e_1^{\dagger}(k_+)$ are calculated at such momenta ($k_+ \geq \frac{\kappa}{2} \ln (2)$) that there is no way to write them as a product of the form $e_1(p)e_2(q)$. In other terms, there is no element of the group $\mathcal{G}_{\kappa}^N$ that corresponds to the algebraic expression  $e_2(k_+)e_1^{\dagger}(k_+)$ appearing in  the last line of~(\ref{eq:nopauli}). 
As we will show now, actually this holds only for the elements of $\mathcal{G}_{\kappa}^N$  that can be expressed as an exponential with the prescribed ordering, \textit{i.e.} the component of $\mathcal{G}_{\kappa}^N$ that is connected to the identity. The way out of this \textit{empasse} is to bring into the picture the other connected components of the group.

\section{New-type plane waves}
We proved that, already at the classical level, we do not have enough ingredients to construct a field theory on $\kappa$-Minkowski. More specifically, we are not able to write the Pauli-Jordan function for our theory, because of the presence of a non-Lorentz-invariant boundary of momentum space. This issue was known, in one form or the other, for quite some time in the $\kappa$-Minkowski/noncommutative geometry literature~\cite{SMajid_1988,Majid:2006xn,Freidel:2007hk}. In some cases, it has been interpreted as an outright breaking of Lorentz symmetry~\cite{Freidel:2007hk}, and various mechanisms have been proposed to circumvent it~\cite{Arzano:2009ci,Mercati:2018hlc,Arzano:2020jro} (notice however that all of the cited papers are concerned with the \emph{timelike} version of the  $\kappa$-Minkowski algebra, $v^{\mu} v_{\mu} < 0$ in our notation). Of particular interest for us is the recent~\cite{Arzano:2020jro}, which proposes to extend momentum space beyond the border of the coordinate patch that is needed to parametrize plane waves (\emph{i.e.}  the component of the $\mathcal{G}_{\kappa}^1$ Lie group that is connected to the identity). 
They then consider a certain element of the $\kappa$-Minkowski algebra that acts like a certain reflection matrix in a certain finite-dimensional representation of $\mathcal{A}$, \footnote{The representation used in~\cite{Arzano:2020jro} is one of the finite-dimensional representations mentioned at the end of Section~\ref{sec:repr}. These representations cannot be Hermitian (the representation of the associated noncompact Lie group cannot be unitary). The algebra element that realizes the reflection transformation acts in a different way in different representations. In particular, it is idempotent in the representation of~\cite{Arzano:2020jro}, while it behaves differently in other representations. In our case, we have good physical reasons to refer to the Hermitian, infinite-dimensional representation (see Sec.~\ref{sec:repr}).} and sends ordinary plane waves into elements of a component of $\mathcal{G}_{\kappa}^1$ that is not connected to the identity. This component of the group acts like a ``second'' momentum space, which is coordinatized by momenta with a constant imaginary part. This interesting observation inspired us to study with more attention what happens when a plane wave is pushed by a $\kappa$-Lorentz boost through the boundary~$\xi_+ =- 1/2$ of momentum space.

Let us recall how momenta transform under $\kappa$-Lorentz transformations:
\begin{equation}
    \lambda(k,\omega)_-=e^{-\omega}k_-, \quad \lambda(k,\omega)_+=\frac{1}{2}\ln[1+e^{\omega}(e^{2k_+}-1)] \,.
\end{equation}
For $k_+<0 \text{ and } e^{\omega}>\frac{1}{1-e^{2k_+}}$,  the argument of the logarithm in 
$\lambda(k,\omega)_+$ becomes negative. In this case the logarithm, considered as a complex function, becomes multi-valued, with infinite branches spaced at an (imaginary) distance $\pi$ from each other. In formulas:
\begin{equation}
     \lambda(k,\omega)_+=\frac{1}{2}\ln[e^{\omega}(1-e^{2k_+})-1] + i \frac{\pi}{2}+ n i \pi \,.
\end{equation}
Here $n$ is an arbitrary (integer) branch number. Plane waves transform accordingly:
\begin{equation} \label{eq:newplane}
    E_a[k] \longrightarrow \mathcal{E}_a \big{[} e^{-\omega}k_-, {\textstyle \frac{1}{2}} \ln[e^{\omega}(1-e^{2k_+})-1] \big{]} e^{-n\pi x^+_a},
\end{equation}
where we introduced a new type of ``plane waves"
\begin{equation} \label{eq:new2}
    \mathcal{E}_a[q]\equiv E_a[q]e^{-\frac{\pi}{2} x^+_a}=e^{iq_-x^-_a}e^{i(q_++i\frac{\pi}{2})x^+_a},
\end{equation}
which, having a complex phase, look like exponentially damped waves. As we said, these plane waves are elements of the algebra $\mathcal{A}$ (they can be written as a power series of the algebra generators), but they are not obtained from the standard exponential map of the generators. To parametrize all new-type plane waves of the form~(\ref{eq:new2}), we need a coordinate patch $(q_-,q_+)\in \mathbb{R}^2$. As we will discuss in Section~\ref{sec:2pnt}, this coordinate patch covers exactly the ``part" of Minkowski space that is not covered by ordinary plane-wave momenta, thereby restoring the closure of momentum space under Lorentz transformations.
We will prove that using these new-type plane waves together with the ordinary ones, we can write the Hermitian of the Wightman function (\ref{eq:wigh}) and construct the Pauli-Jordan one. We will further prove that no 2-point function depends on the choice of the branch number $n$, so that we can define them uniquely.

As we remarked in Section \ref{sec:repr}, we take the representation (\ref{eq:repr}) to be the definition of our noncommutative coordinates. In 1+1 dimensions, said representation simplifies to
\begin{equation} \begin{aligned}
    &\hat{x}^+_{cm}=2ix^-_{cm}\frac{\partial}{\partial {x^-_{cm}}}+i+2i\sum_{a=1}^{N-1}\bigg(y_a^+\frac{\partial}{\partial y_a^+}-y_a^-\frac{\partial}{\partial y_a^-}\bigg), \quad
  &\hat{x}^-_{cm}=x^-_{cm}, \quad \hat{y}_a^+=y_a^+, \quad \hat{y}_a^-=y_a^-, 
\end{aligned}\end{equation}
so that

\begin{equation} \label{eq:action+}
  e^{i\alpha \hat{x}^+_{cm}}f(x^-_{cm},y_a^+,y_a^-)=e^{-\alpha}f(e^{-2\alpha}x^-_{cm},e^{2\alpha}y_a^+,e^{-2\alpha}y_a^-)  .
\end{equation}
We now want to apply these results to the new operators $e^{-n\pi {x}^+_a}$ and $e^{-\frac{\pi}{2}{x}^+_a}$ , appearing in Equations (\ref{eq:newplane}) and (\ref{eq:new2}). First of all, to isolate the $x^+_{cm}$ coordinate we can use a relation derived in \cite{PhysRevD.103.126009}:
\begin{equation}
    e^{ik_+x_a^+}=e^{ik_+(x^+_{cm}+y_a^+)}=e^{i(\frac{e^{2k_+}-1}{2})y_a^+}e^{ik_+x^+_{cm}}.
\end{equation}
 So, for any value of $y_a^+$, we have $e^{-n\pi {x}^+_a}  =e^{-n\pi {x}^+_{cm}}$, therefore
 \begin{equation} \label{eq:repprx+}
 e^{-n\pi {x}^+_a}f(x^-_{cm},y_a^+,y_a^-)=e^{-in\pi}f(e^{-2in\pi }x^-_{cm},e^{2in\pi }y_a^+,e^{-2in\pi }y_a^-)=(-1)^nf(x^-_{cm},y_a^+,y_a^-).
 \end{equation}
The operator $e^{-n\pi x_a^+}$ is trivial- it simply acts as a phase. We will discuss the role of the $(-1)^n$ in Appendix~\ref{appendixC}, where it is shown that all terms of the kind $(-1)^n$ that appear in our calculations can only have even $n$ and are therefore equal to one.

 The operator $e^{-\frac{\pi}{2}{x}^+_{a}}$ in~(\ref{eq:new2}) can be written as $e^{-\frac{\pi}{2}{x}^+_{a}}=e^{-iy_a^+}e^{-\frac{\pi}{2}{x}^+_{cm}}$ and it acts as a reflection times a $y_a^+$-dependent phase:
\begin{equation}
    \begin{aligned} \label{eq:operatorrefl}
    e^{-\frac{\pi}{2}\hat{x}^+_a}f(x^-_{cm},y_a^+,y_a^-)=e^{-iy_a^+}e^{-i\frac{\pi}{2}}f(e^{-i\pi }x^-_{cm},e^{i\pi }y_a^+,e^{-i\pi }y_a^-)=ie^{-iy_a^+}f(-x^-_{cm},-y_a^+,-y_a^-).
    \end{aligned}
\end{equation}
This is the operator that, when multiplied by an ordinary plane wave $E_a[k]$, turns it into a new-type plane wave $\mathcal{E}_a[k]$.
\subsection{Two-point new-type plane waves} \label{sec:2pnt}
Equipped with our new-type plane waves $\mathcal{E}_a[q]\equiv E_a[q]e^{-\frac{\pi}{2} x^+_a}$, we can briefly retrace the steps of the analysis of~\cite{PhysRevD.103.126009}, highlighting similarities and differences. When we now write a function of $N$ points in Fourier transform, we decompose it in ``usual" Fourier modes (corresponding to ordinary plane waves) plus ``new" Fourier modes (clearly corresponding to new-type plane waves), so that it can be invariant under finite $\kappa$-Lorentz transformations:
\begin{equation} \label{eq:fournew}
  f(x_a^{\mu})=\int d^4k^1\dots d^4k^N\tilde{f}(k^a_{\mu})E_1[k^1] \dots E_N[k^N]+\int d^4q^1\dots d^4q^N\breve{f}(q^a_{\mu})\mathcal{E}_1[k^1] \dots \mathcal{E}_N[k^N].
\end{equation}
In the second integral, we are essentially shifting the $k_+$ of ordinary plane waves by a factor $i\pi/2$. This clearly does not affect the results of Section \ref{sec:transinv}: $N$-point functions can only depend on coordinate differences. We need to understand which products involving new-type plane waves satisfy this request. There are three possibilities to consider:
\begin{equation} \label{eq:products}
    E_1[k]\mathcal{E}_2[q], \quad \mathcal{E}_1[k]E_2[q], \quad \mathcal{E}_1[k]\mathcal{E}_2[q],
\end{equation}
but they all reduce either to the ordinary products in Equation (\ref{eq:2pplwv}) or to
\begin{equation} \label{eq:2pnplwv}
    \begin{aligned}
    &E_1[(k_-,k_++i\pi/2)]E_2^{\dagger}[(k_-,k_++i\pi/2)]=\mathcal{E}_1[k]\mathcal{E}_2^{\dagger}[k]= e^{i\eta_-(x_1^--x_2^-)}e^{i\eta_+(x_1^+-x_2^+)},\\
&E^{\dagger}_1[(k_-,k_++i\pi/2)]E_2[(k_-,k_++i\pi/2)]=\mathcal{E}_1^{\dagger}[k]\mathcal{E}_2[k]=e^{i\rho_-(x_1^--x_2^-)}e^{i\rho_+(x_1^+-x_2^+)},
    \end{aligned}
\end{equation}
where
\begin{equation}
    \eta_-=k_-, \quad \eta_+=-\frac{e^{2k_+}+1}{2}, \quad \rho_-=e^{2k_+}k_-, \quad \rho_+=-\frac{e^{-2k_+}+1}{2}.
\end{equation}
The calculations for the first product are reported in detail in Appendix~\ref{appendixA} and one can proceed analogously for the other two products.

In order to better understand the meaning of the coordinates $\eta_{\pm}$ introduced in (\ref{eq:2pnplwv}), we want to determine how new-type plane waves transform under $\kappa$-Poincaré. Recall that $\mathcal{E}_a[k]=E_a[k_{\mu}+i\frac{\pi}{2}\delta^+_{\mu}]$ and $E_a'[k]=E_a[e^{-\omega}k_-,\frac{1}{2}\ln[1+e^{\omega}(e^{2k_+}-1)]]\boldsymbol{a}[k]$. This result only relies on the transformation rule (\ref{eq:transf}) and on the algebra (\ref{eq:1+1alg}) and it holds true for any value of $k_+$, real or complex. For this reason, new-type plane waves transform as follows:
\begin{equation}\label{eq:transfnewplanewaves}
    \begin{aligned}
    \mathcal{E}'_a[k]&=\mathcal{E}_a[{\lambda}^*(k,\omega)]\boldsymbol{a}[k]e^{-\frac{\pi}{2}a^+}\\
    {\lambda}^*(k,\omega)&=\bigg(e^{-\omega}k_-,\frac{1}{2}\ln[e^{\omega}(e^{2k_+}+1)-1]\bigg).
    \end{aligned}
\end{equation}
We notice that for both positive and negative values of $k_+$, if $e^{\omega}<\frac{1}{e^{2k_+}+1}$, the argument of the logarithm becomes negative and new-type plane waves transform into ordinary ones. Now, using Equation (\ref{eq:transfnewplanewaves}), we can simply verify that the $\eta_{\pm}$ are linear coordinates on momentum space, \textit{i.e.} they transform linearly under $\kappa$-Poincaré, just like the $\xi_{\pm}$ of Equation~(\ref{eq:xidef}). 
We also notice that $\eta_+<-1/2$. As expected, new-type plane waves cover the second half of momentum space.

To determine how the on-shell condition reads for our new-type plane waves we can proceed as in Section \ref{sec:massh}. The $\kappa$-invariant line element for the second half of momentum space is given by

\begin{equation}
    ds^2=d\eta_-d\eta_+=-e^{2k_+}dk_-dk_+.
\end{equation}
We notice that the determinant of the metric is the same as in the first half of momentum space: $\sqrt{-g(k)}=e^{2k_+}/2$. We then have the following deformed mass:
\begin{equation}
    \breve{C}(k)=\eta_-\eta_+=-\frac{1}{2}k_-(e^{2k_+}+1).
\end{equation}
This is essentially the same as $\tilde C$ in Equation (\ref{eq:casimir}), apart from having sent $k_+\rightarrow k_++i\pi/2$. On-shell new-type plane waves, then, take the form
\begin{equation}
    \epsilon_a(k_+)=\mathcal{E}_a[k_+,\Omega_r(k_+)], \text{ where } \Omega_r(k_+)=-\frac{2m^2}{e^{2k_+}+1} .
\end{equation}

\subsection{\texorpdfstring{$\kappa$}--Klein Gordon equation}
Let us consider again the $\kappa$-Klein Gordon equation~(\ref{eq:kleing}), given that now we decompose the field $\phi(x_a)$ both in ``usual" and ``new" Fourier modes. As we saw, different halves of momentum space have a different expression for the Casimir: if $f(x_a)$ is a Fourier-transformable function,

\begin{equation}
   f(x_a)=\int d^2k\sqrt{-g(k)}\bigg(\tilde{f}(k)E_a[k]+\breve{f}(k)\mathcal{E}_a[k]\bigg), 
\end{equation}
we  extend the action of the Casimir operator $C$ onto the new-type plane waves as follows:
\begin{equation}\begin{aligned}
C \rhd f(x_a)&=\int d^2k\sqrt{-g(k)} \bigg (\tilde{f}(k)\tilde{C}(k)E_a[k]+\breve{f}(k)\breve{C}(k)\mathcal{E}_a[k] \bigg).\end{aligned}
\end{equation}

The solution to the $\kappa$-Klein Gordon equation can be written as

\begin{equation}\begin{aligned}
\phi(x_a)&=\int d^2k\sqrt{-g(k)}(\delta(\tilde C(k)-m^2)\tilde{\phi}(k)E_a[k]+\delta(\breve{C}(k)-m^2) \breve{\phi}(k)\mathcal{E}_a[k])\\
&=\int d^2k\sqrt{-g(k)}2\Big(\frac{\delta(k_--\omega_r(k_+))}{|e^{2k_+}-1|}\tilde{\phi}(k)E_a[k]+\frac{\delta(k_--\Omega_r(k_+))}{|e^{2k_+}+1|}\breve{\phi}(k)\mathcal{E}_a[k]\Big).
\end{aligned}\end{equation}

We can now split $\tilde{\phi}(k)$ e $\breve{\phi}(k)$ in positive and negative frequency modes:

\begin{equation}
    \begin{aligned}
    \tilde{\phi}(k)&=a(k_+)\theta(k_+)+b^*(-k_+)\theta(-k_+), \quad \breve{\phi}(k)=\alpha(k_+)\theta(k_+)+\beta^*(-k_+)\theta(-k_+) \Longrightarrow\\
\phi(x_a)&=\int_0^{+\infty}dk_+e^{2k_+}\bigg(\frac{a(k_+)e_a(k_+)}{e^{2k_+}-1}+\frac{\alpha(k_+)\epsilon_a(k_+)}{e^{2k_+}+1}\bigg)\\&+\int_{-\infty}^0dk_+e^{2k_+}\bigg(\frac{b^*(-k_+)e_a(k_+)}{1-e^{2k_+}}+\frac{\beta^*(-k_+)\epsilon_a(k_+)}{e^{2k_+}+1}\bigg),
    \end{aligned}
\end{equation}

where we recall that $e_a(k_+)=E_a[\omega_r(k_+),k_+]$ and  $\epsilon_a(k_+)=\mathcal{E}_a[\Omega_r(k_+),k_+]$. As we know, $e_a(-k_+)=e_a^{\dagger}(k_+)$ and the same is true for $\epsilon_a(-k_+)$ (see Appendix~\ref{appendixB}).

Sending $k_+\longrightarrow -k_+$ in the second integral, we find
\begin{equation} \label{eq:fields}
    \begin{aligned}
    \phi(x_a)&=\int_0^{+\infty}dk_+ \frac{e^{2k_+}}{e^{2k_+}-1}(a(k_+)e_a(k_+)+e^{-2k_+}b^*(k_+)e^{\dagger}_a(k_+))\\
    &+\int_0^{+\infty}dk_+\frac{e^{2k_+}}{e^{2k_+}+1}(\alpha(k_+)\epsilon_a(k_+)+e^{-2k_+}\beta^*(k_+)\epsilon^{\dagger}_a(k_+)).
    \end{aligned}
\end{equation}

Before discussing two-point functions, we report a useful relation on how to commute on-shell plane waves of different points (the derivation can be found in Appendix~\ref{appendixC}):
\begin{equation}\label{eq:scambio}\begin{aligned}
    E_1[k]E_2[q]=&E_2\bigg[\frac{1}{2}\ln(e^{2k_+}(e^{2q_+}-1)+1),e^{-2k_+}q_-\bigg]\\
    &E_1\bigg[k_++q_+-\frac{1}{2}\ln(e^{2k_+}(e^{2q_+}-1)+1),(e^{2k_+}(e^{2q_+}-1)+1)k_-\bigg].
\end{aligned}\end{equation}
Here, $k$ and $q$ can be any complex numbers, so that the result applies also to new-type plane waves. One can simply verify that if we commute two on-shell plane waves, we obtain again on-shell plane waves.

\subsection{Pauli-Jordan function}
We can now extend the two-point function of Section \ref{sec:twopoint}, using the new-type plane waves of Equation (\ref{eq:2pnplwv}):
\begin{equation}\label{eq:newtwopoint} \begin{aligned}
    &\Delta(x_1-x_2)=F(x_1-x_2)+H(x_1-x_2),\\
    &F(x_1-x_2)=\int d^2k\sqrt{-g(k)}E_1[k]E_2^{\dagger}[k]f(k)\delta(\tilde C(k)-m^2),\\
    &H(x_1-x_2)=\int d^2k \sqrt{-g(k)}\mathcal E_1[k] \mathcal E_2^{\dagger}[k]h(k)\delta(\breve{C}(k)-m^2).
\end{aligned}\end{equation}
As we did for $f(k)$ in Section \ref{sec:twopoint}, we can assume $h(k)$ to be constant on the forward and backward light-cones: 
\begin{equation}\label{eq:backfor}
    h(k)=h_-\Theta(-k_+)+h_+\Theta(k_+).
\end{equation}
Under this condition (and restoring $\kappa$), $H(x_1-x_2)$ simplifies to
\begin{equation} \begin{aligned}
    H(x_1-x_2)&=\frac{1}{2}h_-\int_{m\sinh(\ln(\frac{\kappa}{2m}))}^{m\sinh(\ln(\frac{\kappa}{m}))}\frac{dp}{\sqrt{p^2+m^2}}e^{-2i[\sqrt{p^2+m^2}(x_1^0-x_2^0)+p(x_1^1-x_2^1)]} + \\
    &+\frac{1}{2}h_+\int_{m\sinh(\ln(\frac{\kappa}{m}))}^{+\infty}\frac{dp}{\sqrt{p^2+m^2}}e^{-2i[\sqrt{p^2+m^2}(x_1^0-x_2^0)+p(x_1^1-x_2^1)]}.
\end{aligned}\end{equation}
The calculation is reported in Appendix~\ref{appendixD}, where we also comment on the possibility to use the other invariant two-point plane waves, like $E_1^{\dagger}[k]E_2[k]$ and $\mathcal E_1^{\dagger}[k]\mathcal E_2[k]$, which turns out to be redundant. We immediately notice that for $h_-=h_+$, we get an integral that is complementary to the second integral of (\ref{eq:twopointfunction}). If we put $f_+=2A$ and $h_-=h_+=f_-=2B$, the two-point function becomes the same as in the commutative case:

\begin{equation}\label{eq:def2point}
    \begin{aligned}
  &\Delta (x_1-x_2)=A\int d^2k \sqrt{-g(k)}\Theta(k_+)\delta(\tilde C[k]-m^2)E_1[k]E_2^{\dagger}[k]+  \\
  &+B\int d^2k\sqrt{-g(k)}(\Theta(-k_+)\delta(\tilde C[k]-m^2)E_1[k]E_2^{\dagger}[k]+\delta(\breve{C}[k]-m^2)\mathcal E_1[k]\mathcal E_2^{\dagger}[k])=\\
  &=A\int_{-\infty}^{+\infty}\frac{dp}{\sqrt{p^2+m^2}}e^{2i[\sqrt{p^2+m^2}(x_1^0-x_2^0)+p(x_1^1-x_2^1)]}+B\int_{-\infty}^{+\infty}\frac{dp}{\sqrt{p^2+m^2}}e^{-2i[\sqrt{p^2+m^2}(x_1^0-x_2^0)+p(x_1^1-x_2^1)]}
    \end{aligned}
\end{equation}

The choice $A=1$ and $B=0$ corresponds to the Wightman function $\Delta_W$ defined in Equation (\ref{eq:wigh}).
We can now write the Pauli-Jordan function as its anti-Hermitian part:
\begin{equation}\label{eq:PauliJ}\begin{aligned}
   \Delta_{\text{PJ}}&=\int d^2k \sqrt{-g(k)}\Theta(k_+)\delta(\tilde C[k]-m^2)(E_1[k]E_2^{\dagger}[k]-E_2[k]E_1^{\dagger}[k])= \\
   &=\int_0^{+\infty}dk_+\frac{e^{2k_+}}{e^{2k_+}-1}(e_1(k_+)e_2^{\dagger}(k_+)-e_2(k_+)e_1^{\dagger}(k_+))
\end{aligned}\end{equation}

Using the property (\ref{eq:scambio}) and as anticipated in Equation~(\ref{eq:anticipo}), in the region $k_+\in \big]0,\frac{1}{2}\ln(2)\big[$, we can write
\begin{equation}
   e_2(k_+)e_1^{\dagger}(k_+)=e_1\bigg(\frac{1}{2}\ln(2-e^{2k_+})\bigg)e_2\bigg(-\frac{1}{2}\ln(2-e^{2k_+})\bigg). 
\end{equation}
We define, for this region, the change of variables $k'_+=\frac{1}{2}\ln(2-e^{2k_+}), \,\,k'_+\in]-\infty,0[$. Analogously, in the region $k_+>\frac{1}{2}\ln(2)$, if we commute $ e_2(k_+)e_1^{\dagger}(k_+)$, we obtain a product of new-type plane waves:

\begin{equation}
    e_2(k_+)e_1^{\dagger}(k_+)=-\epsilon_1\bigg(\frac{1}{2}\ln(e^{2k_+}-2)\bigg)\epsilon_2\bigg(-\frac{1}{2}\ln(e^{2k_+}-2)\bigg).
\end{equation}

We can define the change of variables $k''_+=\frac{1}{2}\ln(e^{2k_+}-2)$, but this time $k''_+\in]-\infty,+\infty[$. So if we perform these changes of variables in the integral (\ref{eq:PauliJ}) and we send  $k''_+\longrightarrow -k''_+$ in the integral between $-\infty$ and $0$, the Pauli-Jordan function simplifies to 
\begin{equation} \label{eq:PJ}
    \begin{aligned}
    \Delta_{\text{PJ}}(x_1-x_2)&=\int_0^{+\infty}dk_+\frac{e^{2k_+}}{e^{2k_+}-1}e_1(k_+)e_2^{\dagger}(k_+)-\int_0^{+\infty} dk_+ \frac{1}{e^{2k_+}-1}e_1^{\dagger}(k_+)e_2(k_+)+\\
    &-\int_0^{+\infty}dk_+\frac{e^{2k_+}}{e^{2k_+}+1}\epsilon_1(k_+)\epsilon_2^{\dagger}(k_+)-\int_0^{+\infty} dk_+ \frac{1}{e^{2k_+}+1}\epsilon_1^{\dagger}(k_+)\epsilon_2(k_+),
    \end{aligned}
\end{equation}

which corresponds to the choice $A=-B=1$ in the two point function (\ref{eq:def2point}). As we expected, having extended momentum space beyond its non-Lorentz-invariant boundary, we are now able to write the Pauli-Jordan function of our theory. In particular, the algebraic expression $e_2(k_+)e_1^{\dagger}(k_+)$, in the region $k_+>\frac 1 2\ln(2)$, corresponds to an element of $\mathcal{G}_{\kappa}^2$ which can be obtained through multiplication by the reflection operator $e^{-\frac{\pi}{2}x_a^+}$. This element, therefore, does not belong to the component of $\mathcal{G}_{\kappa}^2$ connected to the identity, and it is parametrized by our new-type plane waves with the chosen ordering ($x_2^+$ to-the-right).

\section{Field quantization} \label{sec:algebra}
The Pauli-Jordan function (\ref{eq:PJ}) can be used to define a quantization as follows:
\begin{equation} \label{eq:commrel}
    [\hat{\phi}(x_1),\hat{\phi}^{\dagger}(x_2)]=i \Delta_{\text{PJ}}(x_1-x_2), \quad [\hat{\phi}(x_1),\hat{\phi}(x_2)]=[\hat{\phi}^{\dagger}(x_1),\hat{\phi}^{\dagger}(x_2)]=0.
\end{equation} 
In the commutative case, such a quantization maps a classical field $\phi\in\mathbb{C}[\text{Minkowski}]$ to a quantum field $\hat{\phi}$ which is an operator-valued function of Minkowski spacetime. In our noncommutative case, the quantization is a map from the algebra $\mathcal{A}$ to the tensor product between $\mathcal{A}$ and the algebra of operators on a Hilbert space $\mathcal{H}$:
\begin{equation}
    \mathcal{A}\longrightarrow \mathcal{A}\otimes Op[\mathcal{H}].
\end{equation}

This means that the quantized fields $\hat{\phi}(x_a)$ are those given in (\ref{eq:fields}), in which, however, the Fourier coefficients of on-shell plane waves are promoted to operators, which commute with the $x_a$:

\begin{equation}
    \begin{aligned}
    \hat{\phi}(x_a)&=\int_0^{+\infty}dk_+\frac{e^{2k_+}}{e^{2k_+}-1}(\hat{a}(k_+)e_a(k_+)+e^{-2k_+}\hat{b}^{\dagger}(k_+)e_a^{\dagger}(k_+))\\
    &+\int_0^{+\infty}dk_+\frac{e^{2k_+}}{e^{2k_+}+1}(\hat{\alpha}(k_+)\epsilon_a(k_+)+e^{-2k_+}\hat{\beta}^{\dagger}(k_+)\epsilon_a^{\dagger}(k_+)).
    \end{aligned}
\end{equation}
The Hermitian conjugate fields will be

\begin{equation}
    \begin{aligned}
   \hat{\phi}^{\dagger}(x_a)&=\int_0^{+\infty}dk_+\frac{e^{2k_+}}{e^{2k_+}-1}(\hat{a}^{\dagger}(k_+)e_a^{\dagger}(k_+)+e^{-2k_+}\hat{b}(k_+)e_a(k_+))\\
   &+\int_0^{+\infty}dk_+\frac{e^{2k_+}}{e^{2k_+}+1}(\hat{\alpha}^{\dagger}(k_+)\epsilon^{\dagger}_a(k_+)+e^{-2k_+}\hat{\beta}(k_+)\epsilon_a(k_+)).
    \end{aligned}
\end{equation}
In Appendix~\ref{appendixE} we show how to deduce the algebra of these Fourier coefficients from the commutation relations (\ref{eq:commrel}).
Here we report the final results of the calculation.  There are $36$ independent commutation relations between the eight annihilation and creation operators $\hat a$, $\hat b$, $\hat \alpha$,  $\hat \beta$, $\hat a^\dagger$, $\hat b^\dagger$, $\hat \alpha^\dagger$ and  $\hat \beta^\dagger$. Moreover, the form of the commutation relations might  depend on the values of the (on-shell) momenta at which we are evaluating the two operators. Let us call $k_+$ the momentum of the operator on the left and $q_+$ the other one. The form of the commutation relations will depend on which region of the quadrant $(k_+,q_+) \in \mathbb{R}^2_+$ we are considering. The $36$ commutation relations divide into eight groups, which partition the $(k_+,q_+)$ quadrant in different ways. 

\subsection*{Products of the form \texorpdfstring{$\hat a (k_+) \, \hat a^\dagger (q_+)$,  $\hat a (k_+) \, \hat b^\dagger (q_+)$}- and  \texorpdfstring{$\hat b (k_+)\, \hat a^\dagger (q_+)$}}

\begin{figure}[ht]
	\centering 	\includegraphics[width=0.45\textwidth]{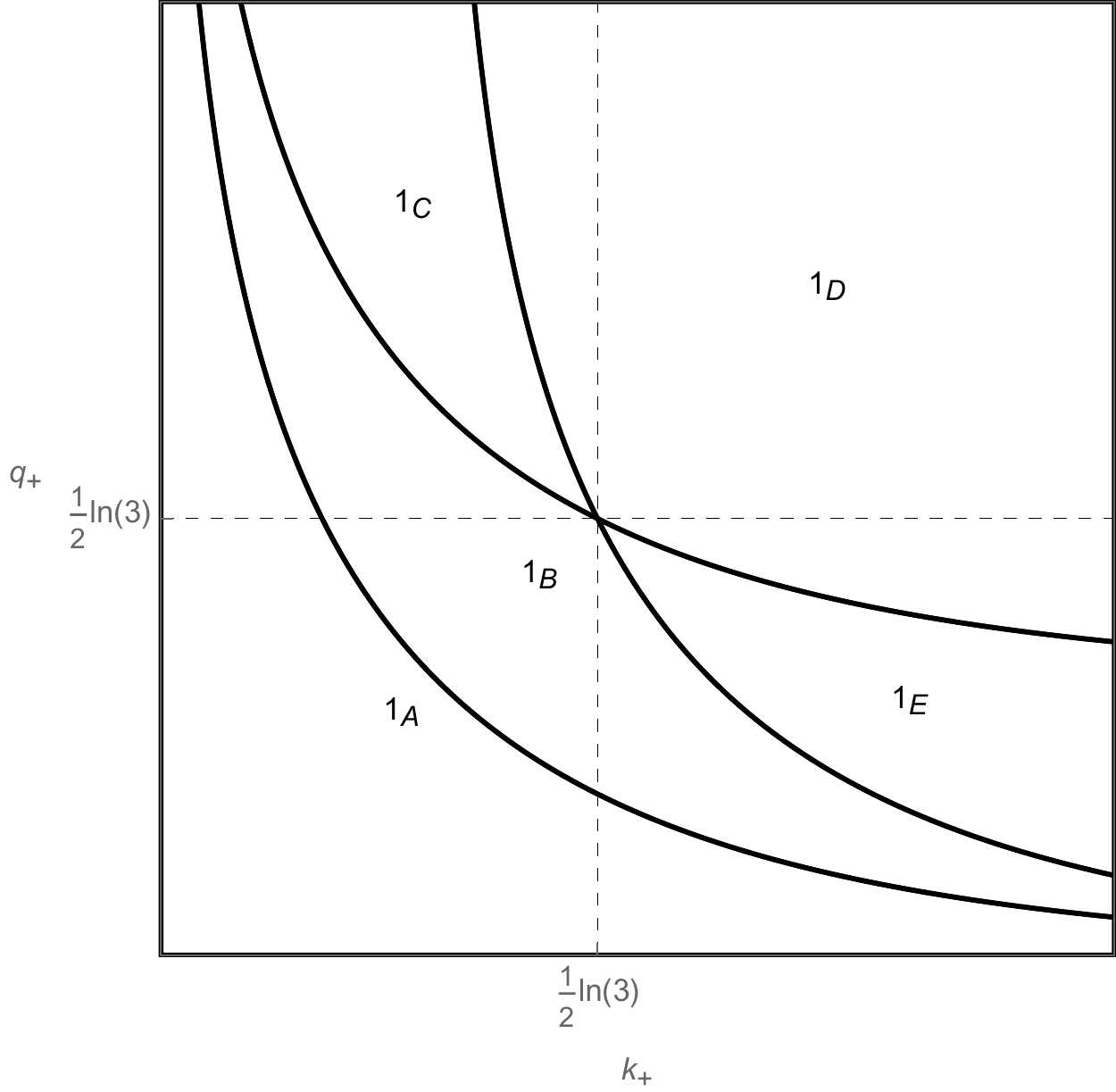}
	\caption{The products of the form $\hat a (k_+) \, \hat a^\dagger (q_+)$,  $\hat a (k_+) \, \hat b^\dagger (q_+)$ and $\hat b (k_+)\, \hat a^\dagger (q_+)$ take different forms in the different regions shown here. These regions correspond to the following integrals in Appendix~\ref{appendixE}:  $1_A\rightarrow I_2^{(i)}$,  $1_B\rightarrow J_2^{(i)}$,  $1_C\rightarrow J_6^{(i)}$, $1_D \rightarrow J_8^{(i)}$ and $1_E\rightarrow J_4^{(i)}$.\label{e1e2dagger}}
\end{figure}

The commutation relations that involve these products take five different forms, according to the region of $\mathbb{R}^2_+$ to which the two momenta $k_+$ and $q_+$ belong. The five regions are represented in Fig.~\ref{e1e2dagger}.

\begin{itemize}
    \item Region $1_A$: $k_+\in]0,+\infty[$ and $ q_+\in\big]0,-\frac{1}{2}\ln(1-e^{-2k_+})\big[$ 
\begin{equation} \label{eq:prima}
    \begin{aligned}
     &\hat{a}(k_+)\hat{a}^{\dagger}(q_+)-\frac{1}{e^{2k_+}+e^{2q_+}-e^{2(k_++q_+)}}\hat{a}^{\dagger}\bigg(\frac{1}{2}\ln\bigg[\frac{e^{2q_+}}{e^{2k_+}+e^{2q_+}-e^{2(k_++q_+)}}\bigg]\bigg)\\&\hat{a}\bigg(\frac{1}{2}\ln\bigg[\frac{e^{2k_+}}{e^{2k_+}+e^{2q_+}-e^{2(k_++q_+)}}\bigg]\bigg)
     =i\delta(q_+-k_+)\frac{(e^{2k_+}-1)}{e^{2k_+}},\\
     &\hat{a}(k_+)\hat{b}^{\dagger}(q_+)=\hat{b}^{\dagger}\bigg(\frac{1}{2}\ln\bigg[\frac{e^{2q_+}}{e^{2k_+}+e^{2q_+}-e^{2(k_++q_+)}}\bigg]\bigg) \hat{a}\bigg(\frac{1}{2}\ln\bigg[\frac{e^{2k_+}}{e^{2k_+}+e^{2q_+}-e^{2(k_++q_+)}}\bigg]\bigg),\\
     &\hat{b}(k_+)\hat{a}^{\dagger}(q_+)=\hat{a}^{\dagger}\bigg(\frac{1}{2}\ln\bigg[\frac{e^{2q_+}}{e^{2k_+}+e^{2q_+}-e^{2(k_++q_+)}}\bigg]\bigg) \hat{b}\bigg(\frac{1}{2}\ln\bigg[\frac{e^{2k_+}}{e^{2k_+}+e^{2q_+}-e^{2(k_++q_+)}}\bigg]\bigg).
    \end{aligned}
\end{equation}
    \item  Region $1_B$: $k_+\in\big]0,\frac{1}{2}\ln(3)\big[$ and $q_+\in\big]-\frac{1}{2}\ln(1-e^{-2k_+}),-\frac{1}{2}\ln(\frac{1-e^{-2k_+}}{2})\big[$ and for $k_+\in\big]\frac{1}{2}\ln(3),+\infty\big[$ and $ q_+\in\big]-\frac{1}{2}\ln(1-e^{-2k_+}),-\frac{1}{2}\ln(1-2e^{-2k_+})\big[$
\begin{equation}
    \begin{aligned}
    &\hat{a}(k_+)\hat{a}^{\dagger}(q_+)+\frac{1}{e^{2(k_++q_+)}-e^{2k_+}-e^{2q_+}}\hat{\alpha}^{\dagger}\bigg(\frac{1}{2}\ln\bigg[\frac{e^{2q_+}}{e^{2(k_++q_+)}-e^{2k_+}-e^{2q_+}}\bigg]\bigg)\\&\hat{\alpha}\bigg(\frac{1}{2}\ln\bigg[\frac{e^{2k_+}}{e^{2(k_++q_+)}-e^{2k_+}-e^{2q_+}}\bigg]\bigg) 
    =i\delta(q_+-k_+)\frac{(e^{2k_+}-1)}{e^{2k_+}},\\
    &\hat{a}(k_+)\hat{b}^{\dagger}(q_+)=-\hat{\beta}^{\dagger}\bigg(\frac{1}{2}\ln\bigg[\frac{e^{2q_+}}{e^{2(k_++q_+)}-e^{2k_+}-e^{2q_+}}\bigg]\bigg) \hat{\alpha}\bigg(\frac{1}{2}\ln\bigg[\frac{e^{2k_+}}{e^{2(k_++q_+)}-e^{2k_+}-e^{2q_+}}\bigg]\bigg),\\
    &\hat{b}(k_+)\hat{a}^{\dagger}(q_+)=-\hat{\alpha}^{\dagger}\bigg(\frac{1}{2}\ln\bigg[\frac{e^{2q_+}}{e^{2(k_++q_+)}-e^{2k_+}-e^{2q_+}}\bigg]\bigg) \hat{\beta}\bigg(\frac{1}{2}\ln\bigg[\frac{e^{2k_+}}{e^{2(k_++q_+)}-e^{2k_+}-e^{2q_+}}\bigg]\bigg).
    \end{aligned}
\end{equation}
    \item Region $1_C$: $k_+\in\big]0,\frac{1}{2}\ln(2)\big[$ and $ q_+\in\big]-\frac{1}{2}\ln(\frac{1-e^{-2k_+}}{2}),+\infty\big[$ and for $k_+\in\big]\frac{1}{2}\ln(2),\frac{1}{2}\ln(3)\big[$ and $ q_+\in\big]-\frac{1}{2}\ln(\frac{1-e^{-2k_+}}{2}),-\frac{1}{2}\ln(1-2e^{-2k_+})\big[$ 
\begin{equation}
    \begin{aligned} 
     &\hat{a}(k_+)\hat{a}^{\dagger}(q_+)+\frac{1}{e^{2(k_++q_+)}-e^{2k_+}-e^{2q_+}}\hat{\alpha}^{\dagger}\bigg(\frac{1}{2}\ln\bigg[\frac{e^{2q_+}}{e^{2(k_++q_+)}-e^{2k_+}-e^{2q_+}}\bigg]\bigg)\\&\hat{\beta}^{\dagger}\bigg(-\frac{1}{2}\ln\bigg[\frac{e^{2k_+}}{e^{2(k_++q_+)}-e^{2k_+}-e^{2q_+}}\bigg]\bigg)=0,\\
     &\hat{a}(k_+)\hat{b}^{\dagger}(q_+)=-\hat{\beta}^{\dagger}\bigg(\frac{1}{2}\ln\bigg[\frac{e^{2q_+}}{e^{2(k_++q_+)}-e^{2k_+}-e^{2q_+}}\bigg]\bigg) \hat{\beta}^{\dagger}\bigg(-\frac{1}{2}\ln\bigg[\frac{e^{2k_+}}{e^{2(k_++q_+)}-e^{2k_+}-e^{2q_+}}\bigg]\bigg),\\
     &\hat{b}(k_+)\hat{a}^{\dagger}(q_+)=-\hat{\alpha}^{\dagger}\bigg(\frac{1}{2}\ln\bigg[\frac{e^{2q_+}}{e^{2(k_++q_+)}-e^{2k_+}-e^{2q_+}}\bigg]\bigg) \hat{\alpha}^{\dagger}\bigg(-\frac{1}{2}\ln\bigg[\frac{e^{2k_+}}{e^{2(k_++q_+)}-e^{2k_+}-e^{2q_+}}\bigg]\bigg).
    \end{aligned}
\end{equation}
    \item Region $1_D$: $k_+\in\big]\frac{1}{2}\ln(2),\frac{1}{2}\ln(3)\big[$ and $ q_+\in\big]-\frac{1}{2}\ln(1-2e^{-2k_+}),+\infty\big[$ and for $k_+\in\big]\frac{1}{2}\ln(3),+\infty\big[$ and $ q_+\in\big]-\frac{1}{2}\ln(\frac{1-e^{-2k_+}}{2}),+\infty\big[$ 
\begin{equation}
    \begin{aligned}
     &\hat{a}(k_+)\hat{a}^{\dagger}(q_+)+\frac{1}{e^{2(k_++q_+)}-e^{2k_+}-e^{2q_+}}\hat{\beta}\bigg(-\frac{1}{2}\ln\bigg[\frac{e^{2q_+}}{e^{2(k_++q_+)}-e^{2k_+}-e^{2q_+}}\bigg]\bigg)\\&\hat{\beta}^{\dagger}\bigg(-\frac{1}{2}\ln\bigg[\frac{e^{2k_+}}{e^{2(k_++q_+)}-e^{2k_+}-e^{2q_+}}\bigg]\bigg)=i \delta (q_+-k_+)\frac{e^{2k_+}-1}{e^{2k_+}},\\
     &\hat{a}(k_+)\hat{b}^{\dagger}(q_+)=-\hat{\alpha}\bigg(-\frac{1}{2}\ln\bigg[\frac{e^{2q_+}}{e^{2(k_++q_+)}-e^{2k_+}-e^{2q_+}}\bigg]\bigg) \hat{\beta}^{\dagger}\bigg(-\frac{1}{2}\ln\bigg[\frac{e^{2k_+}}{e^{2(k_++q_+)}-e^{2k_+}-e^{2q_+}}\bigg]\bigg),\\
     &\hat{b}(k_+)\hat{a}^{\dagger}(q_+)=-\hat{\beta}\bigg(-\frac{1}{2}\ln\bigg[\frac{e^{2q_+}}{e^{2(k_++q_+)}-e^{2k_+}-e^{2q_+}}\bigg]\bigg) \hat{\alpha}^{\dagger}\bigg(-\frac{1}{2}\ln\bigg[\frac{e^{2k_+}}{e^{2(k_++q_+)}-e^{2k_+}-e^{2q_+}}\bigg]\bigg).
    \end{aligned}
\end{equation}
    \item Region $1_E$: $k_+\in\big]\frac{1}{2}\ln(3),+\infty\big[$ and $ q_+\in\big]-\frac{1}{2}\ln(1-2e^{-2k_+}),-\frac{1}{2}\ln(\frac{1-e^{-2k_+}}{2})\big[$ 
\begin{equation}
    \begin{aligned}
     &\hat{a}(k_+)\hat{a}^{\dagger}(q_+)+\frac{1}{e^{2(k_++q_+)}-e^{2k_+}-e^{2q_+}}\hat{\beta}\bigg(-\frac{1}{2}\ln\bigg[\frac{e^{2q_+}}{e^{2(k_++q_+)}-e^{2k_+}-e^{2q_+}}\bigg]\bigg)\\&\hat{\alpha}\bigg(\frac{1}{2}\ln\bigg[\frac{e^{2k_+}}{e^{2(k_++q_+)}-e^{2k_+}-e^{2q_+}}\bigg]\bigg)=0,\\
     &\hat{a}(k_+)\hat{b}^{\dagger}(q_+)=-\hat{\alpha}\bigg(-\frac{1}{2}\ln\bigg[\frac{e^{2q_+}}{e^{2(k_++q_+)}-e^{2k_+}-e^{2q_+}}\bigg]\bigg) \hat{\alpha}\bigg(\frac{1}{2}\ln\bigg[\frac{e^{2k_+}}{e^{2(k_++q_+)}-e^{2k_+}-e^{2q_+}}\bigg]\bigg),\\
     &\hat{b}(k_+)\hat{a}^{\dagger}(q_+)=-\hat{\beta}\bigg(-\frac{1}{2}\ln\bigg[\frac{e^{2q_+}}{e^{2(k_++q_+)}-e^{2k_+}-e^{2q_+}}\bigg]\bigg) \hat{\beta}\bigg(\frac{1}{2}\ln\bigg[\frac{e^{2k_+}}{e^{2(k_++q_+)}-e^{2k_+}-e^{2q_+}}\bigg]\bigg).
    \end{aligned}
\end{equation}
\end{itemize}

\subsection*{Products of the form \texorpdfstring{$\hat a (k_+) \, \hat \alpha^\dagger (q_+)$,  $\hat a (k_+) \, \hat \beta ^\dagger (q_+)$}- and  \texorpdfstring{$\hat b (k_+)\, \hat \alpha ^\dagger (q_+)$}}
\begin{figure}[ht]
	\centering 	\includegraphics[width=0.45\textwidth]{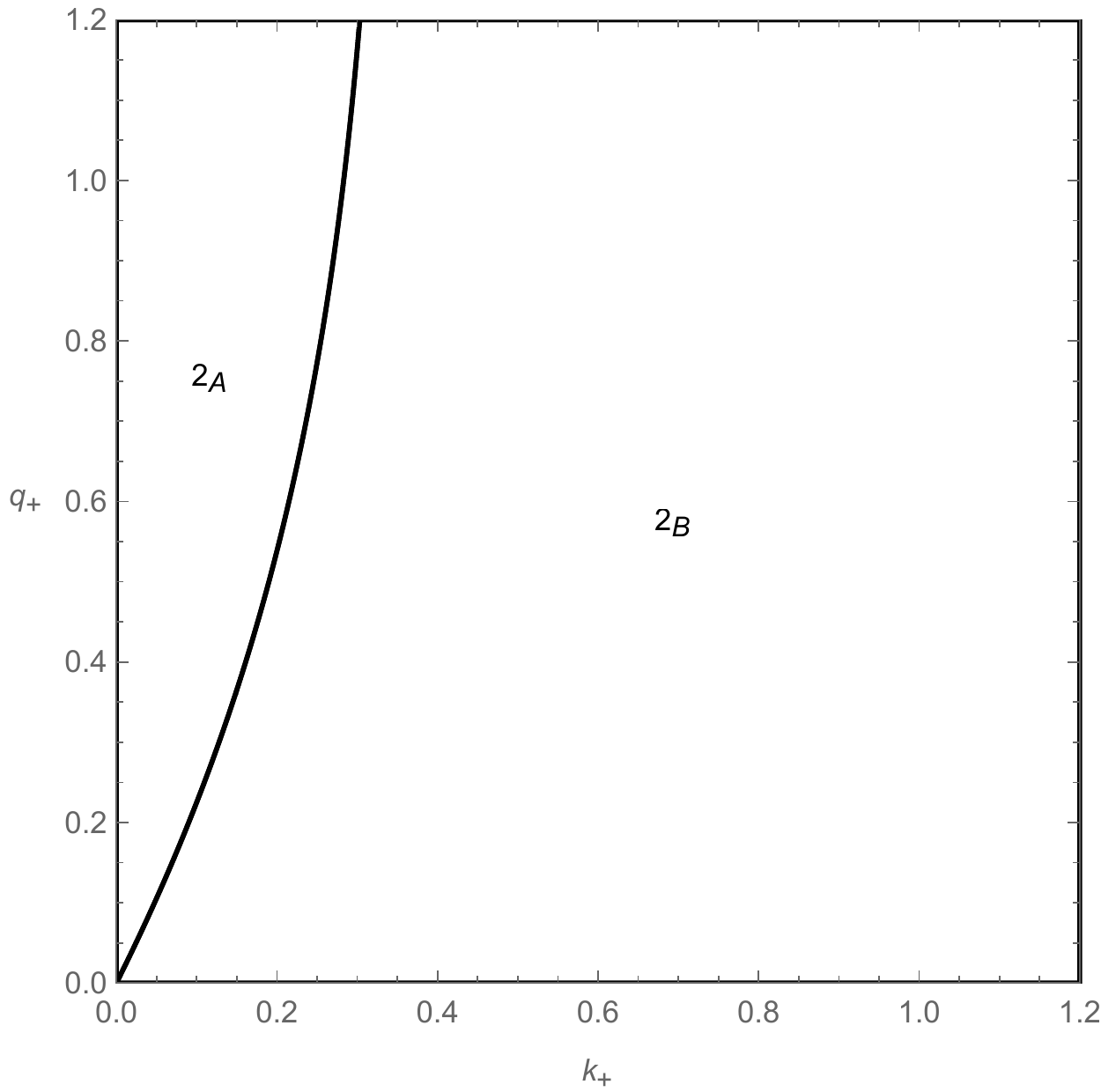}
\caption{Regions of the products of the form $\hat a (k_+) \, \hat \alpha^\dagger (q_+)$,  $\hat a (k_+) \, \hat \beta ^\dagger (q_+)$ and $\hat b (k_+)\, \hat \alpha ^\dagger (q_+)$. Corresponding integrals in Appendix~\ref{appendixE}:  $2_A\rightarrow H_{6}^{(i)}$ and $2_B\rightarrow H_{8}^{(i)}$.\label{e1epsilon2dagger}}
\end{figure}

The commutation relations that involve these products take two different forms, according to the region of $\mathbb{R}^2_+$ to which the two momenta $k_+$ and $q_+$ belong. The two regions are represented in Fig.~\ref{e1epsilon2dagger}.

\begin{itemize}
    \item Region $2_A$: $q_+\in ]0,+\infty[$ and $ k_+ \in \big]0,-\frac{1}{2}\ln(\frac{1+e^{-2q_+}}{2})\big[$
 \begin{equation}
     \begin{aligned}
      &\hat{a}(k_+)\hat{\alpha}^{\dagger}(q_+)=\frac{1}{e^{2(k_++q_+)}+e^{2k_+}-e^{2q_+}}\hat{\alpha}^{\dagger}\bigg(\frac{1}{2}\ln\bigg[\frac{e^{2q_+}}{e^{2(k_++q_+)}+e^{2k_+}-e^{2q_+}}\bigg]\bigg)\\
      &\hat{b}^{\dagger}\bigg(\frac{1}{2}\ln\bigg[\frac{e^{2(k_++q_+)}+e^{2k_+}-e^{2q_+}}{e^{2k_+}}\bigg]\bigg),\\
      &\hat{a}(k_+)\hat{\beta}^{\dagger}(q_+)= \hat{\beta}^{\dagger}\bigg(\frac{1}{2}\ln\bigg[\frac{e^{2q_+}}{e^{2(k_++q_+)}+e^{2k_+}-e^{2q_+}}\bigg]\bigg)\hat{b}^{\dagger}\bigg(\frac{1}{2}\ln\bigg[\frac{e^{2(k_++q_+)}+e^{2k_+}-e^{2q_+}}{e^{2k_+}}\bigg]\bigg),\\
       &\hat{b}(k_+)\hat{\alpha}^{\dagger}(q_+)= \hat{\alpha}^{\dagger}\bigg(\frac{1}{2}\ln\bigg[\frac{e^{2q_+}}{e^{2(k_++q_+)}+e^{2k_+}-e^{2q_+}}\bigg]\bigg)\hat{a}^{\dagger}\bigg(\frac{1}{2}\ln\bigg[\frac{e^{2(k_++q_+)}+e^{2k_+}-e^{2q_+}}{e^{2k_+}}\bigg]\bigg).
     \end{aligned}
 \end{equation}
\item  Region $2_B$: $q_+\in ]0,+\infty[$ and $ k_+ \in \big]-\frac{1}{2}\ln(\frac{1+e^{-2q_+}}{2}),+\infty\big[$ 
 \begin{equation}
     \begin{aligned}
      &\hat{a}(k_+)\hat{\alpha}^{\dagger}(q_+)=\frac{1}{e^{2(k_++q_+)}+e^{2k_+}-e^{2q_+}}\hat{\beta}\bigg(\frac{1}{2}\ln\bigg[\frac{e^{2(k_++q_+)}+e^{2k_+}-e^{2q_+}}{e^{2q_+}}\bigg]\bigg)\\&\hat{b}^{\dagger}\bigg(\frac{1}{2}\ln\bigg[\frac{e^{2(k_++q_+)}+e^{2k_+}-e^{2q_+}}{e^{2k_+}}\bigg]\bigg),\\
      &\hat{a}(k_+)\hat{\beta}^{\dagger}(q_+)=\hat{\alpha}\bigg(\frac{1}{2}\ln\bigg[\frac{e^{2(k_++q_+)}+e^{2k_+}-e^{2q_+}}{e^{2q_+}}\bigg]\bigg)\hat{b}^{\dagger}\bigg(\frac{1}{2}\ln\bigg[\frac{e^{2(k_++q_+)}+e^{2k_+}-e^{2q_+}}{e^{2k_+}}\bigg]\bigg),\\
       &\hat{b}(k_+)\hat{\alpha}^{\dagger}(q_+)=\hat{\beta}\bigg(\frac{1}{2}\ln\bigg[\frac{e^{2(k_++q_+)}+e^{2k_+}-e^{2q_+}}{e^{2q_+}}\bigg]\bigg)\hat{a}^{\dagger}\bigg(\frac{1}{2}\ln\bigg[\frac{e^{2(k_++q_+)}+e^{2k_+}-e^{2q_+}}{e^{2k_+}}\bigg]\bigg).
     \end{aligned}
 \end{equation}
 \end{itemize}

\subsection*{Products of the form \texorpdfstring{$\hat b^\dagger (k_+)\, \hat \alpha^\dagger (q_+)$,  $\hat b^\dagger (k_+)\, \hat \beta ^\dagger (q_+)$}- and  \texorpdfstring{$\hat a^\dagger (k_+) \, \hat \alpha ^\dagger (q_+)$ } }
\begin{figure}[ht]
	\centering 	\includegraphics[width=0.45\textwidth]{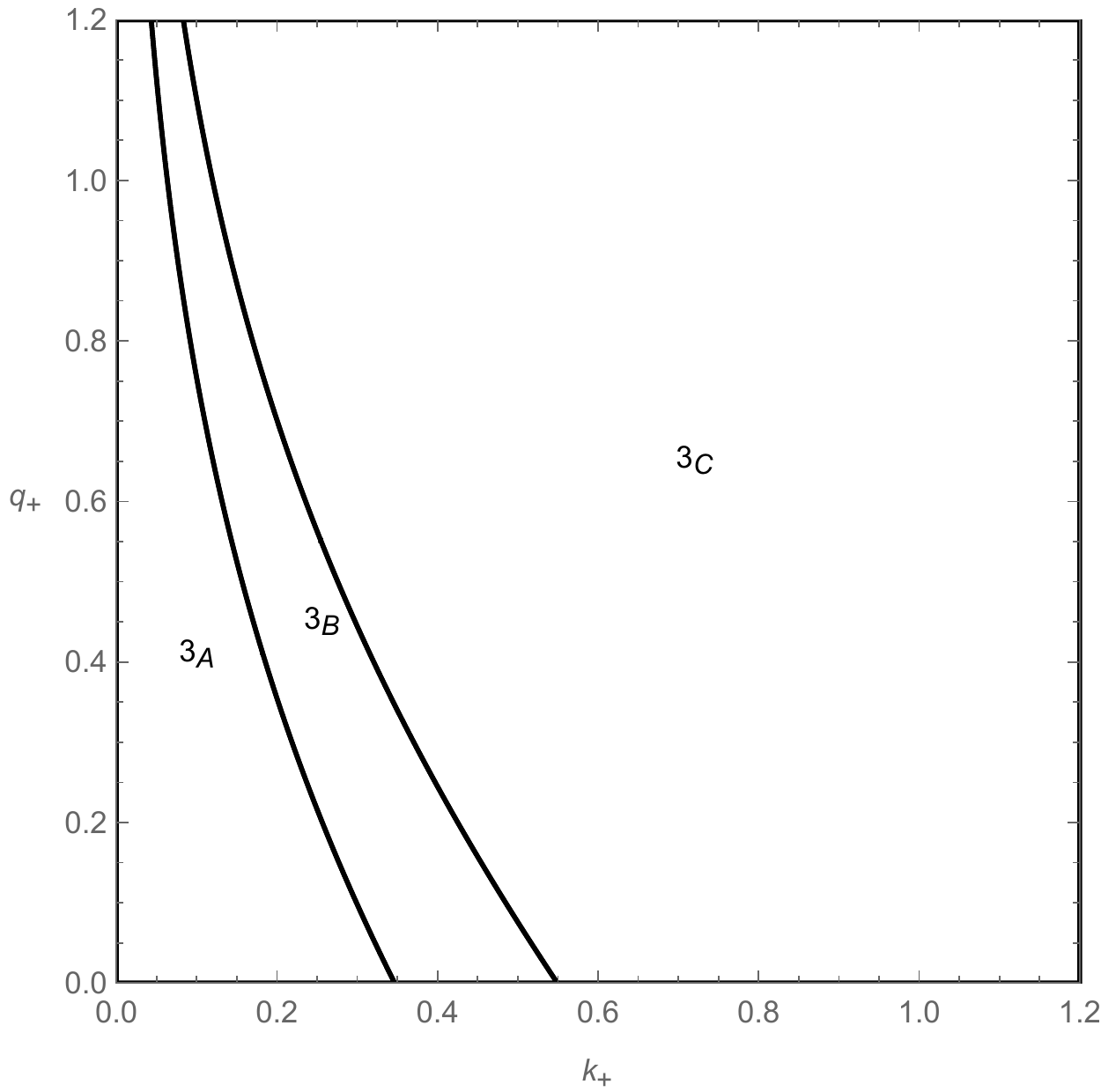}
\caption{Regions of the products of the form $\hat b^\dagger (k_+)\, \hat \alpha^\dagger (q_+)$,  $\hat b^\dagger (k_+)\, \hat \beta ^\dagger (q_+)$ and  $\hat a^\dagger (k_+) \, \hat \alpha ^\dagger (q_+)$. Corresponding integrals in Appendix~\ref{appendixE}: $3_A\rightarrow H_{2}^{(i)}$, $3_B\rightarrow H_{10}^{(i)}$ and $3_C\rightarrow H_{14}^{(i)}$.\label{e1daggerepsilon2dagger}}
\end{figure}

The commutation relations that involve these products take three different forms, according to the region of $\mathbb{R}^2_+$ to which the two momenta $k_+$ and $q_+$ belong. The three regions are represented in Fig.~\ref{e1daggerepsilon2dagger}.
\begin{itemize}
    \item Region $3_A$: $q_+\in]0,+\infty[$ and $k_+ \in \big]0,\frac{1}{2}\ln(e^{-2q_+}+1)\big[$ 
 \begin{equation}
     \begin{aligned}
     &\hat{b}^{\dagger}(k_+)\hat{\alpha}^{\dagger}(q_+)=\\&\frac{e^{2k_+}}{1+e^{2q_+}-e^{2(q_++k_+)}}\hat{\alpha}^{\dagger}\bigg(\frac{1}{2}\ln\bigg[\frac{e^{2(q_++k_+)}}{1+e^{2q_+}-e^{2(q_++k_+)}}\bigg]\bigg)\hat{a}\bigg(\frac{1}{2}\ln\bigg[\frac{1}{1+e^{2q_+}-e^{2(q_++k_+)}}\bigg]\bigg),\\
     &\hat{b}^{\dagger}(k_+)\hat{\beta}^{\dagger}(q_+)=\hat{\beta}^{\dagger}\bigg(\frac{1}{2}\ln\bigg[\frac{e^{2(q_++k_+)}}{1+e^{2q_+}-e^{2(q_++k_+)}}\bigg]\bigg)\hat{a}\bigg(\frac{1}{2}\ln\bigg[\frac{1}{1+e^{2q_+}-e^{2(q_++k_+)}}\bigg]\bigg),\\
     &\hat{a}^{\dagger}(k_+)\hat{\alpha}^{\dagger}(q_+)=\hat{\alpha}^{\dagger}\bigg(\frac{1}{2}\ln\bigg[\frac{e^{2(q_++k_+)}}{1+e^{2q_+}-e^{2(q_++k_+)}}\bigg]\bigg)\hat{b}\bigg(\frac{1}{2}\ln\bigg[\frac{1}{1+e^{2q_+}-e^{2(q_++k_+)}}\bigg]\bigg).
     \end{aligned}
 \end{equation}
 \item Region $3_B$: $q_+\in]0,+\infty[$ and $k_+\in\big]\frac{1}{2}\ln(e^{-2q_+}+1),\frac{1}{2}\ln(2e^{-2q_+}+1)\big[$
\begin{equation}\begin{aligned}
&\hat{b}^{\dagger}(k_+)\hat{\alpha}^{\dagger}(q_+)=\\&\frac{e^{2k_+}}{e^{2(q_++k_+)}+e^{2q_+}-1}\hat{a}^{\dagger}\bigg(\frac{1}{2}\ln\bigg[\frac{e^{2(k_++q_+)}}{e^{2(k_++q_+)}-e^{2q_+}-1}\bigg]\bigg)\hat{\alpha}\bigg(\frac{1}{2}\ln\bigg[\frac{1}{e^{2(k_++q_+)}-e^{2q_+}-1}\bigg]\bigg) ,\\
&\hat{b}^{\dagger}(k_+)\hat{\beta}^{\dagger}(q_+)= \hat{b}^{\dagger}\bigg(\frac{1}{2}\ln\bigg[\frac{e^{2(k_++q_+)}}{e^{2(k_++q_+)}-e^{2q_+}-1}\bigg]\bigg)\hat{\alpha}\bigg(\frac{1}{2}\ln\bigg[\frac{1}{e^{2(k_++q_+)}-e^{2q_+}-1}\bigg]\bigg) ,\\
&\hat{a}^{\dagger}(k_+)\hat{\alpha}^{\dagger}(q_+)= \hat{a}^{\dagger}\bigg(\frac{1}{2}\ln\bigg[\frac{e^{2(k_++q_+)}}{e^{2(k_++q_+)}-e^{2q_+}-1}\bigg]\bigg)\hat{\beta}\bigg(\frac{1}{2}\ln\bigg[\frac{1}{e^{2(k_++q_+)}-e^{2q_+}-1}\bigg]\bigg) .
\end{aligned}\end{equation} 
 \item Region $3_C$: $q_+\in]0,+\infty[$ and $k_+\in\big]\frac{1}{2}\ln(2e^{-2q_+}+1),+\infty\big[$ 
 \begin{equation}
 \begin{aligned}
  &\hat{b}^{\dagger}(k_+)\hat{\alpha}^{\dagger}(q_+)=\\&\frac{e^{2k_+}}{e^{2(q_++k_+)}-e^{2q_+}-1}\hat{a}^{\dagger}\bigg(\frac{1}{2}\ln\bigg[\frac{e^{2(q_++k_+)}}{e^{2(q_++k_+)}-e^{2q_+}-1}\bigg]\bigg)\hat{\beta}^{\dagger}\bigg(\frac{1}{2}\ln\bigg[{e^{2(q_++k_+)}-e^{2q_+}-1}\bigg]\bigg),\\
  &\hat{b}^{\dagger}(k_+)\hat{\beta}^{\dagger}(q_+)=\hat{b}^{\dagger}\bigg(\frac{1}{2}\ln\bigg[\frac{e^{2(q_++k_+)}}{e^{2(q_++k_+)}-e^{2q_+}-1}\bigg]\bigg)\hat{\beta}^{\dagger}\bigg(\frac{1}{2}\ln\bigg[{e^{2(q_++k_+)}-e^{2q_+}-1}\bigg]\bigg),\\
  &\hat{a}^{\dagger}(k_+)\hat{\alpha}^{\dagger}(q_+)=\hat{a}^{\dagger}\bigg(\frac{1}{2}\ln\bigg[\frac{e^{2(q_++k_+)}}{e^{2(q_++k_+)}-e^{2q_+}-1}\bigg]\bigg)\hat{\alpha}^{\dagger}\bigg(\frac{1}{2}\ln\bigg[{e^{2(q_++k_+)}-e^{2q_+}-1}\bigg]\bigg).
 \end{aligned}    
 \end{equation}
 \end{itemize}
 
\subsection*{Products of the form \texorpdfstring{$\hat b^\dagger (k_+) \, \hat \beta (q_+)$,  $\hat b^\dagger (k_+) \, \hat \alpha (q_+)$}- and  \texorpdfstring{$\hat a^\dagger (k_+)\, \hat \beta (q_+)$}}
\begin{figure}[ht]
	\centering 	\includegraphics[width=0.45\textwidth]{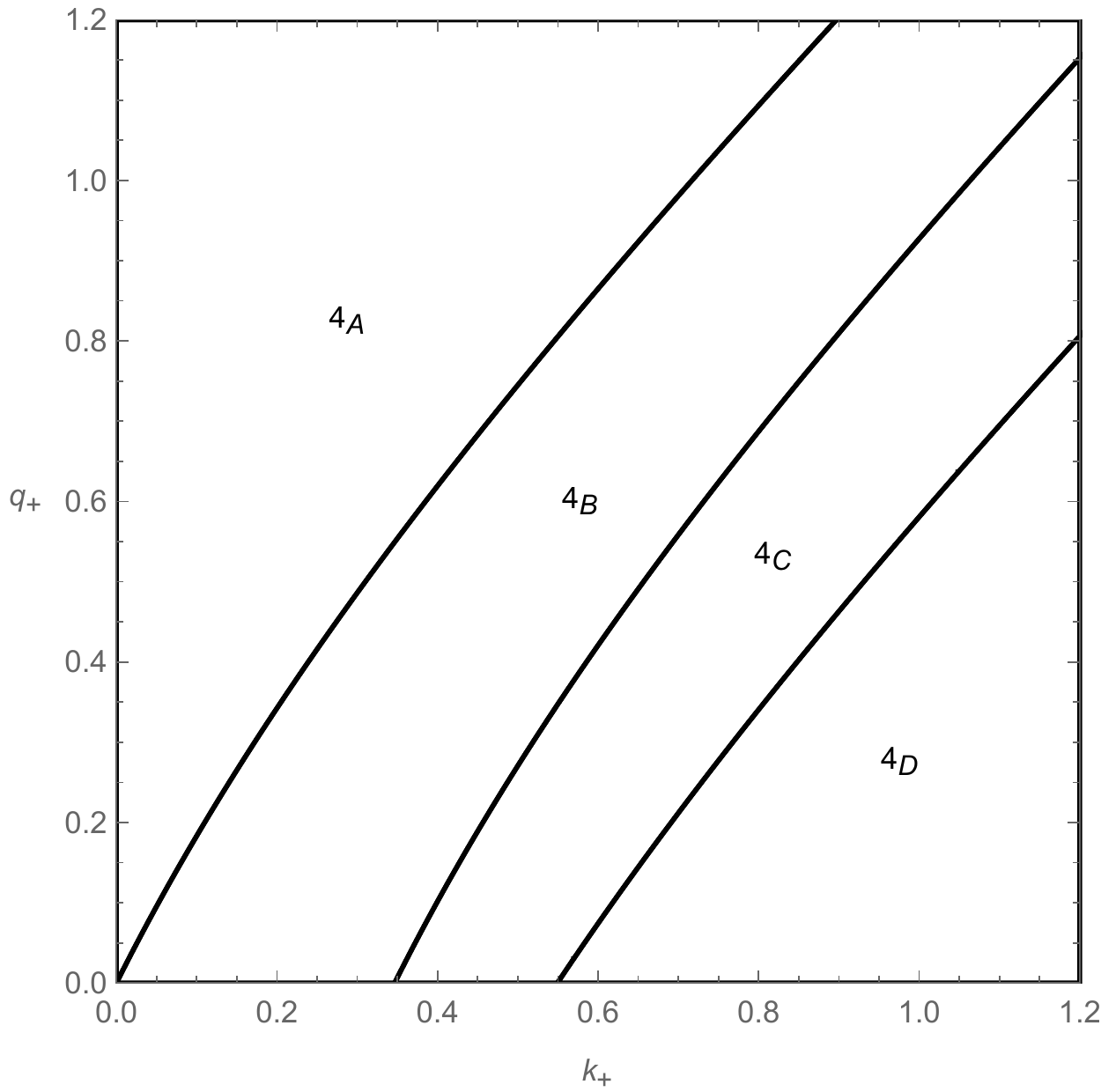}
\caption{Regions of the products of the form $\hat b^\dagger (k_+) \, \hat \beta (q_+)$,  $\hat b^\dagger (k_+) \, \hat \alpha (q_+)$ and $\hat a^\dagger (k_+)\, \hat \beta (q_+)$. Corresponding integrals in Appendix~\ref{appendixE}: $4_A\rightarrow H_{4}^{(i)}$, $4_B\rightarrow H_{2}^{(i)}$, $4_C\rightarrow H_{10}^{(i)}$ and $4_D\rightarrow H_{14}^{(i)}$.\label{e1daggerepsilon2}}
\end{figure}

The commutation relations that involve these products take four different forms, according to the region of $\mathbb{R}^2_+$ to which the two momenta $k_+$ and $q_+$ belong. The four regions are represented in Fig.~\ref{e1daggerepsilon2}.

\begin{figure}[ht]
	\centering
	\includegraphics[width=0.45\textwidth]{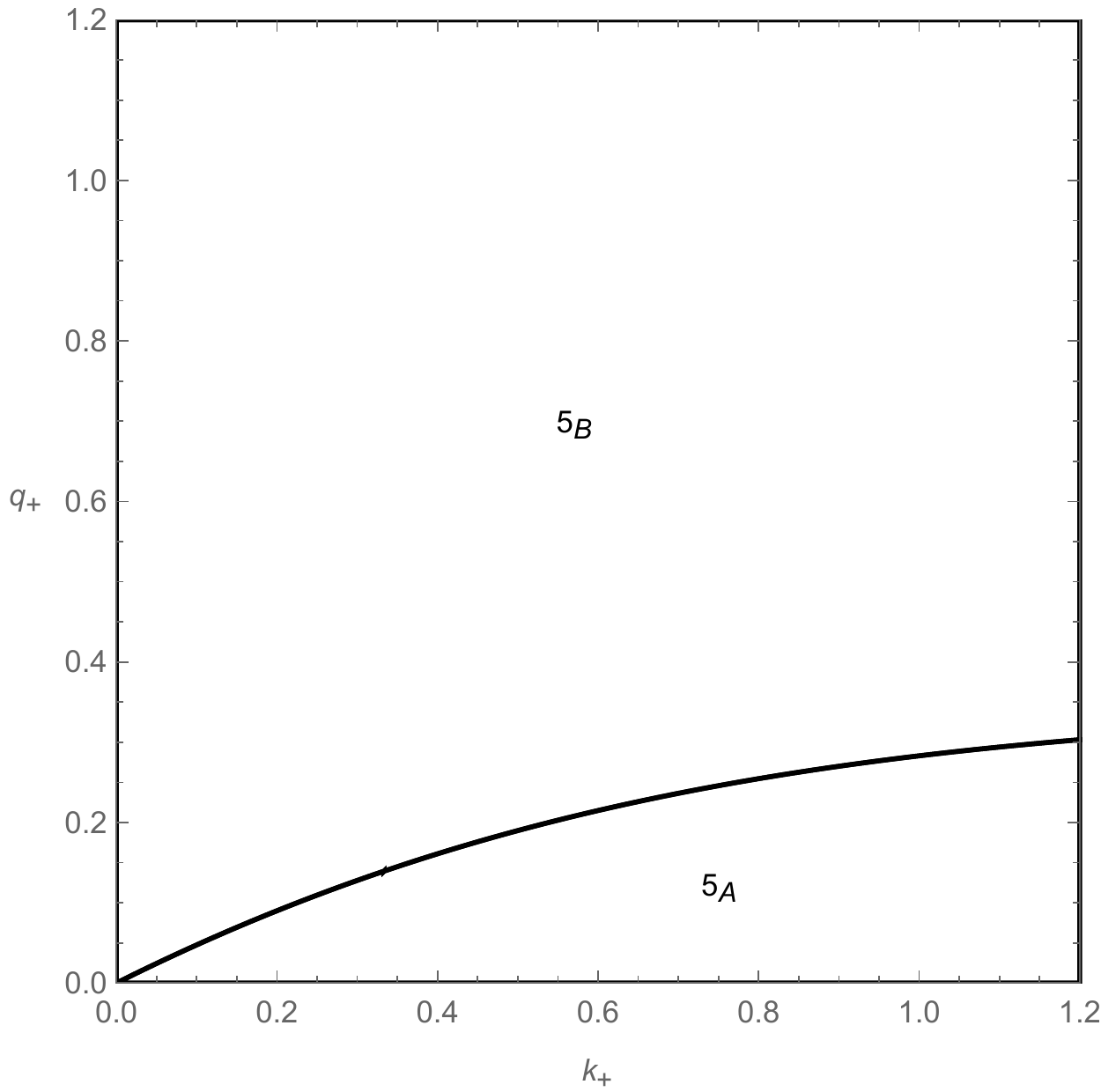}
\caption{Regions of the products of the form $\hat \alpha (k_+) \, \hat a^\dagger (q_+)$,  $\hat \alpha (k_+)\, \hat b^\dagger (q_+)$ and $\hat \beta (k_+)\, \hat a^\dagger (q_+)$. Corresponding integrals in Appendix~\ref{appendixE}: $5_A \rightarrow H_{12}^{(i)}$ and $5_B\rightarrow H_{16}^{(i)}$.\label{epsilon1e2dagger}}
\end{figure}

\begin{itemize}
\item Region $4_A$: $q_+\in]0,+\infty[$ and $k_+\in\big]0,\frac{1}{2}\ln(\frac{e^{2q_+}+1}{2})\big[$ 
\begin{equation}
    \begin{aligned}
     &\hat{b}^{\dagger}(k_+)\hat{\beta}(q_+)=\frac{e^{2(k_++q_+)}}{e^{2q_+}+1-e^{2k_+}}\hat{\beta}\bigg(\frac{1}{2}\ln\bigg[\frac{e^{2q_+}+1-e^{2k_+}}{e^{2k_+}}\bigg]\bigg)\hat{a}\bigg(\frac{1}{2}\ln\bigg[\frac{e^{2q_+}}{e^{2q_+}+1-e^{2k_+}}\bigg]\bigg) ,\\
     &\hat{b}^{\dagger}(k_+)\hat{\alpha}(q_+)=\hat{\alpha}\bigg(\frac{1}{2}\ln\bigg[\frac{e^{2q_+}+1-e^{2k_+}}{e^{2k_+}}\bigg]\bigg)\hat{a}\bigg(\frac{1}{2}\ln\bigg[\frac{e^{2q_+}}{e^{2q_+}+1-e^{2k_+}}\bigg]\bigg) ,\\
     &\hat{a}^{\dagger}(k_+)\hat{\beta}(q_+)=\hat{\beta}\bigg(\frac{1}{2}\ln\bigg[\frac{e^{2q_+}+1-e^{2k_+}}{e^{2k_+}}\bigg]\bigg)\hat{b}\bigg(\frac{1}{2}\ln\bigg[\frac{e^{2q_+}}{e^{2q_+}+1-e^{2k_+}}\bigg]\bigg) .
    \end{aligned}
\end{equation}
\item Region $4_B$: $q_+\in]0,+\infty[$ and $k_+\in\big]\frac{1}{2}\ln(\frac{e^{2q_+}+1}{2}),\frac{1}{2}\ln(e^{2q_+}+1)\big[$ 
 \begin{equation}
     \begin{aligned}
     &\hat{b}^{\dagger}(k_+)\hat{\beta}(q_+)=\frac{e^{2(k_++q_+)}}{e^{2q_+}+1-e^{2k_+}}\hat{\alpha}^{\dagger}\bigg(\frac{1}{2}\ln\bigg[\frac{e^{2k_+}}{e^{2q_+}+1-e^{2k_+}}\bigg]\bigg)\hat{a}\bigg(\frac{1}{2}\ln\bigg[\frac{e^{2q_+}}{e^{2q_+}+1-e^{2k_+}}\bigg]\bigg) ,\\
     &\hat{b}^{\dagger}(k_+)\hat{\alpha}(q_+)=\hat{\beta}^{\dagger}\bigg(\frac{1}{2}\ln\bigg[\frac{e^{2k_+}}{e^{2q_+}+1-e^{2k_+}}\bigg]\bigg)\hat{a}\bigg(\frac{1}{2}\ln\bigg[\frac{e^{2q_+}}{e^{2q_+}+1-e^{2k_+}}\bigg]\bigg) ,\\
     &\hat{a}^{\dagger}(k_+)\hat{\beta}(q_+)=\hat{\alpha}^{\dagger}\bigg(\frac{1}{2}\ln\bigg[\frac{e^{2k_+}}{e^{2q_+}+1-e^{2k_+}}\bigg]\bigg)\hat{b}\bigg(\frac{1}{2}\ln\bigg[\frac{e^{2q_+}}{e^{2q_+}+1-e^{2k_+}}\bigg]\bigg).
     \end{aligned}
 \end{equation}
\item Region $4_C$: $q_+\in]0,+\infty[$ and $k_+\in\big]\frac{1}{2}\ln(e^{2q_+}+1),\frac{1}{2}\ln(2e^{2q_+}+1)\big[$ 
 \begin{equation}
     \begin{aligned}
      &\hat{b}^{\dagger}(k_+)\hat{\beta}(q_+)=\frac{e^{2(k_++q_+)}}{e^{2k_+}-1-e^{2q_+}}\hat{a}^{\dagger}\bigg(\frac{1}{2}\ln\bigg[\frac{e^{2k_+}}{e^{2k_+}-1-e^{2q_+}}\bigg]\bigg)\hat{\alpha}\bigg(\frac{1}{2}\ln\bigg[\frac{e^{2q_+}}{e^{2k_+}-1-e^{2q_+}}\bigg]\bigg),\\
      &\hat{b}^{\dagger}(k_+)\hat{\alpha}(q_+)=\hat{b}^{\dagger}\bigg(\frac{1}{2}\ln\bigg[\frac{e^{2k_+}}{e^{2k_+}-1-e^{2q_+}}\bigg]\bigg)\hat{\alpha}\bigg(\frac{1}{2}\ln\bigg[\frac{e^{2q_+}}{e^{2k_+}-1-e^{2q_+}}\bigg]\bigg),\\
      &\hat{a}^{\dagger}(k_+)\hat{\beta}(q_+)=\hat{a}^{\dagger}\bigg(\frac{1}{2}\ln\bigg[\frac{e^{2k_+}}{e^{2k_+}-1-e^{2q_+}}\bigg]\bigg)\hat{\beta}\bigg(\frac{1}{2}\ln\bigg[\frac{e^{2q_+}}{e^{2k_+}-1-e^{2q_+}}\bigg]\bigg).
     \end{aligned}
 \end{equation}
\item Region $4_D$: $q_+\in]0,+\infty[$ and $k_+\in\big]\frac{1}{2}\ln(2e^{2q_+}+1),+\infty\big[$ 
\begin{equation}
    \begin{aligned}
     &\hat{b}^{\dagger}(k_+)\hat{\beta}(q_+)=\frac{e^{2(k_++q_+)}}{e^{2k_+}-1-e^{2q_+}}\hat{a}^{\dagger}\bigg(\frac{1}{2}\ln\bigg[\frac{e^{2k_+}}{e^{2k_+}-1-e^{2q_+}}\bigg]\bigg)\hat{\beta}^{\dagger}\bigg(\frac{1}{2}\ln\bigg[\frac{e^{2k_+}-1-e^{2q_+}}{e^{2q_+}}\bigg]\bigg) ,\\
     &\hat{b}^{\dagger}(k_+)\hat{\alpha}(q_+)=\hat{b}^{\dagger}\bigg(\frac{1}{2}\ln\bigg[\frac{e^{2k_+}}{e^{2k_+}-1-e^{2q_+}}\bigg]\bigg)\hat{\beta}^{\dagger}\bigg(\frac{1}{2}\ln\bigg[\frac{e^{2k_+}-1-e^{2q_+}}{e^{2q_+}}\bigg]\bigg) ,\\
     &\hat{a}^{\dagger}(k_+)\hat{\beta}(q_+)=\hat{a}^{\dagger}\bigg(\frac{1}{2}\ln\bigg[\frac{e^{2k_+}}{e^{2k_+}-1-e^{2q_+}}\bigg]\bigg)\hat{\alpha}^{\dagger}\bigg(\frac{1}{2}\ln\bigg[\frac{e^{2k_+}-1-e^{2q_+}}{e^{2q_+}}\bigg]\bigg).
    \end{aligned}
\end{equation}

\subsection*{Products of the form \texorpdfstring{$\hat \alpha (k_+) \, \hat a^\dagger (q_+)$,  $\hat \alpha (k_+)\, \hat b^\dagger (q_+)$}- and  \texorpdfstring{$\hat \beta (k_+)\, \hat a^\dagger (q_+)$}}
The commutation relations that involve these products take two different forms, according to the region of $\mathbb{R}^2_+$ to which the two momenta $k_+$ and $q_+$ belong. The two regions are represented in Fig.~\ref{epsilon1e2dagger}.
\end{itemize}
\begin{itemize}
    \item Region $5_A$: $k_+\in]0,+\infty[$ and $ q_+\in\big]0,-\frac{1}{2}\ln(\frac{1+e^{-2k_+}}{2})\big[$ 
 \begin{equation}
     \begin{aligned}
      &\hat{\alpha}(k_+)\hat{a}^{\dagger}(q_+)=\frac{1}{e^{2(k_++q_+)}+e^{2q_+}-e^{2k_+}}\hat{b}\bigg(\frac{1}{2}\ln\bigg[\frac{e^{2q_+}+e^{2(k_++q_+)}-e^{2k_+}}{e^{2q_+}}\bigg]\bigg)\\&\hat{\alpha}\bigg(\frac{1}{2}\ln\bigg[\frac{e^{2k_+}}{e^{2q_+}+e^{2(k_++q_+)}-e^{2k_+}}\bigg]\bigg) ,\\
      &\hat{\alpha}(k_+)\hat{b}^{\dagger}(q_+)=\hat{a}\bigg(\frac{1}{2}\ln\bigg[\frac{e^{2q_+}+e^{2(k_++q_+)}-e^{2k_+}}{e^{2q_+}}\bigg]\bigg)\hat{\alpha}\bigg(\frac{1}{2}\ln\bigg[\frac{e^{2k_+}}{e^{2q_+}+e^{2(k_++q_+)}-e^{2k_+}}\bigg]\bigg) ,\\
      &\hat{\beta}(k_+)\hat{a}^{\dagger}(q_+)=\hat{b}\bigg(\frac{1}{2}\ln\bigg[\frac{e^{2q_+}+e^{2(k_++q_+)}-e^{2k_+}}{e^{2q_+}}\bigg]\bigg)\hat{\beta}\bigg(\frac{1}{2}\ln\bigg[\frac{e^{2k_+}}{e^{2q_+}+e^{2(k_++q_+)}-e^{2k_+}}\bigg]\bigg).
     \end{aligned}
 \end{equation}
\item Region $5_B$: $k_+\in]0,+\infty[$ and $ q_+\in\big]-\frac{1}{2}\ln(\frac{1+e^{-2k_+}}{2}),+\infty\big[$ 
 \begin{equation} \begin{aligned}
     &\hat{\alpha}(k_+)\hat{a}^{\dagger}(q_+)=\frac{1}{e^{2(k_++q_+)}+e^{2q_+}-e^{2k_+}}\hat{b}\bigg(\frac{1}{2}\ln\bigg[\frac{e^{2q_+}+e^{2(k_++q_+)}-e^{2k_+}}{e^{2q_+}}\bigg]\bigg)\\
     &\hat{\beta}^{\dagger}\bigg(\frac{1}{2}\ln\bigg[\frac{e^{2q_+}+e^{2(k_++q_+)}-e^{2k_+-q}}{e^{2k_+}}\bigg]\bigg) ,\\
     &\hat{\alpha}(k_+)\hat{b}^{\dagger}(q_+)=\hat{a}\bigg(\frac{1}{2}\ln\bigg[\frac{e^{2q_+}+e^{2(k_++q_+)}-e^{2k_+}}{e^{2q_+}}\bigg]\bigg)\hat{\beta}^{\dagger}\bigg(\frac{1}{2}\ln\bigg[\frac{e^{2q_+}+e^{2(k_++q_+)}-e^{2k_+-q}}{e^{2k_+}}\bigg]\bigg) ,\\
     &\hat{\beta}(k_+)\hat{a}^{\dagger}(q_+)=\hat{b}\bigg(\frac{1}{2}\ln\bigg[\frac{e^{2q_+}+e^{2(k_++q_+)}-e^{2k_+}}{e^{2q_+}}\bigg]\bigg)\hat{\alpha}^{\dagger}\bigg(\frac{1}{2}\ln\bigg[\frac{e^{2q_+}+e^{2(k_++q_+)}-e^{2k_+-q}}{e^{2k_+}}\bigg]\bigg).
 \end{aligned}\end{equation}
 \end{itemize}
 
 \subsection*{Products of the form \texorpdfstring{$\hat \alpha (k_+) \, \hat b (q_+)$,  $\hat \alpha (k_+) \, \hat a (q_+)$}- and  \texorpdfstring{$\hat \beta (k_+) \, \hat b (q_+)$}}
\begin{figure}[ht]
	\centering 	\includegraphics[width=0.45\textwidth]{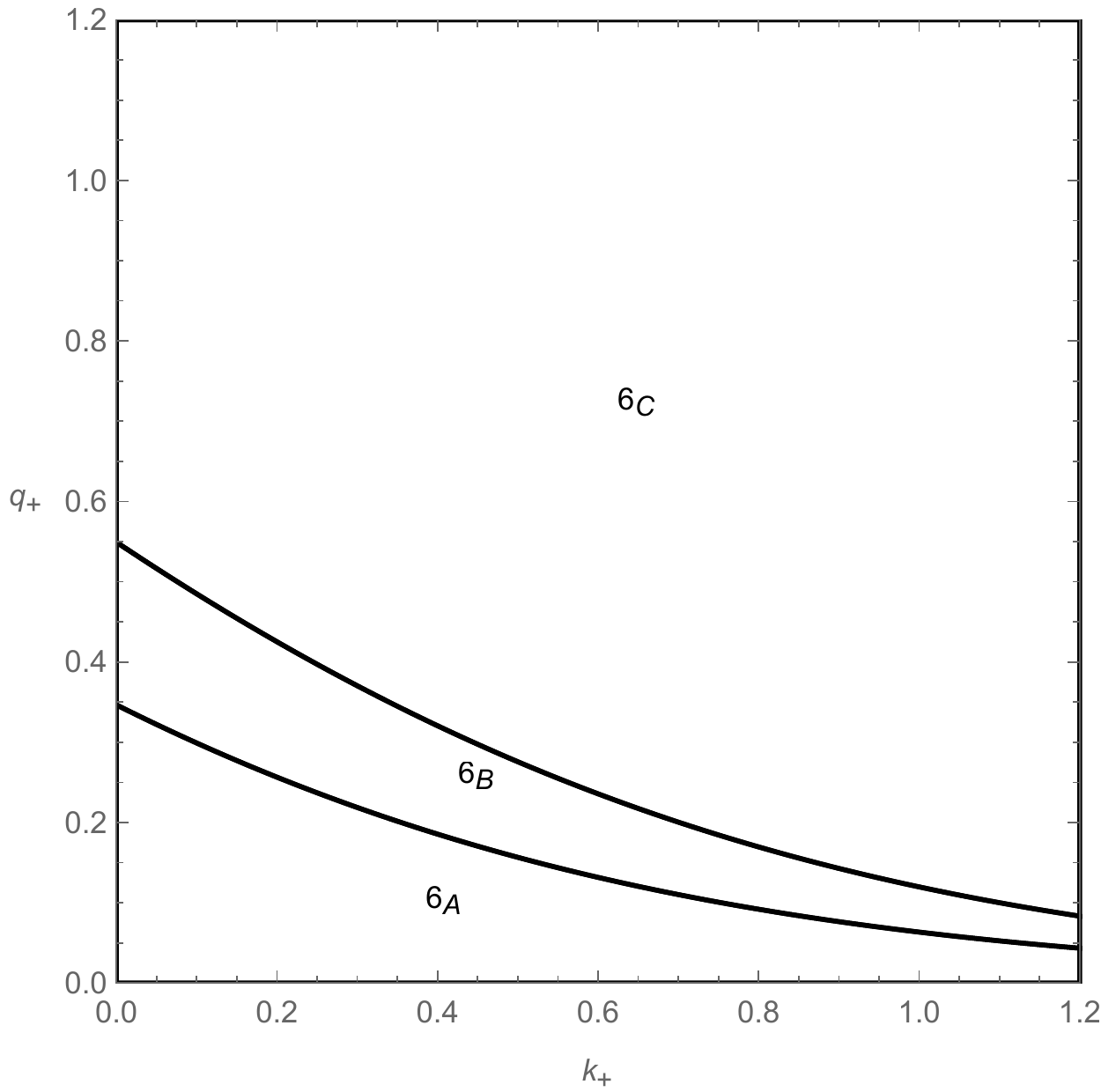}
	\caption{Regions of the products of the form $\hat \alpha (k_+) \, \hat b (q_+)$,  $\hat \alpha (k_+) \, \hat a (q_+)$ and $\hat \beta (k_+) \, \hat b (q_+)$. Corresponding integrals in Appendix~\ref{appendixE}: $6_A\rightarrow H_{10}^{(i)}$, $6_B\rightarrow H_{2}^{(i)}$ and $6_C\rightarrow H_{4}^{(i)}$.\label{epsilon1e2}}
\end{figure}

 The commutation relations that involve these products take three different forms, according to the region of $\mathbb{R}^2_+$ to which the two momenta $k_+$ and $q_+$ belong. The three regions are represented in Fig.~\ref{epsilon1e2}.

\begin{itemize}
    \item  Region $6_A$: $k_+\in]0,+\infty[$ and $ q_+\in\big]0,\frac{1}{2}\ln(e^{-2k_+}+1)\big[$ 
 \begin{equation}
     \begin{aligned}
      &\hat{\alpha}(k_+)\hat{b}(q_+)=\\&\frac{e^{2q_+}}{1+e^{2k_+}-e^{2(k_++q_+)}}\hat{a}^{\dagger}\bigg(\frac{1}{2}\ln\bigg[\frac{1}{1+e^{2k_+}-e^{2(k_++q_+)}}\bigg]\bigg)\hat{\alpha}\bigg(\frac{1}{2}\ln\bigg[\frac{e^{2(k_++q_+)}}{1+e^{2k_+}-e^{2(k_++q_+)}}\bigg]\bigg),\\
      &\hat{\alpha}(k_+)\hat{a}(q_+)=\hat{b}^{\dagger}\bigg(\frac{1}{2}\ln\bigg[\frac{1}{1+e^{2k_+}-e^{2(k_++q_+)}}\bigg]\bigg)\hat{\alpha}\bigg(\frac{1}{2}\ln\bigg[\frac{e^{2(k_++q_+)}}{1+e^{2k_+}-e^{2(k_++q_+)}}\bigg]\bigg),\\
      &\hat{\beta}(k_+)\hat{b}(q_+)=\hat{a}^{\dagger}\bigg(\frac{1}{2}\ln\bigg[\frac{1}{1+e^{2k_+}-e^{2(k_++q_+)}}\bigg]\bigg)\hat{\beta}\bigg(\frac{1}{2}\ln\bigg[\frac{e^{2(k_++q_+)}}{1+e^{2k_+}-e^{2(k_++q_+)}}\bigg]\bigg).
     \end{aligned}
 \end{equation}
 \item Region $6_B$: $k_+\in]0,+\infty[$ and $ q_+\in\big]\frac{1}{2}\ln(e^{-2k_+}+1),\frac{1}{2}\ln(2e^{-2k_+}+1)\big[$
 \begin{equation}
     \begin{aligned}
    &\hat{\alpha}(k_+)\hat{b}(q_+)=\\&\frac{e^{2q_+}}{e^{2(k_++q_+)}-1-e^{2k_+}}\hat{\alpha}^{\dagger}\bigg(\frac{1}{2}\ln\bigg[\frac{1}{e^{2(k_++q_+)}-e^{2k_+}-1}\bigg]\bigg)\hat{a}\bigg(\frac{1}{2}\ln\bigg[\frac{e^{2(k_++q_+)}}{e^{2(k_++q_+)}-e^{2k_+}-1}\bigg]\bigg),\\
    &\hat{\alpha}(k_+)\hat{a}(q_+)= \hat{\beta}^{\dagger}\bigg(\frac{1}{2}\ln\bigg[\frac{1}{e^{2(k_++q_+)}-e^{2k_+}-1}\bigg]\bigg)\hat{a}\bigg(\frac{1}{2}\ln\bigg[\frac{e^{2(k_++q_+)}}{e^{2(k_++q_+)}-e^{2k_+}-1}\bigg]\bigg),\\
    &\hat{\beta}(k_+)\hat{b}(q_+)= \hat{\alpha}^{\dagger}\bigg(\frac{1}{2}\ln\bigg[\frac{1}{e^{2(k_++q_+)}-e^{2k_+}-1}\bigg]\bigg)\hat{b}\bigg(\frac{1}{2}\ln\bigg[\frac{e^{2(k_++q_+)}}{e^{2(k_++q_+)}-e^{2k_+}-1}\bigg]\bigg).
     \end{aligned}
 \end{equation}
\item Region $6_C$: $k_+\in]0,+\infty[$ and $ q_+\in\big]\frac{1}{2}\ln(2e^{-2k_+}+1),+\infty\big[$ 
 \begin{equation}
     \begin{aligned}
      &\hat{\alpha}(k_+)\hat{b}(q_+)=\\&\frac{e^{2q_+}}{e^{2(k_++q_+)}-1-e^{2k_+}}\hat{\beta}\bigg(\frac{1}{2}\ln\bigg[e^{2(q_++k_+)}-e^{2k_+}-1\bigg]\bigg)\hat{a}\bigg(\frac{1}{2}\ln\bigg[\frac{e^{2(q_++k_+)}}{e^{2(q_++k_+)}-e^{2k_+}-1}\bigg]\bigg),\\
      &\hat{\alpha}(k_+)\hat{a}(q_+)= \hat{\alpha}\bigg(\frac{1}{2}\ln\bigg[e^{2(q_++k_+)}-e^{2k_+}-1\bigg]\bigg)\hat{a}\bigg(\frac{1}{2}\ln\bigg[\frac{e^{2(q_++k_+)}}{e^{2(q_++k_+)}-e^{2k_+}-1}\bigg]\bigg),\\
      &\hat{\beta}(k_+)\hat{b}(q_+)= \hat{\beta}\bigg(\frac{1}{2}\ln\bigg[e^{2(q_++k_+)}-e^{2k_+}-1\bigg]\bigg)\hat{b}\bigg(\frac{1}{2}\ln\bigg[\frac{e^{2(q_++k_+)}}{e^{2(q_++k_+)}-e^{2k_+}-1}\bigg]\bigg).
     \end{aligned}
 \end{equation}
\end{itemize} 

\subsection*{Products of the form \texorpdfstring{$\hat \beta^\dagger (k_+) \, \hat b (q_+)$,  $\hat \beta^\dagger (k_+) \, \hat a (q_+)$}- and  \texorpdfstring{$\hat \alpha^\dagger (k_+) \, \hat b (q_+)$}}
\begin{figure}[ht]
	\centering
	\includegraphics[width=0.45\textwidth]{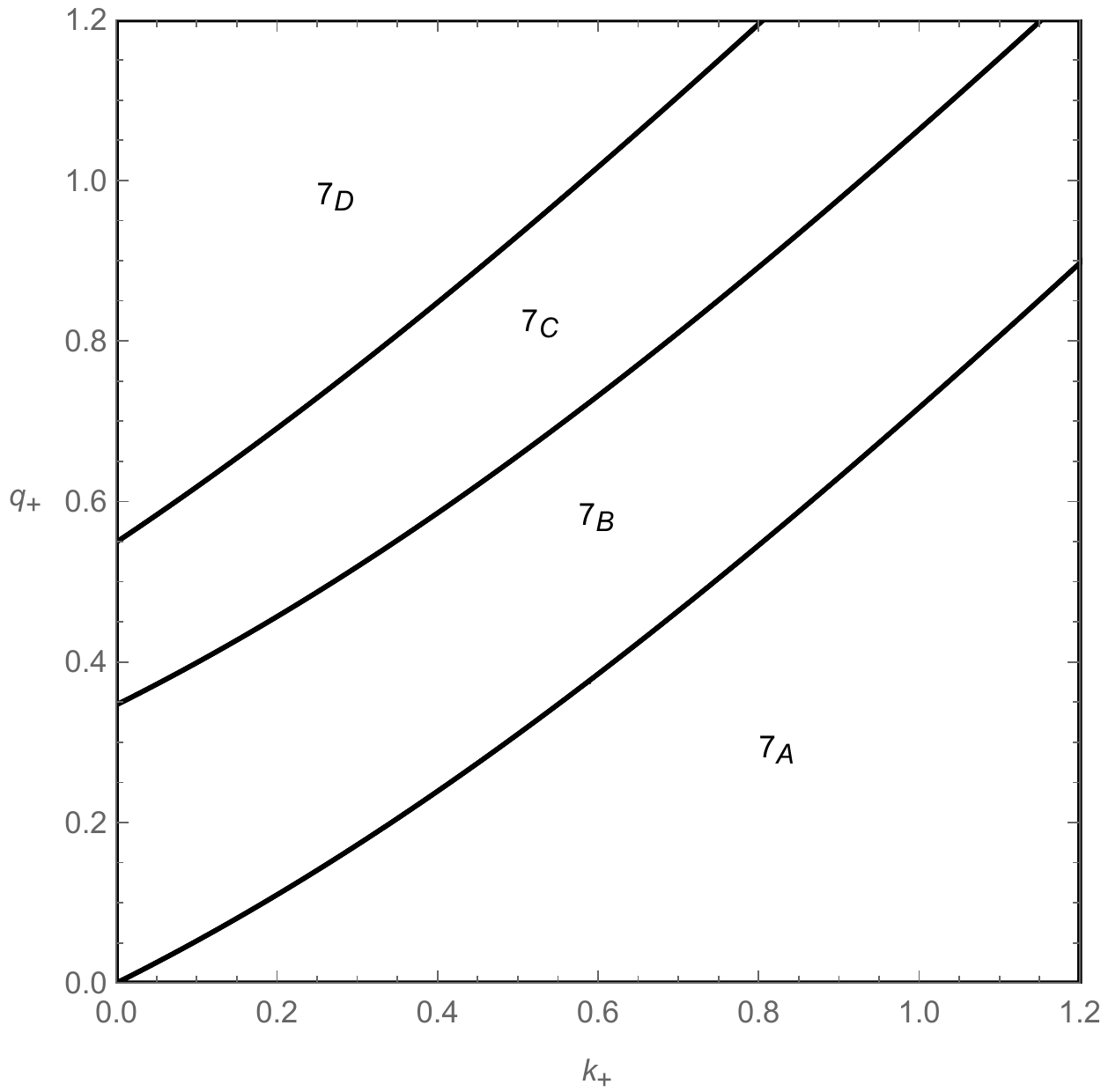}
\caption{Regions of the products of the form $\hat \beta^\dagger (k_+) \, \hat b (q_+)$,  $\hat \beta^\dagger (k_+) \, \hat a (q_+)$ and $\hat \alpha^\dagger (k_+) \, \hat b (q_+)$. Corresponding integrals in Appendix~\ref{appendixE}:  $7_A\rightarrow H_{14}^{(i)}$, $7_B\rightarrow H_{10}^{(i)}$, $7_C\rightarrow H_{2}^{(i)}$ and $7_D\rightarrow H_{4}^{(i)}$.\label{epsilon1daggere2}}
\end{figure}

 The commutation relations that involve these products take four different forms, according to the region of $\mathbb{R}^2_+$ to which the two momenta $k_+$ and $q_+$ belong. The four regions are represented in Fig.~\ref{epsilon1daggere2}.
\begin{itemize}
    \item Region $7_A$: $k_+\in]0,+\infty[$ and $ q_+\in \big]0,\frac{1}{2}\ln(\frac{e^{2k_+}+1}{2})\big[$ 
\begin{equation}
    \begin{aligned}
     &\hat{\beta}^{\dagger}(k_+)\hat{b}(q_+)=\frac{e^{2(q_++k_+)}}{e^{2k_+}+1-e^{2q_+}}\hat{a}^{\dagger}\bigg(\frac{1}{2}\ln\bigg[\frac{e^{2k_+}}{e^{2k_+}+1-e^{2q_+}}\bigg]\bigg)\hat{\beta}^{\dagger}\bigg(\frac{1}{2}\ln\bigg[\frac{e^{2k_+}+1-e^{2q_+}}{e^{2q_+}}\bigg]\bigg),\\
     &\hat{\beta}^{\dagger}(k_+)\hat{a}(q_+)= \hat{b}^{\dagger}\bigg(\frac{1}{2}\ln\bigg[\frac{e^{2k_+}}{e^{2k_+}+1-e^{2q_+}}\bigg]\bigg)\hat{\beta}^{\dagger}\bigg(\frac{1}{2}\ln\bigg[\frac{e^{2k_+}+1-e^{2q_+}}{e^{2q_+}}\bigg]\bigg),\\
     &\hat{\alpha}^{\dagger}(k_+)\hat{b}(q_+)= \hat{a}^{\dagger}\bigg(\frac{1}{2}\ln\bigg[\frac{e^{2k_+}}{e^{2k_+}+1-e^{2q_+}}\bigg]\bigg)\hat{\alpha}^{\dagger}\bigg(\frac{1}{2}\ln\bigg[\frac{e^{2k_+}+1-e^{2q_+}}{e^{2q_+}}\bigg]\bigg).
    \end{aligned}
\end{equation}
\item Region $7_B$: $k_+\in]0,+\infty[$ and $ q_+\in \big]\frac{1}{2}\ln(\frac{e^{2k_+}+1}{2}),\frac{1}{2}\ln(e^{2k_+}+1)\big[$ 
 \begin{equation}
     \begin{aligned}
      &\hat{\beta}^{\dagger}(k_+)\hat{b}(q_+)=\frac{e^{2(q_++k_+)}}{e^{2k_+}+1-e^{2q_+}}\hat{a}^{\dagger}\bigg(\frac{1}{2}\ln\bigg[\frac{e^{2k_+}}{e^{2k_+}+1-e^{2q_+}}\bigg]\bigg)\hat{\alpha}\bigg(\frac{1}{2}\ln\bigg[\frac{e^{2q_+}}{e^{2k_+}+1-e^{2q_+}}\bigg]\bigg),\\
      &\hat{\beta}^{\dagger}(k_+)\hat{a}(q_+)=\hat{b}^{\dagger}\bigg(\frac{1}{2}\ln\bigg[\frac{e^{2k_+}}{e^{2k_+}+1-e^{2q_+}}\bigg]\bigg)\hat{\alpha}\bigg(\frac{1}{2}\ln\bigg[\frac{e^{2q_+}}{e^{2k_+}+1-e^{2q_+}}\bigg]\bigg),\\
      &\hat{\alpha}^{\dagger}(k_+)\hat{b}(q_+)=\hat{a}^{\dagger}\bigg(\frac{1}{2}\ln\bigg[\frac{e^{2k_+}}{e^{2k_+}+1-e^{2q_+}}\bigg]\bigg)\hat{\beta}\bigg(\frac{1}{2}\ln\bigg[\frac{e^{2q_+}}{e^{2k_+}+1-e^{2q_+}}\bigg]\bigg).
     \end{aligned}
 \end{equation}
\item Region $7_C$: $k_+\in]0,+\infty[$ and $ q_+\in\big ]\frac{1}{2}\ln(e^{2k_+}+1),\frac{1}{2}\ln(2e^{2k_+}+1)\big[$ 
 \begin{equation}
     \begin{aligned}
      &\hat{\beta}^{\dagger}(k_+)\hat{b}(q_+)=\frac{e^{2(q_++k_+)}}{e^{2q_+}-1-e^{2k_+}}\hat{\alpha}^{\dagger}\bigg(\frac{1}{2}\ln\bigg[\frac{e^{2k_+}}{e^{2q_+}-1-e^{2k_+}}\bigg]\bigg)\hat{a}\bigg(\frac{1}{2}\ln\bigg[\frac{e^{2q_+}}{e^{2q_+}-1-e^{2k_+}}\bigg]\bigg),\\
      &\hat{\beta}^{\dagger}(k_+)\hat{a}(q_+)=\hat{\beta}^{\dagger}\bigg(\frac{1}{2}\ln\bigg[\frac{e^{2k_+}}{e^{2q_+}-1-e^{2k_+}}\bigg]\bigg)\hat{a}\bigg(\frac{1}{2}\ln\bigg[\frac{e^{2q_+}}{e^{2q_+}-1-e^{2k_+}}\bigg]\bigg),\\
      &\hat{\alpha}^{\dagger}(k_+)\hat{b}(q_+)=\hat{\alpha}^{\dagger}\bigg(\frac{1}{2}\ln\bigg[\frac{e^{2k_+}}{e^{2q_+}-1-e^{2k_+}}\bigg]\bigg)\hat{b}\bigg(\frac{1}{2}\ln\bigg[\frac{e^{2q_+}}{e^{2q_+}-1-e^{2k_+}}\bigg]\bigg).
     \end{aligned}
 \end{equation}
\item Region $7_D$: $k_+\in]0,+\infty[$ and $ q_+\in \big]\frac{1}{2}\ln(2e^{2k_+}+1),+\infty\big[$ 
 \begin{equation}
     \begin{aligned}
      &\hat{\beta}^{\dagger}(k_+)\hat{b}(q_+)=\frac{e^{2(q_++k_+)}}{e^{2q_+}-1-e^{2k_+}}\hat{\beta}\bigg(\frac{1}{2}\ln\bigg[\frac{e^{2q_+}-1-e^{2k_+}}{e^{2k_+}}\bigg]\bigg)\hat{a}\bigg(\frac{1}{2}\ln\bigg[\frac{e^{2q_+}}{e^{2q_+}-1-e^{2k_+}}\bigg]\bigg),\\
      &\hat{\beta}^{\dagger}(k_+)\hat{a}(q_+)=\hat{\alpha}\bigg(\frac{1}{2}\ln\bigg[\frac{e^{2q_+}-1-e^{2k_+}}{e^{2k_+}}\bigg]\bigg)\hat{a}\bigg(\frac{1}{2}\ln\bigg[\frac{e^{2q_+}}{e^{2q_+}-1-e^{2k_+}}\bigg]\bigg),\\
      &\hat{\alpha}^{\dagger}(k_+)\hat{b}(q_+)=\hat{\beta}\bigg(\frac{1}{2}\ln\bigg[\frac{e^{2q_+}-1-e^{2k_+}}{e^{2k_+}}\bigg]\bigg)\hat{b}\bigg(\frac{1}{2}\ln\bigg[\frac{e^{2q_+}}{e^{2q_+}-1-e^{2k_+}}\bigg]\bigg).
     \end{aligned}
 \end{equation}
 
\end{itemize}
 
 \subsection*{Remaining products, valid in all of $\mathbb{R}^2_+$}
\allowdisplaybreaks
For $k_+\in]0,+\infty[$ and $q_+\in]0,+\infty[$
\begin{align} 
    &\begin{aligned}[t]&\hat{a}(k_+)\hat{b}(q_+)=
    \\&\frac{e^{2q_+}}{e^{2(q_++k_+)}-e^{2k_+}+1}\hat{b}\bigg(\frac{1}{2}\ln\bigg[e^{2(q_++k_+)}-e^{2k_+}+1\bigg]\bigg)\hat{a}\bigg(\frac{1}{2}\ln\bigg[\frac{e^{2(q_++k_+)}}{e^{2(q_++k_+)}-e^{2k_+}+1}\bigg]\bigg)
    \end{aligned}\\
    &\begin{aligned}[t]&\hat{b}^{\dagger}(k_+)\hat{a}^{\dagger}(q_+)=\\&\frac{e^{2k_+}}{e^{2(q_++k_+)}-e^{2q_+}+1}\hat{a}^{\dagger}\bigg(\frac{1}{2}\ln\bigg[\frac{e^{2(q_++k_+)}}{e^{2(q_++k_+)}-e^{2q_+}+1}\bigg]\bigg)\hat{b}^{\dagger}\bigg(\frac{1}{2}\ln\bigg[e^{2(q_++k_+)}-e^{2q_+}+1\bigg]\bigg),\end{aligned}
    \\
    &\begin{aligned}[t]&\hat{b}^{\dagger}(k_+)\hat{b}(q_+)-\frac{e^{2(q_++k_+)}}{e^{2q_+}+e^{2k_+}-1}\hat{b}\bigg(-\frac{1}{2}\ln\bigg[\frac{e^{2k_+}}{e^{2q_+}+e^{2k_+}-1}\bigg]\bigg)\hat{b}^{\dagger}\bigg(-\frac{1}{2}\ln\bigg[\frac{e^{2q_+}}{e^{2q_+}+e^{2k_+}-1}\bigg]\bigg)\\
    &=-i\delta(q_+-k_+){(e^{2k_+}-1)},\end{aligned}
    \\
    &\begin{aligned}[t]&\hat{\alpha}(k_+)\hat{\alpha}^{\dagger}(q_+)+\frac{1}{e^{2(q_++k_+)}+e^{2q_+}+e^{2k_+}}\hat{b}\bigg(-\frac{1}{2}\ln\bigg[\frac{e^{2q_+}}{e^{2(q_++k_+)}+e^{2q_+}+e^{2k_+}}\bigg]\bigg)\\&\hat{b}^{\dagger}\bigg(-\frac{1}{2}\ln\bigg[\frac{e^{2k_+}}{e^{2(q_++k_+)}+e^{2q_+}+e^{2k_+}}\bigg]\bigg)=i\delta(q_+-k_+)\frac{(e^{2k_+}+1)}{e^{2k_+}},\end{aligned} \\
    &\begin{aligned}[t]&\hat{\alpha}(k_+)\hat{\beta}(q_+)=\\&-\frac{ e^{2q_+}}{e^{2(q_++k_+)}+e^{2k_+}+1}\hat{b}\bigg(\frac{1}{2}\ln\bigg[e^{2(q_++k_+)}+e^{2k_+}+1\bigg]\bigg)\hat{b}^{\dagger}\bigg(-\frac{1}{2}\ln\bigg[\frac{e^{2(q_++k_+)}}{e^{2(q_++k_+)}+e^{2k_+}+1}\bigg]\bigg),\end{aligned}\\
    &\begin{aligned}[t]&\hat{\beta}^{\dagger}(k_+)\hat{\alpha}^{\dagger}(q_+)=\\&-\frac{ e^{2k_+}}{e^{2(q_++k_+)}+e^{2q_+}+1}\hat{b}\bigg(-\frac{1}{2}\ln\bigg[\frac{e^{2(q_++k_+)}}{e^{2(q_++k_+)}+e^{2q_+}+1}\bigg]\bigg)\hat{b}^{\dagger}\bigg(\frac{1}{2}\ln\bigg[{e^{2(q_++k_+)}+e^{2q_+}+1}\bigg]\bigg),\end{aligned}\\
    &\begin{aligned}[t]&\hat{\beta}^{\dagger}(k_+)\hat{\beta}(q_+)+\frac{e^{2(q_++k_+)}}{(e^{2q_+}+e^{2k_+}+1)}\hat{b}\bigg(-\frac{1}{2}\ln\bigg[\frac{e^{2k_+}}{e^{2q_+}+e^{2k_+}+1}\bigg]\bigg)\\&\hat{b}^{\dagger}\bigg(-\frac{1}{2}\ln\bigg[\frac{e^{2q_+}}{e^{2q_+}+e^{2k_+}+1}\bigg]\bigg)=i\delta(q_+-k_+){(e^{2k_+}+1)},\end{aligned}\\
    &\hat{a}(k_+)\hat{a}(q_+)=\hat{a}\bigg(\frac{1}{2}\ln\bigg[e^{2(q_++k_+)}-e^{2k_+}+1\bigg]\bigg)\hat{a}\bigg(\frac{1}{2}\ln\bigg[\frac{e^{2(q_++k_+)}}{e^{2(q_++k_+)}-e^{2k_+}+1}\bigg]\bigg),\\
    &\hat{b}^{\dagger}(k_+)\hat{b}^{\dagger}(q_+)=\hat{b}^{\dagger}\bigg(\frac{1}{2}\ln\bigg[\frac{e^{2(q_++k_+)}}{e^{2(q_++k_+)}-e^{2q_+}+1}\bigg]\bigg)\hat{b}^{\dagger}\bigg(\frac{1}{2}\ln\bigg[e^{2(q_++k_+)}-e^{2q_+}+1\bigg]\bigg),\\
    &\hat{b}^{\dagger}(k_+)\hat{a}(q_+)=\hat{a}\bigg(-\frac{1}{2}\ln\bigg[\frac{e^{2k_+}}{e^{2q_+}+e^{2k_+}-1}\bigg]\bigg)\hat{b}^{\dagger}\bigg(-\frac{1}{2}\ln\bigg[\frac{e^{2q_+}}{e^{2q_+}+e^{2k_+}-1}\bigg]\bigg),\\
    &\hat{\alpha}(k_+)\hat{\beta}^{\dagger}(q_+)=\hat{a}\bigg(-\frac{1}{2}\ln\bigg[\frac{e^{2q_+}}{e^{2(q_++k_+)}+e^{2q_+}+e^{2k_+}}\bigg]\bigg)\hat{b}^{\dagger}\bigg(-\frac{1}{2}\ln\bigg[\frac{e^{2k_+}}{e^{2(q_++k_+)}+e^{2q_+}+e^{2k_+}}\bigg]\bigg),\\
    &\hat{\alpha}(k_+)\hat{\alpha}(q_+)=-\hat{a}\bigg(\frac{1}{2}\ln\bigg[e^{2(q_++k_+)}+e^{2k_+}+1\bigg]\bigg)\hat{b}^{\dagger}\bigg(-\frac{1}{2}\ln\bigg[\frac{e^{2(q_++k_+)}}{e^{2(q_++k_+)}+e^{2k_+}+1}\bigg]\bigg),\\
    &\hat{\beta}^{\dagger}(k_+)\hat{\beta}^{\dagger}(q_+)=-\hat{a}\bigg(-\frac{1}{2}\ln\bigg[\frac{e^{2(q_++k_+)}}{e^{2(q_++k_+)}+e^{2q_+}+1}\bigg]\bigg)\hat{b}^{\dagger}\bigg(\frac{1}{2}\ln\bigg[{e^{2(q_++k_+)}+e^{2q_+}+1}\bigg]\bigg),\\
    &\hat{\beta}^{\dagger}(k_+)\hat{\alpha}(q_+)=-\hat{a}\bigg(-\frac{1}{2}\ln\bigg[\frac{e^{2k_+}}{e^{2q_+}+e^{2k_+}+1}\bigg]\bigg)\hat{b}^{\dagger}\bigg(-\frac{1}{2}\ln\bigg[\frac{e^{2q_+}}{e^{2q_+}+e^{2k_+}+1}\bigg]\bigg),\\
    &\hat{b}(k_+)\hat{b}(q_+)=\hat{b}\bigg(\frac{1}{2}\ln\bigg[e^{2(q_++k_+)}-e^{2k_+}+1\bigg]\bigg)\hat{b}\bigg(\frac{1}{2}\ln\bigg[\frac{e^{2(q_++k_+)}}{e^{2(q_++k_+)}-e^{2k_+}+1}\bigg]\bigg),\\
    &\hat{a}^{\dagger}(k_+)\hat{a}^{\dagger}(q_+)=\hat{a}^{\dagger}\bigg(\frac{1}{2}\ln\bigg[\frac{e^{2(q_++k_+)}}{e^{2(q_++k_+)}-e^{2q_+}+1}\bigg]\bigg)\hat{a}^{\dagger}\bigg(\frac{1}{2}\ln\bigg[e^{2(q_++k_+)}-e^{2q_+}+1\bigg]\bigg),\\
    &\hat{a}^{\dagger}(k_+)\hat{b}(q_+)=\hat{b}\bigg(-\frac{1}{2}\ln\bigg[\frac{e^{2k_+}}{e^{2q_+}+e^{2k_+}-1}\bigg]\bigg)\hat{a}^{\dagger}\bigg(-\frac{1}{2}\ln\bigg[\frac{e^{2q_+}}{e^{2q_+}+e^{2k_+}-1}\bigg]\bigg),\\
    &\hat{\beta}(k_+)\hat{\alpha}^{\dagger}(q_+)=\hat{b}\bigg(-\frac{1}{2}\ln\bigg[\frac{e^{2q_+}}{e^{2(q_++k_+)}+e^{2q_+}+e^{2k_+}}\bigg]\bigg)\hat{a}^{\dagger}\bigg(-\frac{1}{2}\ln\bigg[\frac{e^{2k_+}}{e^{2(q_++k_+)}+e^{2q_+}+e^{2k_+}}\bigg]\bigg),\\
    &\hat{\beta}(k_+)\hat{\beta}(q_+)=-\hat{b}\bigg(\frac{1}{2}\ln\bigg[e^{2(q_++k_+)}+e^{2k_+}+1\bigg]\bigg)\hat{a}^{\dagger}\bigg(-\frac{1}{2}\ln\bigg[\frac{e^{2(q_++k_+)}}{e^{2(q_++k_+)}+e^{2k_+}+1}\bigg]\bigg),\\
    &\hat{\alpha}^{\dagger}(k_+)\hat{\alpha}^{\dagger}(q_+)=-\hat{b}\bigg(-\frac{1}{2}\ln\bigg[\frac{e^{2(q_++k_+)}}{e^{2(q_++k_+)}+e^{2q_+}+1}\bigg]\bigg)\hat{a}^{\dagger}\bigg(\frac{1}{2}\ln\bigg[{e^{2(q_++k_+)}+e^{2q_+}+1}\bigg]\bigg),\\
    &\hat{\alpha}^{\dagger}(k_+)\hat{\beta}(q_+)=-\hat{b}\bigg(-\frac{1}{2}\ln\bigg[\frac{e^{2k_+}}{e^{2q_+}+e^{2k_+}+1}\bigg]\bigg)\hat{a}^{\dagger}\bigg(-\frac{1}{2}\ln\bigg[\frac{e^{2q_+}}{e^{2q_+}+e^{2k_+}+1}\bigg]\bigg),
    \\
      &\begin{aligned}[t]&\hat{a}(k_+)\hat{\beta}(q_+)=\\&\frac{e^{2q_+}}{e^{2(k_++q_+)}+e^{2k_+}-1}\hat{\beta}\bigg(\frac{1}{2}\ln\bigg[e^{2(k_++q_+)}+e^{2k_+}-1\bigg]\bigg)\hat{b}^{\dagger}\bigg(\frac{1}{2}\ln\bigg[\frac{e^{2(k_++q_+)}+e^{2k_+}-1}{e^{2(q_++k_+)}}\bigg]\bigg),\end{aligned}\\
     &\hat{a}(k_+)\hat{\alpha}(q_+)= \hat{\alpha}\bigg(\frac{1}{2}\ln\bigg[e^{2(k_++q_+)}+e^{2k_+}-1\bigg]\bigg)\hat{b}^{\dagger}\bigg(\frac{1}{2}\ln\bigg[\frac{e^{2(k_++q_+)}+e^{2k_+}-1}{e^{2(q_++k_+)}}\bigg]\bigg),\\
     &\hat{b}(k_+)\hat{\beta}(q_+)= \hat{\beta}\bigg(\frac{1}{2}\ln\bigg[e^{2(k_++q_+)}+e^{2k_+}-1\bigg]\bigg)\hat{a}^{\dagger}\bigg(\frac{1}{2}\ln\bigg[\frac{e^{2(k_++q_+)}+e^{2k_+}-1}{e^{2(q_++k_+)}}\bigg]\bigg),\\
     &\begin{aligned}[t]&\hat{\beta}^{\dagger}(k_+)\hat{a}^{\dagger}(q_+)=\\&\frac{e^{2k_+}}{e^{2(k_++q_+)}+e^{2q_+}-1}\hat{b}\bigg(\frac{1}{2}\ln\bigg[\frac{e^{2(k_++q_+)}+e^{2q_+}-1}{e^{2(k_++q_+)}}\bigg]\bigg)\hat{\beta}^{\dagger}\bigg(\frac{1}{2}\ln\bigg[{e^{2(k_++q_+)}+e^{2q_+}-1}\bigg]\bigg),\end{aligned}\\
     &\hat{\beta}^{\dagger}(k_+)\hat{b}^{\dagger}(q_+)= \hat{a}\bigg(\frac{1}{2}\ln\bigg[\frac{e^{2(k_++q_+)}+e^{2q_+}-1}{e^{2(k_++q_+)}}\bigg]\bigg)\hat{\beta}^{\dagger}\bigg(\frac{1}{2}\ln\bigg[{e^{2(k_++q_+)}+e^{2q_+}-1}\bigg]\bigg),\\
     &\hat{\alpha}^{\dagger}(k_+)\hat{a}^{\dagger}(q_+)= \hat{b}\bigg(\frac{1}{2}\ln\bigg[\frac{e^{2(k_++q_+)}+e^{2q_+}-1}{e^{2(k_++q_+)}}\bigg]\bigg)\hat{\alpha}^{\dagger}\bigg(\frac{1}{2}\ln\bigg[{e^{2(k_++q_+)}+e^{2q_+}-1}\bigg]\bigg).
\end{align}

\section{Conclusions}
In our previous work~\cite{PhysRevD.103.126009}, we proved that one can define a QFT on $\kappa$-Minkowski in terms of $N$-point functions, only in the so-called \textit{lightlike} case: $g_{\mu\nu}v^{\mu}v^{\nu}=0$. This condition is in fact needed to construct a $\kappa$-Poincaré-covariant algebra of functions of more than more point. However, the momentum space of such theory turns out to have a boundary, which can be crossed with a finite $\kappa$-Poincaré transformation of momenta.
In this work we showed how this boundary represents an obstruction to the definition of the Pauli-Jordan function and hence to the quantization procedure itself. Inspired by~\cite{Arzano:2020jro}, we proposed an ``extension" of momentum space beyond said boundary through the introduction of what we called new-type plane waves. These arise naturally from the observation that momentum space is a group manifold and the action of $\kappa$-Lorentz on it brings one out of the component connected to the identity, into a disconnected component~\cite{Arzano:2020jro}. We then studied explicitly all possible two-point functions, which can be written in terms of ordinary as well as new-type plane waves. 
We proved that, if we require invariance under finite $\kappa$-Poincaré transformations, we end up with the commutative, undeformed, two-point function. The fact that two-point functions are undeformed suggests that all $N$-point functions of the free classical field theory might be so. However, we leave the investigation of this conjecture to future works, together with the study of interacting theories. 

Using the undeformed Pauli-Jordan function~(\ref{eq:PJ}), we imposed the quantization rules for free complex $\kappa$-Klein-Gordon fields. We obtained a deformed bosonic oscillator algebra for the creation and annihilation operators, similarly to other results in the $\kappa$-QFT literature (see for instance~\cite{Arzano:2007ef,Daszkiewicz:2007az}).

Future investigations will need to address the issue of the fate of discrete symmetries, \textit{i.e.} $\mathcal{C}, \,\, \mathcal{P}$ and $\mathcal{T}$, in the theory, their (necessarily deformed) action on the bosonic oscillator algebra and their interactions with the $\kappa$-Poincaré transformations. With the results at our disposal at this point, we can only conjecture that these symmetries might not be broken in this theory, because they certainly are present at the level of the two-point functions.

There is a number of directions in which this line of investigation might develop in the future: the first ones are the already-mentioned study of $N$-point functions and of interacting scalar field theories. Further down the road, gauge theories and fermions might be explored.

\section*{Acknowledgements}
F.M.\  has been partially supported by Agencia Estatal de Investigaci\'on (Spain)  under grant  PID2019-106802GB-I00/AEI/10.13039/501100011033.

\bibliographystyle{utphys}

\bibliography{BraidedBib}

\appendix

\section*{Appendices}

\section{Two-point new-type plane waves} \label{appendixA}
Using the identity derived in \cite{PhysRevD.103.126009}:
\begin{equation}
    e^{ir_-x_1^-}e^{ir_+x_1^+}e^{is_-x_2^-}e^{is_+x_2^+}= e^{i(r_-x^-_1+e^{-2r_+}s_-x_2^-)}e^{i\frac{r_++s_+}{1-e^{2(r_++s_+)}} [(1-e^{2r_+})x_1^++e^{2r_+}(1-e^{2s_+})x_2^+]},
\end{equation}
we can re-order the product $E_1[k]\mathcal{E}_2[q]$ in Equation (\ref{eq:products}) as 
\begin{equation} \label{eq:prefact}
E_1[k]\mathcal{E}_2[q] =E_1[k]E_2[q]e^{-\frac{\pi}{2}x_2^+}= e^{i(k_-x^-_1+e^{-2k_+}q_-x_2^-)}e^{i\frac{k_++q_++i\pi/2}{1+e^{2(k_++q_+)}} [(1-e^{2k_+})x_1^++e^{2k_+}(1+e^{2q_+})x_2^+]}.
\end{equation}
There are two cases in which this expression only depends upon $x_1^{\mu}-x_2^{\mu}$, according to whether one solves the corresponding conditions for $k_{\mu}$ or $q_{\mu}$. The first case is when
\begin{equation}
\begin{cases}
    q_-=-e^{2k_+}k_-\\ e^{2(k_++q_+)}=-1 
    \end{cases}
    \Longrightarrow
    \begin{cases}  q_-=S(k)_-\\
    q_+=-k_+-i\frac{\pi}{2}+ni\pi=S(k)_+-i\frac{\pi}{2} +ni\pi, \quad n\in\mathbb{Z}.
    \end{cases}
\end{equation}
Notice that, if the second equation is satisfied, the denominator of the prefactor $\frac{k_++q_++i\pi/2}{1+e^{2(k_++q_+)}}$ in~(\ref{eq:prefact}) vanishes. This means that the prefactor itself can only be finite if $n=0$. In this case:
\begin{equation} \label{eq:case1}
     \begin{cases}  q_-=S(k)_-\\
    q_+=-k_+-i\frac{\pi}{2}=S(k)_+-i\frac{\pi}{2} .
    \end{cases}
\end{equation}
The second solution again requires $k_++q_+=-i\pi/2$ and takes the following form:
\begin{equation} \label{eq:case2}
\begin{cases}
k_-=e^{2q_+}q_-=S(q_-,q_++i\pi/2)_-\\
k_+=-q_+-i\frac{\pi}{2}=S(q_-,q_++i\pi/2)_+.
\end{cases}
\end{equation}
In the first case~(\ref{eq:case1}), the product simply reduces to $E_1[k]E_2^{\dagger}[k]$ as in Equation (\ref{eq:2pplwv}). In the second case~(\ref{eq:case2}), instead, the product becomes 
\begin{equation} \label{eq:red1}
   E_1[S(q_-,q_++i\pi/2)]E_2[(q_-,q_++i\pi/2)]= E_1^{\dagger}[(q_-,q_++i\pi/2)]E_2[(q_-,q_++i\pi/2)] =\mathcal{E}_1^{\dagger}[q]\mathcal{E}_2[q].
\end{equation}
The same reasoning can be applied to the product $\mathcal{E}_1[k]E_2[q]$ in (\ref{eq:products}), which either reduces to  $E_1^{\dagger}[q]E_2[q]$ or to 
\begin{equation} \label{eq:red2}
 E_1[(k_-,k_++i\pi/2)]E_2^{\dagger}[(k_-,k_++i\pi/2)]=\mathcal{E}_1[k]\mathcal{E}_2^{\dagger}[k].   
\end{equation}
 Lastly, the product $\mathcal{E}_1[k]\mathcal{E}_2[q]$ reduces either to~(\ref{eq:red1}) or to~(\ref{eq:red2}).

\section{Hermitian conjugate of on-shell plane waves} \label{appendixB}
We want to prove that $\epsilon_a(-k_+)=\epsilon^{\dagger}_a(k_+)$. First of all notice that
\begin{equation}
    \epsilon_a^{\dagger}(k_+)=e^{-\frac{\pi}{2}x_a^+}e^{-ik_+x_a^+}e^{-i\Omega_r(k_+)x_a^-}=e^{-\pi x_a^+}\epsilon_a^{-1}(k_+)=-\epsilon^{-1}_a(k_+),
\end{equation}
where we used the relation $e^{-\pi x^+_a}=-1$ (see Equation (\ref{eq:repprx+})). We will now verify that $\epsilon_a(-k_+)=-\epsilon_a^{-1}(k_+)$.
Using the commutator $[x_a^+,x_a^-]=2ix^-_a$ of Equation (\ref{eq:1+1alg}), we can show that $e^{-(\pi/2)x_a^+}x_a^- =-x_a^-e^{-(\pi/2)x_a^+}$:
\begin{equation}
    \begin{aligned}
    &x^+_ax^-_a=x^-_ax^+_a+2ix^-_a=x^-_a(x^+_a+2i) \Longrightarrow (x_a^+)^nx^-_a=x^-_a(x^+_a+2i)^n \\
    \Longrightarrow \,&e^{-\frac{\pi}{2}x_a^+}x_a^- =\sum_{n=0}^{+\infty}\bigg(-\frac{\pi}{2}\bigg)^n(x_a^+)^nx^-_a=x_a^-\sum_{n=0}^{+\infty}\bigg(-\frac{\pi}{2}\bigg)^n(x^+_a+2i)^n=e^{-i\pi}x_a^-e^{-\frac{\pi}{2}x_a^+}.
    \end{aligned}
\end{equation}
This implies that the product of two new-type plane waves of the same point gives an ordinary plane wave:
\begin{equation}\label{eq:prodnpw} \begin{aligned}
    \mathcal{E}_a[k]\mathcal{E}_a[q]&=E_a[k]e^{-\frac{\pi}{2}x_a^+}E_a[q]e^{-\frac{\pi}{2}x_a^+}=E_a[k]E_a[(-q_-,q_+)]e^{-\pi x^+_a}\\
    &=-E_a[k]E_a[(-q_-,q_+)]=-E_a[k\oplus(-q_-,q_+)]=-E_a[(k_--e^{-2k_+}q_-,k_++q_+)].
\end{aligned}\end{equation}

Setting the momenta in Equation~(\ref{eq:prodnpw}) on-shell, we can finally compute the product $\epsilon_a(-k_+)\epsilon_a(k_+)$:
\begin{equation} \begin{aligned}
    &\epsilon_a(-k_+)\epsilon_a(k_+)=\mathcal{E}_a\bigg[-\frac{2m^2}{e^{-2k_+}+1},-k_+\bigg]\mathcal{E}_a\bigg[-\frac{2m^2}{e^{2k_+}+1},k_+\bigg]\\
    &=-E_a\bigg[\bigg(-\frac{2m^2}{e^{-2k_+}+1}+e^{2k_+}\frac{2m^2}{e^{2k_+}+1},-k_++k_+\bigg)\bigg]=-1.
\end{aligned}\end{equation}
In the same way, we can verify that $\epsilon_a(k_+)\epsilon_a(-k_+)=-1$.

To recap, we proved the following chain of identities:
\begin{equation}
    \epsilon^{\dagger}_a(k_+)=-\epsilon^{-1}_a(k_+)=\epsilon_a(-k_+).
\end{equation}

\section{Flipping new-type plane waves} \label{appendixC}
In \cite{PhysRevD.103.126009}, we proved that, given the commutator $[A,B]=icB$ (with $c$ being some constant), then 
\begin{equation}
    e^{ip_0A}e^{ip_1B}=e^{ie^{-cp_0}p_1B}e^{ip_0A}\,\, \text{ and }\, \, e^{i(1-e^{-cp_0})p_1B}e^{ip_0A}=e^{ip_0(A+cp_1B)}.
\end{equation}

Recalling the commutation relations (\ref{eq:1+1alg}) and the result in Equation (\ref{eq:action+}), we can write how the product of two plane waves acts on a function $f(x_{cm}^-,y^{\pm}_a)$:

\begin{equation} \label{eq:granlavoro}\begin{aligned}
  E_1[k]E_2[q]f(x_{cm}^-,y^{\pm}_a)=&e^{-(k_++q_+)}e^{i\big[k_-\big(x^-_{cm}+y_1^-\big)+\big(\frac{e^{2k_+}-1}{2}\big)y_1^+\big]}  \\& e^{i\big[e^{-2k_+}q_-\big(x_{cm}^-+y_2^-\big)+e^{2k_+}\big(\frac{e^{2q_+}-1}{2}\big)y_2^+\big]}f(e^{-2(k_++q_+)}x^-_{cm},e^{\pm2(k_++q_+)}y^{\pm}_a),
\end{aligned}\end{equation}

for any complex value of $k$ and $q$. This result is obtained by moving the operator $e^{i(k_++q_+)x^+_{cm}}$ to the right.
If we now ask $E_1[k]E_2[q]=E_2[r]E_1[p]$, we obtain the following conditions:

\begin{equation}
    \begin{cases}
    e^{2r_+}=e^{2k_+}(e^{2q_+}-1)+1 \\
    r_-=e^{-2k_+}q_- \\
    e^{2p_+}=\frac{e^{2(k_++q_+)}}{e^{2k_+}(e^{2q_+}-1)+1}\\
    p_-=(e^{2k_+}(e^{2q_+}-1)+1)k_- \\
    r_++p_+=k_++q_++2li\pi
    \end{cases}
    \Longrightarrow
    \begin{cases}
    r_+=\frac{1}{2}\ln(e^{2k_+}(e^{2q_+}-1)+1)+ni\pi \\
    r_-=e^{-2k_+}q_-\\
    p_+=k_++q_+-\frac{1}{2}\ln(e^{2k_+}(e^{2q_+}-1)+1)+mi\pi\\
     p_-=(e^{2k_+}(e^{2q_+}-1)+1)k_-\\
     r_++p_+=k_++q_++2li\pi,
    \end{cases}
\end{equation}
where $n,m,l\in\mathbb{Z}$. Replacing the first and third equations into the last we get the constraint $n+m=2l$, \textit{i.e.} $n+m$ has to be even. From Equation (\ref{eq:granlavoro}), we notice that $e_1[k^{\mu}+in\pi\delta^{\mu}_+]e_2[q^{\mu}+im\pi\delta^{\mu}_+]=e^{-\pi(nx_1^++mx_2^+)}e_1[k]e_2[q]=e_1[k]e_2[q]$  $\forall m,n:\!(m+n)$ is even. We conclude that, although reordering plane waves requires us to invert some multi-valued functions, the appearance of the corresponding branch numbers does not introduce any ambiguity because they cancel out from the final expression.

\section{Two-point function with new-type plane waves} \label{appendixD}
Let us compute the term in the two-point function (\ref{eq:newtwopoint}) that involves new-type plane waves:
\begin{equation}
    \begin{aligned}
    H(x_1-x_2)&=\int_{-\infty}^{+\infty}dk_+\frac{e^{2k_+}}{e^{2k_+}+1}e^{-\frac{2im^2}{e^{2k_+}+1}(x_1^--x_2^-)}e^{-i\frac{e^{2k_+}-1}{2}(x_1^+-x_2^+)}h\bigg(k_+,-\frac{2m^2}{e^{2k_+}+1}\bigg)\\
    &=\frac{1}{2}\int_0^{+\infty}\frac{dy}{y+1}e^{-\frac{2im^2}{y+1}(x_1^--x_2^-)}e^{-i\frac{y+1}{2}(x_1^+-x_2^+)}h\bigg(\frac{1}{2}\ln(y),-\frac{2m^2}{y+1}\bigg)\\
    &=\frac{1}{2}\int_1^{+\infty}\frac{dz}{z}e^{-\frac{2im^2}{z}(x_1^--x_2^-)}e^{-i\frac{z}{2}(x_1^+-x_2^+)}h\bigg(\frac{1}{2}\ln(y),-\frac{2m^2}{y+1}\bigg)\\
    &=\frac{1}{2}\int_1^{+\infty}\frac{dz}{z}e^{-\frac{2im^2}{z}(x_1^--x_2^-)}e^{-i\frac{z}{2}(x_1^+-x_2^+)}h\bigg(\frac{1}{2}\ln(z-1),-\frac{2m^2}{z}\bigg).
    \end{aligned}
\end{equation}
If we now apply the condition (\ref{eq:backfor}) and restore $\kappa$, we get
\begin{equation}
    \begin{aligned}
    &H(x_1-x_2)=\frac{1}{2}h_-\int_1^2\frac{dz}{z}e^{-\frac{2im^2}{\kappa z}(x_1^--x_2^-)}e^{-i\frac{\kappa z}{2}(x_1^+-x_2^+)}+\frac{1}{2}h_+\int_2^{+\infty}\frac{dz}{z}e^{-\frac{2im^2}{\kappa z}(x_1^--x_2^-)}e^{-i\frac{\kappa z}{2}(x_1^+-x_2^+)}\\
    &=\frac{1}{2}h_-\int_{\frac{\kappa}{2m}}^{\frac{\kappa}{m}}\frac{du}{u}e^{-i\frac{m}{u}(x_1^--x_2^-)}e^{-imu(x_1^+-x_2^+)}+\frac{1}{2}h_+\int_{\frac{\kappa}{m}}^{+\infty}\frac{du}{u}e^{-i\frac{m}{u}(x_1^--x_2^-)}e^{-imu(x_1^+-x_2^+)}\\
    &=\frac{1}{2}h_-\int_{\ln(\frac{\kappa}{2m})}^{\ln(\frac{\kappa}{m})}d\chi e^{-ime^{-\chi}(x_1^--x_2^-)}e^{-ime^{\chi}(x_1^+-x_2^+)}+\frac{1}{2}h_+\int_{\ln(\frac{\kappa}{m})}^{+\infty}d\chi e^{-ime^{-\chi}(x_1^--x_2^-)}e^{-ime^{\chi}(x_1^+-x_2^+)}\\
    &=\frac{1}{2}h_-\int_{m\sinh(\ln(\frac{\kappa}{2m}))}^{m\sinh(\ln(\frac{\kappa}{m}))}\frac{dp}{\sqrt{p^2+m^2}} e^{-i(\sqrt{p^2+m^2}-p)(x_1^--x_2^-)}e^{-i(\sqrt{p^2+m^2}+p)(x_1^+-x_2^+)}+\\
    &+\frac{1}{2}h_+\int_{m\sinh(\ln(\frac{\kappa}{m}))}^{+\infty}\frac{dp}{\sqrt{p^2+m^2}} e^{-i(\sqrt{p^2+m^2}-p)(x_1^--x_2^-)}e^{-i(\sqrt{p^2+m^2}+p)(x_1^+-x_2^+)}\\
    &=\frac{1}{2}h_-\int_{m\sinh(\ln(\frac{\kappa}{2m}))}^{m\sinh(\ln(\frac{\kappa}{m}))}\frac{dp}{\sqrt{p^2+m^2}}e^{-2i(\sqrt{p^2+m^2}(x_1^0-x_2^0)+p(x_1^1-x_2^1))} +\\
    &+\frac{1}{2}h_+\int_{m\sinh(\ln(\frac{\kappa}{m}))}^{+\infty}\frac{dp}{\sqrt{p^2+m^2}}e^{-2i(\sqrt{p^2+m^2}(x_1^0-x_2^0)+p(x_1^1-x_2^1))}.
    \end{aligned}
\end{equation}

In principle, we could also have used the other invariant two-point plane waves to write a different two-point function:
\begin{equation}
    \begin{aligned}
    &\Delta'(x_1-x_2)=L(x_1-x_2)+M(x_1-x_2), \\
    &L(x_1-x_2)=\int d^2k E_1^{\dagger}[k]E_2[k]l(k)\delta(\tilde C(k)-m^2),\\
    &M(x_1-x_2)=\int d^2k \mathcal E_1^{\dagger}[k]\mathcal E_2[k]m (k)\delta(\breve{C}(k)-m^2).
    \end{aligned}
\end{equation}

However, if we assume $l(k)$ and $m(k)$ to be constant on the forward and backward light cones, $l(k)=l_-\Theta(-k_+)+l_+\Theta(k_+)$ and $m(k)=m_-\Theta(-k_+)+m_+\Theta(k_+)$, then a similar calculation gives
\begin{equation}
\begin{aligned}
\Delta '(x_1-x_2)&=l_-\int_{-\infty}^{+\infty}\frac{dp}{\sqrt{p^2+m^2}}e^{2i(\sqrt{p^2+m^2}(x_1^0-x_2^0)+p(x_1^1-x_2^1))}+\\&+l_+\int_{-\infty}^{m\sinh(\ln(\frac{\kappa}{2m}))}\frac{dp}{\sqrt{p^2+m^2}}e^{-2i(\sqrt{p^2+m^2}(x_1^0-x_2^0)+p(x_1^1-x_2^1))}+\\
&+m_+\int_{m\sinh(\ln(\frac{\kappa}{2m}))}^{m\sinh(\ln(\frac{\kappa}{m}))}\frac{dp}{\sqrt{p^2+m^2}}e^{-2i(\sqrt{p^2+m^2}(x_1^0-x_2^0)+p(x_1^1-x_2^1))} + \\&+m_-\int_{m\sinh(\ln(\frac{\kappa}{m}))}^{+\infty}\frac{dp}{\sqrt{p^2+m^2}}e^{-2i(\sqrt{p^2+m^2}(x_1^0-x_2^0)+p(x_1^1-x_2^1))}
\end{aligned}
\end{equation}
So for $m_+=m_-=l_+$ we obtain again the commutative two-point function.

\section{Bosonic oscillator algebra} \label{appendixE}
We want to derive the conditions that the commutators in Equation (\ref{eq:commrel}) imply on the Fourier coefficients of the fields. Let us consider the first equation in~(\ref{eq:commrel}),  $[\hat{\phi}(x_1),\hat{\phi}^{\dagger}(x_2)]=i \Delta_{\text{PJ}}(x_1-x_2)$. The left hand-side can be written as the sum of 32 integrals:
\begin{equation}
    [\hat{\phi}(x_1),\hat{\phi}^{\dagger}(x_2)]=\sum_{i=1}^{8}I^{(1)}_i+\sum_{i=1}^{8}J^{(1)}_i+\sum_{i=1}^{16}H^{(1)}_i \,.
\end{equation}
The first 16 integrals involve products of plane waves of the same type:
\begin{equation}
    \begin{aligned}
    I^{(1)}_1&=\int _0 ^{+\infty}dk_+ \int _0 ^{+\infty} dq_+ \frac{e^{2(k_++q_+)}}{(e^{2k_+}-1)(e^{2q_+}-1)}\hat{a}(k_+)\hat{a}^{\dagger}(q_+)\textcolor{black}{e_1(k_+)e_2^{\dagger}(q_+)},\\
    -I^{(1)}_2&=\int _0 ^{+\infty}dk_+ \int _0 ^{+\infty} dq_+ \frac{e^{2(k_++q_+)}}{(e^{2k_+}-1)(e^{2q_+}-1)}\hat{a}^{\dagger}(q_+)\hat{a}(k_+)e_2^{\dagger}(q_+)e_1(k_+),\\
    I^{(1)}_3&=\int _0 ^{+\infty}dk_+ \int _0 ^{+\infty} dq_+ \frac{e^{2k_+}}{(e^{2k_+}-1)(e^{2q_+}-1)}\hat{a}(k_+)\hat{b}(q_+)\textcolor{black}{e_1(k_+)e_2(q_+)},\\
    -I^{(1)}_4&=\int _0 ^{+\infty}dk_+ \int _0 ^{+\infty} dq_+ \frac{e^{2k_+}}{(e^{2k_+}-1)(e^{2q_+}-1)}\hat{b}(q_+)\hat{a}(k_+)\textcolor{black}{e_2(q_+)e_1(k_+)},\\
    I^{(1)}_5&=\int _0 ^{+\infty}dk_+ \int _0 ^{+\infty} dq_+ \frac{e^{2q_+}}{(e^{2k_+}-1)(e^{2q_+}-1)}\hat{b}^{\dagger}(k_+)\hat{a}^{\dagger}(q_+)\textcolor{black}{e_1^{\dagger}(k_+)e_2^{\dagger}(q_+)},\\
    -I^{(1)}_6&=\int _0 ^{+\infty}dk_+ \int _0 ^{+\infty} dq_+ \frac{e^{2q_+}}{(e^{2k_+}-1)(e^{2q_+}-1)}\hat{a}^{\dagger}(q_+)\hat{b}^{\dagger}(k_+)\textcolor{black}{e_2^{\dagger}(q_+)e_1^{\dagger}(k_+)},\\
    I^{(1)}_7&=\int _0 ^{+\infty}dk_+ \int _0 ^{+\infty} dq_+ \frac{1}{(e^{2k_+}-1)(e^{2q_+}-1)}\hat{b}^{\dagger}(k_+)\hat{b}(q_+)\textcolor{black}{e_1^{\dagger}(k_+)e_2(q_+)},\\
    -I^{(1)}_8&=\int _0 ^{+\infty}dk_+ \int _0 ^{+\infty} dq_+ \frac{1}{(e^{2k_+}-1)(e^{2q_+}-1)}\hat{b}(q_+)\hat{b}^{\dagger}(k_+)\textcolor{black}{e_2(q_+)}e_1^{\dagger}(k_+),\\
    J^{(1)}_1&=\int _0 ^{+\infty}dk_+ \int _0 ^{+\infty} dq_+ \frac{e^{2(k_++q_+)}}{(e^{2k_+}+1)(e^{2q_+}+1)}\hat{\alpha}(k_+)\hat{\alpha}^{\dagger}(q_+)\textcolor{black}{\epsilon_1(k_+)\epsilon_2^{\dagger}(q_+)},\\
    -J^{(1)}_2&=\int _0 ^{+\infty}dk_+ \int _0 ^{+\infty} dq_+ \frac{e^{2(k_++q_+)}}{(e^{2k_+}+1)(e^{2q_+}+1)}\hat{\alpha}^{\dagger}(q_+)\hat{\alpha}(k_+)\epsilon_2^{\dagger}(q_+)\epsilon_1(k_+),\\
    J^{(1)}_3&=\int _0 ^{+\infty}dk_+ \int _0 ^{+\infty} dq_+ \frac{e^{2k_+}}{(e^{2k_+}+1)(e^{2q_+}+1)}\hat{\alpha}(k_+)\hat{\beta}(q_+)\textcolor{black}{\epsilon_1(k_+)\epsilon_2(q_+)},\\
    -J^{(1)}_4&=\int _0 ^{+\infty}dk_+ \int _0 ^{+\infty} dq_+ \frac{e^{2k_+}}{(e^{2k_+}+1)(e^{2q_+}+1)}\hat{\beta}(q_+)\hat{\alpha}(k_+)\textcolor{black}{\epsilon_2(q_+)\epsilon_1(k_+)},\\
    J^{(1)}_5&=\int _0 ^{+\infty}dk_+ \int _0 ^{+\infty} dq_+ \frac{e^{2q_+}}{(e^{2k_+}+1)(e^{2q_+}+1)}\hat{\beta}^{\dagger}(k_+)\hat{\alpha}^{\dagger}(q_+)\textcolor{black}{\epsilon_1^{\dagger}(k_+)\epsilon_2^{\dagger}(q_+)},\\
    -J^{(1)}_6&=\int _0 ^{+\infty}dk_+ \int _0 ^{+\infty} dq_+ \frac{e^{2q_+}}{(e^{2k_+}+1)(e^{2q_+}+1)}\hat{\alpha}^{\dagger}(q_+)\hat{\beta}^{\dagger}(k_+)\textcolor{black}{\epsilon_2^{\dagger}(q_+)\epsilon_1^{\dagger}(k_+)},\\
    J^{(1)}_7&=\int _0 ^{+\infty}dk_+ \int _0 ^{+\infty} dq_+ \frac{1}{(e^{2k_+}+1)(e^{2q_+}+1)}\hat{\beta}^{\dagger}(k_+)\hat{\beta}(q_+)\textcolor{black}{\epsilon_1^{\dagger}(k_+)\epsilon_2(q_+)},\\
    -J^{(1)}_8&=\int _0 ^{+\infty}dk_+ \int _0 ^{+\infty} dq_+ \frac{1}{(e^{2k_+}+1)(e^{2q_+}+1)}\hat{\beta}(q_+)\hat{\beta}^{\dagger}(k_+)\textcolor{black}{\epsilon_2(q_+)}\epsilon_1^{\dagger}(k_+).
    \end{aligned}
\end{equation}
The remaining 16 integrals involve products of plane waves of different types:
\begin{equation}
    \begin{aligned}
    H^{(1)}_1&=\int _0 ^{+\infty}dk_+ \int _0 ^{+\infty} dq_+ \frac{e^{2(k_++q_+)}}{(e^{2k_+}-1)(e^{2q_+}+1)}\hat{a}(k_+)\hat{\alpha}^{\dagger}(q_+)\textcolor{black}{e_1(k_+)\epsilon_2^{\dagger}(q_+)},\\
    -H^{(1)}_2&=\int _0 ^{+\infty}dk_+ \int _0 ^{+\infty} dq_+ \frac{e^{2(k_++q_+)}}{(e^{2k_+}-1)(e^{2q_+}+1)}\hat{\alpha}^{\dagger}(q_+)\hat{a}(k_+)\epsilon_2^{\dagger}(q_+)e_1(k_+),\\
    H^{(1)}_3&=\int _0 ^{+\infty}dk_+ \int _0 ^{+\infty} dq_+ \frac{e^{2k_+}}{(e^{2k_+}-1)(e^{2q_+}+1)}\hat{a}(k_+)\hat{\beta}(q_+)\textcolor{black}{e_1(k_+)\epsilon_2(q_+)},\\
    -H^{(1)}_4&=\int _0 ^{+\infty}dk_+ \int _0 ^{+\infty} dq_+ \frac{e^{2k_+}}{(e^{2k_+}-1)(e^{2q_+}+1)}\hat{\beta}(q_+)\hat{a}(k_+)\textcolor{black}{\epsilon_2(q_+)e_1(k_+)},\\
    H^{(1)}_5&=\int _0 ^{+\infty}dk_+ \int _0 ^{+\infty} dq_+ \frac{e^{2q_+}}{(e^{2k_+}-1)(e^{2q_+}+1)}\hat{b}^{\dagger}(k_+)\hat{\alpha}^{\dagger}(q_+)\textcolor{black}{e_1^{\dagger}(k_+)\epsilon_2^{\dagger}(q_+)},\\
    -H^{(1)}_6&=\int _0 ^{+\infty}dk_+ \int _0 ^{+\infty} dq_+ \frac{e^{2q_+}}{(e^{2k_+}-1)(e^{2q_+}+1)}\hat{\alpha}^{\dagger}(q_+)\hat{b}^{\dagger}(k_+)\textcolor{black}{\epsilon_2^{\dagger}(q_+)e_1^{\dagger}(k_+)},\\
    H^{(1)}_7&=\int _0 ^{+\infty}dk_+ \int _0 ^{+\infty} dq_+ \frac{1}{(e^{2k_+}-1)(e^{2q_+}+1)}\hat{b}^{\dagger}(k_+)\hat{\beta}(q_+)\textcolor{black}{e_1^{\dagger}(k_+)\epsilon_2(q_+)},\\
    -H^{(1)}_8&=\int _0 ^{+\infty}dk_+ \int _0 ^{+\infty} dq_+ \frac{1}{(e^{2k_+}-1)(e^{2q_+}+1)}\hat{\beta}(q_+)\hat{b}^{\dagger}(k_+)\textcolor{black}{\epsilon_2(q_+)}e_1^{\dagger}(k_+),\\
    H^{(1)}_9&=\int _0 ^{+\infty}dk_+ \int _0 ^{+\infty} dq_+ \frac{e^{2(k_++q_+)}}{(e^{2k_+}+1)(e^{2q_+}-1)}\hat{\alpha}(k_+)\hat{a}^{\dagger}(q_+)\textcolor{black}{\epsilon_1(k_+)e_2^{\dagger}(q_+)},\\
    -H^{(1)}_{10}&=\int _0 ^{+\infty}dk_+ \int _0 ^{+\infty} dq_+ \frac{e^{2(k_++q_+)}}{(e^{2k_+}+1)(e^{2q_+}-1)}\hat{a}^{\dagger}(q_+)\hat{\alpha}(k_+)e_2^{\dagger}(q_+)\epsilon_1(k_+),\\
    H^{(1)}_{11}&=\int _0 ^{+\infty}dk_+ \int _0 ^{+\infty} dq_+ \frac{e^{2k_+}}{(e^{2k_+}+1)(e^{2q_+}-1)}\hat{\alpha}(k_+)\hat{b}(q_+)\textcolor{black}{\epsilon_1(k_+)e_2(q_+)},\\
    -H^{(1)}_{12}&=\int _0 ^{+\infty}dk_+ \int _0 ^{+\infty} dq_+ \frac{e^{2k_+}}{(e^{2k_+}+1)(e^{2q_+}-1)}\hat{b}(q_+)\hat{\alpha}(k_+)\textcolor{black}{e_2(q_+)\epsilon_1(k_+)},\\
    H^{(1)}_{13}&=\int _0 ^{+\infty}dk_+ \int _0 ^{+\infty} dq_+ \frac{e^{2q_+}}{(e^{2k_+}+1)(e^{2q_+}-1)}\hat{\beta}^{\dagger}(k_+)\hat{a}^{\dagger}(q_+)\textcolor{black}{\epsilon_1^{\dagger}(k_+)e_2^{\dagger}(q_+)},\\
    -H^{(1)}_{14}&=\int _0 ^{+\infty}dk_+ \int _0 ^{+\infty} dq_+ \frac{e^{2q_+}}{(e^{2k_+}+1)(e^{2q_+}-1)}\hat{a}^{\dagger}(q_+)\hat{\beta}^{\dagger}(k_+)\textcolor{black}{e_2^{\dagger}(q_+)\epsilon_1^{\dagger}(k_+)},\\
    H^{(1)}_{15}&=\int _0 ^{+\infty}dk_+ \int _0 ^{+\infty} dq_+ \frac{1}{(e^{2k_+}+1)(e^{2q_+}-1)}\hat{\beta}^{\dagger}(k_+)\hat{b}(q_+)\textcolor{black}{\epsilon_1^{\dagger}(k_+)e_2(q_+)},\\
    -H^{(1)}_{16}&=\int _0 ^{+\infty}dk_+ \int _0 ^{+\infty} dq_+ \frac{1}{(e^{2k_+}+1)(e^{2q_+}-1)}\hat{b}(q_+)\hat{\beta}^{\dagger}(k_+)\textcolor{black}{e_2(q_+)}\epsilon_1^{\dagger}(k_+).
    \end{aligned}
\end{equation}

In order to compare this commutator to the compact expression of the Pauli-Jordan function (\ref{eq:PJ}), we need to move $x_2$ to the right in the even-numbered integrals. In $I^{(1)}_2$, for instance, we can rewrite the product $e_2^{\dagger}(q_+)e_1(k_+)$ according to (\ref{eq:scambio}):
\begin{equation}
e_2^{\dagger}(q_+)e_1(k_+)=e_1\bigg(\frac{1}{2}\ln[e^{2q_+}+e^{2k_+}-1]-q_+\bigg)e_2^{\dagger}\bigg(\frac{1}{2}\ln[e^{2q_+}+e^{2k_+}-1]-k_+\bigg)    
\end{equation}

If we make a change of variables
\begin{equation}
    k'_+=\frac{1}{2}\ln[e^{2q_+}+e^{2k_+}-1]-q_+,\,\,\,\,q'_+=\frac{1}{2}\ln[e^{2q_+}+e^{2k_+}-1]-k_+ \,,
    \end{equation}
    then 
    \begin{equation}
    k'_+\in\big]0,+\infty \big[, \,\,\,\,q'_+\in\big]0,-\frac{1}{2}\ln(1-e^{-2{k_{+}}})\big[ \,,
\end{equation}
and the integral becomes

\begin{equation}\begin{aligned}
    -I^{(1)}_2=&\int _0 ^{+\infty}dk_+ \int _0 ^{-\frac{1}{2}\ln(1-e^{-2{k_{+}}})} dq_+ \frac{e^{2(k_++q_+)}}{(e^{2k_+}-1)(e^{2q_+}-1)(e^{2k_+}+e^{2q_+}-e^{2(q_++k_+)})}\\
    &\hat{a}^{\dagger}\bigg(\frac{1}{2}\ln\bigg[\frac{e^{2q_+}}{e^{2k_+}+e^{2q_+}-e^{2(q_++k_+)}}\bigg]\bigg)\hat{a}\bigg(\frac{1}{2}\ln\bigg[\frac{e^{2k_+}}{e^{2k_+}+e^{2q_+}-e^{2(q_++k_+)}}\bigg]\bigg)\textcolor{black}{e_1(k_+)e_2^{\dagger}(q_+)}.
\end{aligned}\end{equation}

Analogously we can prove that
\begin{equation}
    \begin{aligned}
-I^{(1)}_4=&\int _0 ^{+\infty}dk_+ \int _0 ^{+\infty} dq_+ \frac{e^{2(k_++q_+)}}{(e^{2k_+}-1)(e^{2q_+}-1)(e^{2(q_++k_+)}-e^{2k_+}+1)}    \\
&\hat{b}\bigg(\frac{1}{2}\ln\bigg[e^{2(q_++k_+)}-e^{2k_+}+1\bigg]\bigg)\hat{a}\bigg(\frac{1}{2}\ln\bigg[\frac{e^{2(q_++k_+)}}{e^{2(q_++k_+)}-e^{2k_+}+1}\bigg]\bigg)\textcolor{black}{e_1(k_+)e_2(q_+)},\\
-I^{(1)}_6=&\int _0 ^{+\infty}dk_+ \int _0 ^{+\infty} dq_+ \frac{e^{2(k_++q_+)}}{(e^{2k_+}-1)(e^{2q_+}-1)(e^{2(q_++k_+)}-e^{2q_+}+1)}\\
&\hat{a}^{\dagger}\bigg(\frac{1}{2}\ln\bigg[\frac{e^{2(q_++k_+)}}{e^{2(q_++k_+)}-e^{2q_+}+1}\bigg]\bigg)\hat{b}^{\dagger}\bigg(\frac{1}{2}\ln\bigg[e^{2(q_++k_+)}-e^{2q_+}+1\bigg]\bigg)\textcolor{black}{e_1^{\dagger}(k_+)e_2^{\dagger}(q_+)}.
    \end{aligned}
\end{equation}

For what concerns $I^{(1)}_8$, again using (\ref{eq:scambio}), we obtain

\begin{equation}
   e_2(q_+)e_1^{\dagger}(k_+)=e_1^{\dagger}\bigg(k_+-\frac{1}{2}\ln[e^{2q_+}(1-e^{2k_+})+e^{2k_+}]\bigg)e_2\bigg(q_+-\frac{1}{2}\ln[e^{2q_+}(1-e^{2k_+})+e^{2k_+}]\bigg). 
\end{equation}

When $q_+>-\frac{1}{2}\ln(1-e^{-2k_+})$, the argument of the logarithm becomes negative and we can write
\begin{equation}
    \begin{aligned}
    e_2(q_+)e_1^{\dagger}(k_+)=&e_1\bigg(-k_++\frac{1}{2}\ln[e^{2q_+}(e^{2k_+}-1)-e^{2k_+}]+i\frac{\pi}{2}+ni\pi\bigg)\\
    & e_2\bigg(q_+-\frac{1}{2}\ln[e^{2q_+}(e^{2k_+}-1)-e^{2k_+}]-i\frac{\pi}{2}+mi\pi\bigg),
    \end{aligned}
\end{equation}

where $m+n$ is even, as we proved in Appendix~\ref{appendixC}, so that the right-hand side can be written as $\epsilon_1^{\dagger}\big(k_+-\frac{1}{2}\ln[e^{2q_+}(e^{2k_+}-1)-e^{2k_+}]\big)\epsilon_2\big(q_+-\frac{1}{2}\ln[e^{2q_+}(e^{2k_+}-1)-e^{2k_+}]\big)$. If we now proceed as for $I^{(1)}_2$, we get

\begin{equation}
    \begin{aligned}
  -I^{(1)}_8=&\int _0 ^{+\infty}dk_+ \int _0 ^{+\infty} dq_+ \frac{e^{2(k_++q_+)}}{(e^{2k_+}-1)(e^{2q_+}-1)(e^{2q_+}+e^{2k_+}-1)}  \\
  &\hat{b}\bigg(-\frac{1}{2}\ln\bigg[\frac{e^{2k_+}}{e^{2q_+}+e^{2k_+}-1}\bigg]\bigg)\hat{b}^{\dagger}\bigg(-\frac{1}{2}\ln\bigg[\frac{e^{2q_+}}{e^{2q_+}+e^{2k_+}-1}\bigg]\bigg)\textcolor{black}{e_1^{\dagger}(k_+)e_2(q_+)}\\
  -&\int _0 ^{+\infty}dk_+ \int _0 ^{+\infty} dq_+ \frac{ e^{2(k_++q_+)}}{(e^{2k_+}+1)(e^{2q_+}+1)(e^{2q_+}+e^{2k_+}+1)}\\
  &\hat{b}\bigg(\frac{1}{2}\ln\bigg[\frac{e^{2q_+}+e^{2k_+}+1}{e^{2k_+}}\bigg]\bigg)\hat{b}^{\dagger}\bigg(\frac{1}{2}\ln\bigg[\frac{e^{2q_+}+e^{2k_+}+1}{e^{2q_+}}\bigg]\bigg)\textcolor{black}{\epsilon_1^{\dagger}(k_+)\epsilon_2(q_+)}\\
  -&\int _0 ^{+\infty}dk_+ \int _0 ^{+\infty} dq_+ \frac{e^{2(-k_++q_+)}}{(e^{-2k_+}+1)(e^{2q_+}+1)(e^{2q_+}+e^{-2k_+}+1)}\\
  &\hat{b}\bigg(\frac{1}{2}\ln\bigg[\frac{e^{2q_+}+e^{-2k_+}+1}{e^{-2k_+}}\bigg]\bigg)\hat{b}^{\dagger}\bigg(\frac{1}{2}\ln\bigg[\frac{e^{2q_+}+e^{-2k_+}+1}{e^{2q_+}}\bigg]\bigg)\textcolor{black}{\epsilon_1(k_+)\epsilon_2(q_+)}\\
  -&\int _0 ^{+\infty}dk_+ \int _0 ^{+\infty} dq_+ \frac{ e^{2(k_+-q_+)}}{(e^{2k_+}+1)(e^{-2q_+}+1)(e^{-2q_+}+e^{2k_+}+1)}\\
  &\hat{b}\bigg(\frac{1}{2}\ln\bigg[\frac{e^{-2q_+}+e^{2k_+}+1}{e^{2k_+}}\bigg]\bigg)\hat{b}^{\dagger}\bigg(\frac{1}{2}\ln\bigg[\frac{e^{-2q_+}+e^{2k_+}+1}{e^{-2q_+}}\bigg]\bigg)\textcolor{black}{\epsilon_1^{\dagger}(k_+)\epsilon_2^{\dagger}(q_+)}\\
  -&\int _0 ^{+\infty}dk_+ \int _0 ^{+\infty} dq_+ \frac{ e^{-2(k_++q_+)}}{(e^{-2k_+}+1)(e^{-2q_+}+1)(e^{-2q_+}+e^{-2k_+}+1)}\\
  &\hat{b}\bigg(\frac{1}{2}\ln\bigg[\frac{e^{-2q_+}+e^{-2k_+}+1}{e^{-2k_+}}\bigg]\bigg)\hat{b}^{\dagger}\bigg(\frac{1}{2}\ln\bigg[\frac{e^{-2q_+}+e^{-2k_+}+1}{e^{-2q_+}}\bigg]\bigg)\textcolor{black}{\epsilon_1(k_+)\epsilon_2^{\dagger}(q_+)}.
    \end{aligned}
\end{equation}
Analogously, the $J^{(1)}_i$ integrals are given by

\begin{equation}
    \begin{aligned}
    -J^{(1)}_2=&-\int_0^{\frac{1}{2}\ln(3)}dk_+ \int_{-\frac{1}{2}\ln(1-e^{-2k_+})}^{-\frac{1}{2}\ln(\frac{1-e^{-2k_+}}{2})}dq_+\frac{e^{2(k_++q_+)}}{(e^{2k_+}-1)(e^{2q_+}-1)(e^{2(k_++q_+)}-e^{2k_+}-e^{2q_+})}\\&\hat{\alpha}^{\dagger}\bigg(-\frac{1}{2}\ln\bigg[e^{2k_+}(1-e^{-2q_+})-1\bigg]\bigg)\hat{\alpha}\bigg(\frac{1}{2}\ln\bigg[\frac{e^{2(k_+-q_+)}}{e^{2k_+}(1-e^{-2q_+})-1}\bigg]\bigg)\textcolor{black}{e_1(k_+)e_2^{\dagger}(q_+)}\\
    -&\int_{\frac{1}{2}\ln(3)}^{+\infty}dk_+ \int_{-\frac{1}{2}\ln(1-e^{-2k_+})}^{-\frac{1}{2}\ln(1-2e^{-2k_+})}dq_+\frac{e^{2(k_++q_+)}}{(e^{2k_+}-1)(e^{2q_+}-1)(e^{2(k_++q_+)}-e^{2k_+}-e^{2q_+})}\\
    &\hat{\alpha}^{\dagger}\bigg(-\frac{1}{2}\ln\bigg[e^{2k_+}(1-e^{-2q_+})-1\bigg]\bigg)\hat{\alpha}\bigg(\frac{1}{2}\ln\bigg[\frac{e^{2(k_+-q_+)}}{e^{2k_+}(1-e^{-2q_+})-1}\bigg]\bigg)\textcolor{black}{e_1(k_+)e_2^{\dagger}(q_+)},\\
    -J^{(1)}_4=&-\int_{\frac{1}{2}\ln(3)}^{+\infty}dk_+\int_{-\frac{1}{2}\ln(1-2e^{-2k_+})}^{\frac{1}{2}\ln(\frac{1-e^{2k_+}}{2})}dq_+\frac{e^{2(q_++k_+)}}{(e^{2k_+}-1)(e^{2q_+}-1)(e^{2(q_++k_+)}-e^{2k_+}-e^{2q_+})}\\
    &\hat{\beta}\bigg(\frac{1}{2}\ln\bigg[\frac{(e^{2(q_++k_+)}-e^{2k_+}-e^{2q_+})}{e^{2q_+}}\bigg]\bigg)\hat{\alpha}\bigg(\frac{1}{2}\ln\bigg[\frac{e^{2k_+}}{e^{2(q_++k_+)}-e^{2k_+}-e^{2q_+}}\bigg]\bigg)\textcolor{black}{e_1(k_+)e_2^{\dagger}(q_+)},\\
    -J^{(1)}_6=&-\int_0^{\frac{1}{2}\ln(2)}dk_+\int_{-\frac{1}{2}\ln(\frac{1-e^{-2k_+}}{2})}^{+\infty}dq_+\frac{e^{2(k_++q_+)}}{(e^{2k_+}-1)(e^{2q_+}-1)(e^{2(k_++q_+)}-e^{2k_+}-e^{2q_+})}\\
    &\hat{\alpha}^{\dagger}\bigg(\frac{1}{2}\ln\bigg[\frac{e^{2q_+}}{e^{2(k_++q_+)}-e^{2k_+}-e^{2q_+}}\bigg]\bigg)\hat{\beta}^{\dagger}\bigg(\frac{1}{2}\ln\bigg[\frac{(e^{2(k_++q_+)}-e^{2k_+}-e^{2q_+})}{e^{2k_+}}\bigg]\bigg)\textcolor{black}{e_1(k_+)e_2^{\dagger}(q_+)}\\
    -&\int_{\frac{1}{2}\ln(2)}^{\frac{1}{2}\ln(3)}dk_+\int_{-\frac{1}{2}\ln(\frac{1-e^{-2k_+}}{2})}^{-\frac{1}{2}\ln(1-2e^{-2k_+})}dq_+\frac{e^{2(k_++q_+)}}{(e^{2k_+}-1)(e^{2q_+}-1)(e^{2(k_++q_+)}-e^{2k_+}-e^{2q_+})}\\
    &\hat{\alpha}^{\dagger}\bigg(\frac{1}{2}\ln\bigg[\frac{e^{2q_+}}{e^{2(k_++q_+)}-e^{2k_+}-e^{2q_+}}\bigg]\bigg)\hat{\beta}^{\dagger}\bigg(\frac{1}{2}\ln\bigg[\frac{(e^{2(k_++q_+)}-e^{2k_+}-e^{2q_+}}{e^{2k_+}})\bigg]\bigg)\textcolor{black}{e_1(k_+)e_2^{\dagger}(q_+)},\\
    -J^{(1)}_8=&-\int_{\frac{1}{2}\ln(2)}^{\frac{1}{2}\ln(3)}dk_+\int_{-\frac{1}{2}\ln(1-2e^{-2k_+})}^{+\infty}dq_+\frac{e^{2(k_++q_+)}}{(e^{2k_+}-1)(e^{2q_+}-1)(e^{2(k_++q_+)}-e^{2q_+}-e^{2k_+})}\\
    &\hat{\beta}\bigg(\frac{1}{2}\ln\bigg[\frac{e^{2(k_++q_+)}-e^{2q_+}-e^{2k_+}}{e^{2q_+}}\bigg]\bigg)\hat{\beta}^{\dagger}\bigg(\frac{1}{2}\ln\bigg[\frac{e^{2(k_++q_+)}-e^{2q_+}-e^{2k_+}}{e^{2k_+}}\bigg]\bigg)\textcolor{black}{e_1(k_+)e_2^{\dagger}(q_+)}\\
    -&\int_{\frac{1}{2}\ln(3)}^{+\infty}dk_+\int_{-\frac{1}{2}\ln(\frac{1-e^{-2k_+}}{2})}^{+\infty}dq_+\frac{e^{2(k_++q_+)}}{(e^{2k_+}-1)(e^{2q_+}-1)(e^{2(k_++q_+)}-e^{2q_+}-e^{2k_+})}\\
    &\hat{\beta}\bigg(\frac{1}{2}\ln\bigg[\frac{e^{2(k_++q_+)}-e^{2q_+}-e^{2k_+}}{e^{2q_+}}\bigg]\bigg)\hat{\beta}^{\dagger}\bigg(\frac{1}{2}\ln\bigg[\frac{e^{2(k_++q_+)}-e^{2q_+}-e^{2k_+}}{e^{2k_+}}\bigg]\bigg)\textcolor{black}{e_1(k_+)e_2^{\dagger}(q_+)}.
    \end{aligned}
\end{equation}

Lastly, the $H^{(1)}_i$ integrals can be written as

\begin{equation}
    \begin{aligned}
    -H^{(1)}_2=&\int_{\frac{1}{2}\ln(3/2)}^{+\infty}dk_+\int_{\frac{1}{2}\ln(2(e^{2k_+}-1))}^{\frac{1}{2}\ln(2e^{2k_+}-1)}dq_+\frac{e^{2(q_+-k_+)}}{(1-e^{-2k_+})(e^{2q_+}+1)(e^{2(q_+-k_+)}+e^{-2k_+}-1)}\\
    &\hat{\alpha}^{\dagger}\bigg(-\frac{1}{2}\ln\bigg[e^{2(q_+-k_+)}+e^{-2k_+}-1\bigg]\bigg)\hat{a}\bigg(\frac{1}{2}\ln\bigg[\frac{e^{2(q_+-k_+)}}{e^{2(q_+-k_+)}+e^{-2k_+}-1}\bigg]\bigg)\textcolor{black}{e_1^{\dagger}(k_+)\epsilon_2(q_+)}\\
    +&\int_0^{\frac{1}{2}\ln(3/2)}dk_+\int_0^{\frac{1}{2}\ln(2e^{2k_+}-1)}dq_+\frac{e^{2(q_+-k_+)}}{(1-e^{-2k_+})(e^{2q_+}+1)(e^{2(q_+-k_+)}+e^{-2k_+}-1)}\\
    &\hat{\alpha}^{\dagger}\bigg(-\frac{1}{2}\ln\bigg[e^{2(q_+-k_+)}+e^{-2k_+}-1\bigg]\bigg)\hat{a}\bigg(\frac{1}{2}\ln\bigg[\frac{e^{2(q_+-k_+)}}{e^{2(q_+-k_+)}+e^{-2k_+}-1}\bigg]\bigg)\textcolor{black}{e_1^{\dagger}(k_+)\epsilon_2(q_+)}\\
    +&\int_0^{\frac{1}{2}\ln(3/2)}dk_+\int_0^{-\frac{1}{2}\ln(2(e^{2k_+}-1))}dq_+\frac{e^{-2(q_++k_+)}}{(1-e^{-2k_+})(e^{-2q_+}+1)(e^{-2(q_++k_+)}+e^{-2k_+}-1)}\\
    &\hat{\alpha}^{\dagger}\bigg(-\frac{1}{2}\ln\bigg[e^{-2(q_++k_+)}+e^{-2k_+}-1\bigg]\bigg)\hat{a}\bigg(\frac{1}{2}\ln\bigg[\frac{e^{-2(q_++k_+)}}{e^{-2(q_++k_+)}+e^{-2k_+}-1}\bigg]\bigg)\textcolor{black}{e_1^{\dagger}(k_+)\epsilon_2^{\dagger}(q_+)}\\
    +&\int_0^{\frac{1}{2}\ln(3/2)}dk_+\int_{-\frac{1}{2}\ln(2(e^{2k_+}-1))}^{-\frac{1}{2}\ln(e^{2k_+}-1)}dq_+\frac{e^{-2(q_++k_+)}}{(1-e^{-2k_+})(e^{-2q_+}+1)(e^{-2(q_++k_+)}+e^{-2k_+}-1)}\\
    &\hat{\alpha}^{\dagger}\bigg(-\frac{1}{2}\ln\bigg[e^{-2(q_++k_+)}+e^{-2k_+}-1\bigg]\bigg)\hat{a}\bigg(\frac{1}{2}\ln\bigg[\frac{e^{-2(q_++k_+)}}{e^{-2(q_++k_+)}+e^{-2k_+}-1}\bigg]\bigg)\textcolor{black}{e_1^{\dagger}(k_+)\epsilon_2^{\dagger}(q_+)}\\
    +&\int_{\frac{1}{2}\ln(3/2)}^{\frac{1}{2}\ln(2)}dk_+\int_{0}^{-\frac{1}{2}\ln(e^{2k_+}-1)}dq_+\frac{e^{-2(q_++k_+)}}{(1-e^{-2k_+})(e^{-2q_+}+1)(e^{-2(q_++k_+)}+e^{-2k_+}-1)}\\
    &\hat{\alpha}^{\dagger}\bigg(-\frac{1}{2}\ln\bigg[e^{-2(q_++k_+)}+e^{-2k_+}-1\bigg]\bigg)\hat{a}\bigg(\frac{1}{2}\ln\bigg[\frac{e^{-2(q_++k_+)}}{e^{-2(q_++k_+)}+e^{-2k_+}-1}\bigg]\bigg)\textcolor{black}{e_1^{\dagger}(k_+)\epsilon_2^{\dagger}(q_+)}\\
    +&\int_{\frac{1}{2}\ln(3/2)}^{\frac{1}{2}\ln(2)}dk_+\int_{0}^{\frac{1}{2}\ln(2(e^{2k_+}-1))}dq_+\frac{e^{2(q_+-k_+)}}{(1-e^{-2k_+})(e^{2q_+}+1)(e^{2(q_+-k_+)}+e^{-2k_+}-1)}\\
    &\hat{\alpha}^{\dagger}\bigg(-\frac{1}{2}\ln\bigg[e^{2(q_+-k_+)}+e^{-2k_+}-1\bigg]\bigg)\hat{a}\bigg(\frac{1}{2}\ln\bigg[\frac{e^{2(q_+-k_+)}}{e^{2(q_+-k_+)}+e^{-2k_+}-1}\bigg]\bigg)\textcolor{black}{e_1^{\dagger}(k_+)\epsilon_2(q_+)}\\
    +&\int_{\frac{1}{2}\ln(2)}^{+\infty}dk_+\int_{\frac{1}{2}\ln(e^{2k_+}-1)}^{\frac{1}{2}\ln(2(e^{2k_+}-1))}dq_+\frac{e^{2(q_+-k_+)}}{(1-e^{-2k_+})(e^{2q_+}+1)(e^{2(q_+-k_+)}+e^{-2k_+}-1)}\\
    &\hat{\alpha}^{\dagger}\bigg(-\frac{1}{2}\ln\bigg[e^{2(q_+-k_+)}+e^{-2k_+}-1\bigg]\bigg)\hat{a}\bigg(\frac{1}{2}\ln\bigg[\frac{e^{2(q_+-k_+)}}{e^{2(q_+-k_+)}+e^{-2k_+}-1}\bigg]\bigg)\textcolor{black}{e_1^{\dagger}(k_+)\epsilon_2(q_+)}\\
    +&\int_0^{+\infty}dk_+\int_{\frac{1}{2}\ln(1+e^{-2k_+})}^{\frac{1}{2}\ln(1+2e^{-2k_+})}dq_+\frac{e^{2(q_++k_+)}}{(e^{2k_+}+1)(e^{2q_+}-1)(e^{2(k_++q_+)}-e^{2k_+}-1)}\\
    &\hat{\alpha}^{\dagger}\bigg(-\frac{1}{2}\ln\bigg[e^{2(k_++q_+)}-e^{2k_+}-1\bigg]\bigg)\hat{a}\bigg(\frac{1}{2}\ln\bigg[\frac{e^{2(k_++q_+)}}{e^{2(k_++q_+)}-e^{2k_+}-1}\bigg]\bigg)\textcolor{black}{\epsilon_1(k_+)e_2(q_+)}\\
    +&\int_0^{+\infty}dk_+\int_{\frac{1}{2}\ln(1+e^{2k_+})}^{\frac{1}{2}\ln(1+2e^{2k_+})}dq_+\frac{e^{2(q_+-k_+)}}{(e^{-2k_+}+1)(e^{2q_+}-1)(e^{2(-k_++q_+)}-e^{-2k_+}-1)}\\
    &\hat{\alpha}^{\dagger}\bigg(-\frac{1}{2}\ln\bigg[e^{2(-k_++q_+)}-e^{-2k_+}-1\bigg]\bigg)\hat{a}\bigg(\frac{1}{2}\ln\bigg[\frac{e^{2(-k_++q_+)}}{e^{2(-k_++q_+)}-e^{-2k_+}-1}\bigg]\bigg)\textcolor{black}{\epsilon_1^{\dagger}(k_+)e_2(q_+)},
    \end{aligned}
\end{equation}
\begin{equation*}
    \begin{aligned}
    -H^{(1)}_4=&\int_0^{+\infty}dk_+\int_{\frac{1}{2}\ln(2e^{2k_+}-1)}^{+\infty}dq_+\frac{e^{2(q_+-k_+)}}{(1-e^{-2k_+})(e^{2q_+}+1)(e^{2(q_+-k_+)}+e^{-2k_+}-1)}\\
    &\hat{\beta}\bigg(\frac{1}{2}\ln\bigg[e^{2(q_+-k_+)}+e^{-2k_+}-1\bigg]\bigg)\hat{a}\bigg(\frac{1}{2}\ln\bigg[\frac{e^{2(q_+-k_+)}}{e^{2(q_+-k_+)}+e^{-2k_+}-1}\bigg]\bigg)\textcolor{black}{e_1^{\dagger}(k_+)\epsilon_2(q_+)}\\
    +&\int_0^{+\infty}dk_+\int_{\frac{1}{2}\ln(2(e^{-2k_+}+1))}^{+\infty}dq_+\frac{e^{2(q_++k_+)}}{(1+e^{2k_+})(e^{2q_+}-1)(e^{2(q_++k_+)}-e^{2k_+}-1)}\\
    &\hat{\beta}\bigg(\frac{1}{2}\ln\bigg[e^{2(q_++k_+)}-e^{2k_+}-1\bigg]\bigg)\hat{a}\bigg(\frac{1}{2}\ln\bigg[\frac{e^{2(q_++k_+)}}{e^{2(q_++k_+)}-e^{2k_+}-1}\bigg]\bigg)\textcolor{black}{\epsilon_1(k_+)e_2(q_+)}\\
    +&\int_0^{+\infty}dk_+\int_{\frac{1}{2}\ln(2(e^{2k_+}+1))}^{+\infty}dq_+\frac{e^{2(q_+-k_+)}}{(1+e^{-2k_+})(e^{2q_+}-1)(e^{2(q_+-k_+)}-e^{-2k_+}-1)}\\
    &\hat{\beta}\bigg(\frac{1}{2}\ln\bigg[e^{2(q_+-k_+)}-e^{-2k_+}-1\bigg]\bigg)\hat{a}\bigg(\frac{1}{2}\ln\bigg[\frac{e^{2(q_+-k_+)}}{e^{2(q_+-k_+)}-e^{-2k_+}-1}\bigg]\bigg)\textcolor{black}{\epsilon^{\dagger}_1(k_+)e_2(q_+)}\\
    +&\int_0^{+\infty}dk_+\int_{\frac{1}{2}\ln(2e^{-2k_+}+1)}^{\frac{1}{2}\ln(2(e^{-2k_+}+1))}dq_+\frac{e^{2(q_++k_+)}}{(1+e^{2k_+})(e^{2q_+}-1)(e^{2(q_++k_+)}-e^{2k_+}-1)}\\
    &\hat{\beta}\bigg(\frac{1}{2}\ln\bigg[e^{2(q_++k_+)}-e^{2k_+}-1\bigg]\bigg)\hat{a}\bigg(\frac{1}{2}\ln\bigg[\frac{e^{2(q_++k_+)}}{e^{2(q_++k_+)}-e^{2k_+}-1}\bigg]\bigg)\textcolor{black}{\epsilon_1(k_+)e_2(q_+)}\\
    +&\int_0^{+\infty}dk_+\int_{\frac{1}{2}\ln(2e^{2k_+}+1)}^{\frac{1}{2}\ln(2(e^{2k_+}+1))}dq_+\frac{e^{2(q_+-k_+)}}{(1+e^{-2k_+})(e^{2q_+}-1)(e^{2(q_+-k_+)}-e^{-2k_+}-1)}\\
    &\hat{\beta}\bigg(\frac{1}{2}\ln\bigg[e^{2(q_+-k_+)}-e^{-2k_+}-1\bigg]\bigg)\hat{a}\bigg(\frac{1}{2}\ln\bigg[\frac{e^{2(q_+-k_+)}}{e^{2(q_+-k_+)}-e^{-2k_+}-1}\bigg]\bigg)\textcolor{black}{\epsilon^{\dagger}_1(k_+)e_2(q_+)},
    \end{aligned}
\end{equation*}

\begin{equation*}
    \begin{aligned}
    -H^{(1)}_6=&\int_0^{\frac{1}{2}\ln(2)}dk_+\int_{-\frac{1}{2}\ln(2e^{-2k_+}-1)}^{+\infty}dq_+\frac{e^{2(k_++q_+)}}{(e^{2k_+}-1)(e^{2q_+}+1)(e^{2(k_++q_+)}+e^{2k_+}-e^{2q_+})}\\
    &\hat{\alpha}^{\dagger}\bigg(\frac{1}{2}\ln\bigg[\frac{e^{2q_+}}{e^{2(k_++q_+)}+e^{2k_+}-e^{2q_+}}\bigg]\bigg)\hat{b}^{\dagger}\bigg(\frac{1}{2}\ln\bigg[\frac{(e^{2(k_++q_+)}+e^{2k_+}-e^{2q_+})}{e^{2k_+}}\bigg]\bigg)\textcolor{black}{e_1(k_+)\epsilon_2^{\dagger}(q_+)},\\
    -H^{(1)}_8=&\int_{\frac{1}{2}\ln(2)}^{+\infty}dk_+\int_0^{+\infty}dq_+\frac{e^{2(k_++q_+)}}{(e^{2k_+}-1)(e^{2q_+}+1)(e^{2(k_++q_+)}+e^{2k_+}-1)}\\
    &\hat{\beta}\bigg(\frac{1}{2}\ln\bigg[e^{2(k_++q_+)}+e^{2k_+}-1\bigg]\bigg)\hat{b}^{\dagger}\bigg(-\frac{1}{2}\ln\bigg[\frac{e^{2(q_++k_+)}}{e^{2(k_++q_+)}+e^{2k_+}-1}\bigg]\bigg)\textcolor{black}{e_1(k_+)\epsilon_2(q_+)}\\
    +&\int_{\frac{1}{2}\ln(2)}^{+\infty}dk_+\int_0^{+\infty}dq_+\frac{e^{2(k_+-q_+)}}{(e^{2k_+}-1)(e^{-2q_+}+1)(e^{2(k_+-q_+)}+e^{2k_+}-1)}\\
    &\hat{\beta}\bigg(\frac{1}{2}\ln\bigg[e^{2(k_+-q_+)}+e^{2k_+}-1\bigg]\bigg)\hat{b}^{\dagger}\bigg(-\frac{1}{2}\ln\bigg[\frac{e^{2(-q_++k_+)}}{e^{2(k_+-q_+)}+e^{2k_+}-1}\bigg]\bigg)\textcolor{black}{e_1(k_+)\epsilon_2^{\dagger}(q_+)}\\
    +&\int_0^{\frac{1}{2}\ln(2)}dk_+\int_0^{+\infty}dq_+\frac{e^{2(k_++q_+)}}{(e^{2k_+}-1)(e^{2q_+}+1)(e^{2(k_++q_+)}+e^{2k_+}-1)}\\
    &\hat{\beta}\bigg(\frac{1}{2}\ln\bigg[e^{2(k_++q_+)}+e^{2k_+}-1\bigg]\bigg)\hat{b}^{\dagger}\bigg(-\frac{1}{2}\ln\bigg[\frac{e^{2(q_++k_+)}}{e^{2(k_++q_+)}+e^{2k_+}-1}\bigg]\bigg)\textcolor{black}{e_1(k_+)\epsilon_2(q_+)}\\
    +&\int_0^{\frac{1}{2}\ln(2)}dk_+\int_0^{-\frac{1}{2}\ln(2e^{-2k_+}-1)}dq_+\frac{e^{2(k_+-q_+)}}{(e^{2k_+}-1)(e^{-2q_+}+1)(e^{2(k_+-q_+)}+e^{2k_+}-1)}\\
    &\hat{\beta}\bigg(\frac{1}{2}\ln\bigg[e^{2(k_+-q_+)}+e^{2k_+}-1\bigg]\bigg)\hat{b}^{\dagger}\bigg(-\frac{1}{2}\ln\bigg[\frac{e^{2(-q_++k_+)}}{e^{2(k_+-q_+)}+e^{2k_+}-1}\bigg]\bigg)\textcolor{black}{e_1(k_+)\epsilon_2^{\dagger}(q_+)},
    \end{aligned}
\end{equation*}

\begin{equation*}
    \begin{aligned}
    -H^{(1)}_{10}=&\int_0^{\frac{1}{2}\ln(2)}dk_+\int_{-\frac{1}{2}\ln(e^{2k_+}-1)}^{-\frac{1}{2}\ln(\frac{e^{2k_+}-1}{2})}dq_+\frac{e^{-2(k_++q_+)}}{(1-e^{-2k_+})(1+e^{-2q_+})(1-e^{-2k_+}-e^{-2(k_++q_+)})}\\
    &\hat{a}^{\dagger}\bigg(-\frac{1}{2}\ln\bigg[1-e^{-2k_+}-e^{-2(k_++q_+)}\bigg]\bigg)\hat{\alpha}\bigg(\frac{1}{2}\ln\bigg[\frac{e^{-2(k_++q_+)}}{1-e^{-2k_+}-e^{-2(k_++q_+)}}\bigg]\bigg)\textcolor{black}{e_1^{\dagger}(k_+)\epsilon_2^{\dagger}(q_+)}\\
    +&\int_{\frac{1}{2}\ln(2)}^{\frac{1}{2}\ln(3)}dk_+\int_0^{-\frac{1}{2}\ln(\frac{e^{2k_+}-1}{2})}dq_+\frac{e^{-2(k_++q_+)}}{(1-e^{-2k_+})(1+e^{-2q_+})(1-e^{-2k_+}-e^{-2(k_++q_+)})}\\
        &\hat{a}^{\dagger}\bigg(-\frac{1}{2}\ln\bigg[1-e^{-2k_+}-e^{-2(k_++q_+)}\bigg]\bigg)\hat{\alpha}\bigg(\frac{1}{2}\ln\bigg[\frac{e^{-2(k_++q_+)}}{1-e^{-2k_+}-e^{-2(k_++q_+)}}\bigg]\bigg)\textcolor{black}{e_1^{\dagger}(k_+)\epsilon_2^{\dagger}(q_+)}\\
    +& \int_{\frac{1}{2}\ln(2)}^{\frac{1}{2}\ln(3)}dk_+\int_0^{\frac{1}{2}\ln(e^{2k_+}-1)}dq_+\frac{e^{-2(k_+-q_+)}}{(1-e^{-2k_+})(1+e^{2q_+})(1-e^{-2k_+}-e^{-2(k_+-q_+)})}\\
    &\hat{a}^{\dagger}\bigg(-\frac{1}{2}\ln\bigg[1-e^{-2k_+}-e^{-2(k_+-q_+)}\bigg]\bigg)\hat{\alpha}\bigg(\frac{1}{2}\ln\bigg[\frac{e^{-2(k_+-q_+)}}{1-e^{-2k_+}-e^{-2(k_+-q_+)}}\bigg]\bigg)\textcolor{black}{e_1^{\dagger}(k_+)\epsilon_2(q_+)}\\
    +&\int_{\frac{1}{2}\ln(3)}^{+\infty}dk_+\int_{\frac{1}{2}\ln(\frac{e^{2k_+}-1}{2})}^{\frac{1}{2}\ln(e^{2k_+}-1)}dq_+\frac{e^{-2(k_+-q_+)}}{(1-e^{-2k_+})(1+e^{2q_+})(1-e^{-2k_+}-e^{-2(k_+-q_+)})}\\
    &\hat{a}^{\dagger}\bigg(-\frac{1}{2}\ln\bigg[1-e^{-2k_+}-e^{-2(k_+-q_+)}\bigg]\bigg)\hat{\alpha}\bigg(\frac{1}{2}\ln\bigg[\frac{e^{-2(k_+-q_+)}}{1-e^{-2k_+}-e^{-2(k_+-q_+)}}\bigg]\bigg)\textcolor{black}{e_1^{\dagger}(k_+)\epsilon_2(q_+)}\\
    +&\int_0^{+\infty}dk_+\int_0^{\frac{1}{2}\ln(1+e^{-2k_+})}dq_+\frac{e^{2(k_++q_+)}}{(e^{2k_+}+1)(e^{2q_+}-1)(1+e^{2k_+}-e^{2(k_++q_+)})}\\
    &\hat{a}^{\dagger}\bigg(\frac{1}{2}\ln\bigg[\frac{1}{1+e^{2k_+}-e^{2(k_++q_+)}}\bigg]\bigg)\hat{\alpha}\bigg(\frac{1}{2}\ln\bigg[\frac{e^{2(k_++q_+)}}{1+e^{2k_+}-e^{2(k_++q_+)}}\bigg]\bigg)\textcolor{black}{\epsilon_1(k_+)e_2(q_+)}\\
    +&\int_0^{+\infty}dk_+\int_{\frac{1}{2}\ln(\frac{1+e^{2k_+}}{2})}^{\frac{1}{2}\ln(1+e^{2k_+})}dq_+\frac{e^{2(-k_++q_+)}}{(e^{-2k_+}+1)(e^{2q_+}-1)(1+e^{-2k_+}-e^{2(-k_++q_+)})}\\
    &\hat{a}^{\dagger}\bigg(\frac{1}{2}\ln\bigg[\frac{1}{1+e^{-2k_+}-e^{2(-k_++q_+)}}\bigg]\bigg)\hat{\alpha}\bigg(\frac{1}{2}\ln\bigg[\frac{e^{2(-k_++q_+)}}{1+e^{-2k_+}-e^{2(-k_++q_+)}}\bigg]\bigg)\textcolor{black}{\epsilon_1^{\dagger}(k_+)e_2(q_+)},\\
    -H^{(1)}_{12}=&\int_0^{+\infty}dk_+\int_0^{-\frac{1}{2}\ln(\frac{1+e^{-2k_+}}{2})}dq_+\frac{e^{2(k_+-q_+)}}{(e^{2k_+}+1)(1-e^{-2q_+})(1+e^{2k_+}-e^{2(k_+-q_+)})}\\
&\hat{b}\bigg(\frac{1}{2}\ln\bigg[1+e^{2k_+}-e^{2(k_+-q_+)}\bigg]\bigg)\hat{\alpha}\bigg(\frac{1}{2}\ln\bigg[\frac{e^{2(k_+-q_+)}}{1+e^{2k_+}-e^{2(k_+-q_+)}}\bigg]\bigg)\textcolor{black}{\epsilon_1(k_+)e_2^{\dagger}(q_+)},
    \end{aligned}
\end{equation*}
\begin{equation*}
    \begin{aligned}
-H^{(1)}_{14}=&\int_{\frac{1}{2}\ln(3)}^{+\infty}dk_+\int_0^{\frac{1}{2}\ln(\frac{e^{2k_+}-1}{2})}dq_+\frac{e^{2(q_+-k_+)}}{(e^{2q_+}+1)(1-e^{-2k_+})(1-e^{-2k_+}-e^{2(q_+-k_+)})}\\
&\hat{a}^{\dagger}\bigg(\frac{1}{2}\ln\bigg[\frac{1}{1-e^{-2k_+}-e^{2(q_+-k_+)}}\bigg]\bigg)\hat{\beta}^{\dagger}\bigg(\frac{1}{2}\ln\bigg[\frac{1-e^{-2k_+}-e^{2(q_+-k_+)}}{e^{2(q_+-k_+)}}\bigg]\bigg)\textcolor{black}{e_1^{\dagger}(k_+)\epsilon_2(q_+)}\\
+&\int_{\frac{1}{2}\ln(3)}^{+\infty}dk_+\int_0^{+\infty}dq_+\frac{e^{-2(q_++k_+)}}{(e^{-2q_+}+1)(1-e^{-2k_+})(1-e^{-2k_+}-e^{-2(q_++k_+)})}\\
&\hat{a}^{\dagger}\bigg(\frac{1}{2}\ln\bigg[\frac{1}{1-e^{-2k_+}-e^{-2(q_++k_+)}}\bigg]\bigg)\hat{\beta}^{\dagger}\bigg(\frac{1}{2}\ln\bigg[\frac{1-e^{-2k_+}-e^{-2(q_++k_+)}}{e^{-2(q_++k_+)}}\bigg]\bigg)\textcolor{black}{e_1^{\dagger}(k_+)\epsilon_2^{\dagger}(q_+)}\\
+&\int_0^{\frac{1}{2}\ln(3)}dk_+\int_{-\frac{1}{2}\ln(\frac{e^{2k_+}-1}{2})}^{+\infty}dq_+\frac{e^{-2(q_++k_+)}}{(e^{-2q_+}+1)(1-e^{-2k_+})(1-e^{-2k_+}-e^{-2(q_++k_+)})}\\
&\hat{a}^{\dagger}\bigg(\frac{1}{2}\ln\bigg[\frac{1}{1-e^{-2k_+}-e^{-2(q_++k_+)}}\bigg]\bigg)\hat{\beta}^{\dagger}\bigg(\frac{1}{2}\ln\bigg[\frac{1-e^{-2k_+}-e^{-2(q_++k_+)}}{e^{-2(q_++k_+)}}\bigg]\bigg)\textcolor{black}{e_1^{\dagger}(k_+)\epsilon_2^{\dagger}(q_+)}\\
+&\int_0^{+\infty}dk_+\int_0^{\frac{1}{2}\ln(\frac{1+e^{2k_+}}{2})}dq_+\frac{e^{2(q_+-k_+)}}{(e^{-2k_+}+1)(e^{2q_+}-1)(1+e^{-2k_+}-e^{2(q_+-k_+)})}\\
&\hat{a}^{\dagger}\bigg(\frac{1}{2}\ln\bigg[\frac{1}{1+e^{-2k_+}-e^{2(q_+-k_+)}}\bigg]\bigg)\hat{\beta}^{\dagger}\bigg(\frac{1}{2}\ln\bigg[\frac{1+e^{-2k_+}-e^{2(q_+-k_+)}}{e^{2(q_+-k_+)}}\bigg]\bigg)\textcolor{black}{\epsilon_1^{\dagger}(k_+)e_2(q_+)},
    \end{aligned}
\end{equation*}
\begin{equation*}
    \begin{aligned}
 -H^{(1)}_{16}=&\int_0^{+\infty}dk_+\int_0^{+\infty}dq_+\frac{e^{-2(k_++q_+)}}{(e^{-2k_+}+1)(1-e^{-2q_+})(1+e^{-2k_+}-e^{-2(k_++q_+)})}\\
 &\hat{b}\bigg(\frac{1}{2}\ln\bigg[1+e^{-2k_+}-e^{-2(k_++q_+)}\bigg]\bigg)\hat{\beta}^{\dagger}\bigg(-\frac{1}{2}\ln\bigg[\frac{e^{-2(k_++q_+)}}{1+e^{-2k_+}-e^{-2(k_++q_+)}}\bigg]\bigg)\textcolor{black}{\epsilon_1^{\dagger}(k_+)e_2^{\dagger}(q_+)}\\
 +&\int_0^{+\infty}dk_+\int_{-\frac{1}{2}\ln(\frac{1+e^{-2k_+}}{2})}^{+\infty}dq_+\frac{e^{2(k_+-q_+)}}{(e^{2k_+}+1)(1-e^{-2q_+})(1+e^{2k_+}-e^{2(k_+-q_+)})}\\
 &\hat{b}\bigg(\frac{1}{2}\ln\bigg[1+e^{2k_+}-e^{2(k_+-q_+)}\bigg]\bigg)\hat{\beta}^{\dagger}\bigg(-\frac{1}{2}\ln\bigg[\frac{e^{2(k_+-q_+)}}{1+e^{2k_+}-e^{2(k_+-q_+)}}\bigg]\bigg)\textcolor{black}{\epsilon_1(k_+)e_2^{\dagger}(q_+)}.
    \end{aligned}
\end{equation*}
Consider now the second equation in~(\ref{eq:commrel}),  $[\hat{\phi}(x_1),\hat{\phi}(x_2)]=0$. Again, the left hand-side can be written as the sum of 32 integrals:
\begin{equation}
    [\hat{\phi}(x_1),\hat{\phi}(x_2)]=\sum_{i=1}^{8}I^{(2)}_i+\sum_{i=1}^{8}J^{(2)}_i+\sum_{i=1}^{16}H^{(2)}_i \,,
\end{equation}
where
\begin{equation}
    \begin{aligned}
        I^{(2)}_1&=\int _0 ^{+\infty}dk_+ \int _0 ^{+\infty} dq_+ \frac{e^{2k_+}}{(e^{2k_+}-1)(e^{2q_+}-1)}\hat{a}(k_+)\hat{b}^{\dagger}(q_+){e_1(k_+)e_2^{\dagger}(q_+)},\\
        -I^{(2)}_2&=\int _0 ^{+\infty}dk_+ \int _0 ^{+\infty} dq_+ \frac{e^{2k_+}}{(e^{2k_+}-1)(e^{2q_+}-1)}\hat{b}^{\dagger}(q_+)\hat{a}(k_+)e_2^{\dagger}(q_+)e_1(k_+)=\\
        &\int _0 ^{+\infty}dk_+ \int _0 ^{-\frac{1}{2}\ln(1-e^{-2{k_{+}}})} dq_+ \frac{e^{2k_+}}{(e^{2k_+}-1)(e^{2q_+}-1)}\\
        &\hat{b}^{\dagger}\bigg(\frac{1}{2}\ln\bigg[\frac{e^{2q_+}}{e^{2k_+}+e^{2q_+}-e^{2(q_++k_+)}}\bigg]\bigg)\hat{a}\bigg(\frac{1}{2}\ln\bigg[\frac{e^{2k_+}}{e^{2k_+}+e^{2q_+}-e^{2(q_++k_+)}}\bigg]\bigg){e_1(k_+)e_2^{\dagger}(q_+)},\\
        I^{(2)}_3&=\int _0 ^{+\infty}dk_+ \int _0 ^{+\infty} dq_+ \frac{e^{2(k_++q_+)}}{(e^{2k_+}-1)(e^{2q_+}-1)}\hat{a}(k_+)\hat{a}(q_+){e_1(k_+)e_2(q_+)},\\
        -I^{(2)}_4&=\int _0 ^{+\infty}dk_+ \int _0 ^{+\infty} dq_+ \frac{e^{2(k_++q_+)}}{(e^{2k_+}-1)(e^{2q_+}-1)}\hat{a}(q_+)\hat{a}(k_+)\textcolor{black}{e_2(q_+)e_1(k_+)}=\\
        &\int _0 ^{+\infty}dk_+ \int _0 ^{+\infty} dq_+ \frac{e^{2(k_++q_+)}}{(e^{2k_+}-1)(e^{2q_+}-1)}\\
        &\hat{a}\bigg(\frac{1}{2}\ln\bigg[e^{2(q_++k_+)}-e^{2k_+}+1\bigg]\bigg)\hat{a}\bigg(\frac{1}{2}\ln\bigg[\frac{e^{2(q_++k_+)}}{e^{2(q_++k_+)}-e^{2k_+}+1}\bigg]\bigg){e_1(k_+)e_2(q_+)},\\
        I^{(2)}_5&=\int _0 ^{+\infty}dk_+ \int _0 ^{+\infty} dq_+ \frac{1}{(e^{2k_+}-1)(e^{2q_+}-1)}\hat{b}^{\dagger}(k_+)\hat{b}^{\dagger}(q_+){e_1^{\dagger}(k_+)e_2^{\dagger}(q_+)},\\
        -I^{(2)}_6&=\int _0 ^{+\infty}dk_+ \int _0 ^{+\infty} dq_+ \frac{1}{(e^{2k_+}-1)(e^{2q_+}-1)}\hat{b}^{\dagger}(q_+)\hat{b}^{\dagger}(k_+)\textcolor{black}{e_2^{\dagger}(q_+)e_1^{\dagger}(k_+)}=\\
        &\int _0 ^{+\infty}dk_+ \int _0 ^{+\infty} dq_+ \frac{1}{(e^{2k_+}-1)(e^{2q_+}-1)}\\
        &\hat{b}^{\dagger}\bigg(\frac{1}{2}\ln\bigg[\frac{e^{2(q_++k_+)}}{e^{2(q_++k_+)}-e^{2q_+}+1}\bigg]\bigg)\hat{b}^{\dagger}\bigg(\frac{1}{2}\ln\bigg[e^{2(q_++k_+)}-e^{2q_+}+1\bigg]\bigg){e_1^{\dagger}(k_+)e_2^{\dagger}(q_+)},\\
        I^{(2)}_7&=\int _0 ^{+\infty}dk_+ \int _0 ^{+\infty} dq_+ \frac{e^{2q_+}}{(e^{2k_+}-1)(e^{2q_+}-1)}\hat{b}^{\dagger}(k_+)\hat{a}(q_+){e_1^{\dagger}(k_+)e_2(q_+)},
    \end{aligned}
\end{equation}
\begin{equation*}
    \begin{aligned}
     -I^{(2)}_8&=\int _0 ^{+\infty}dk_+ \int _0 ^{+\infty} dq_+ \frac{e^{2q_+}}{(e^{2k_+}-1)(e^{2q_+}-1)}\hat{a}(q_+)\hat{b}^{\dagger}(k_+)\textcolor{black}{e_2(q_+)}e_1^{\dagger}(k_+)=\\
    &\int _0 ^{+\infty}dk_+ \int _0 ^{+\infty} dq_+ \frac{e^{2q_+}}{(e^{2k_+}-1)(e^{2q_+}-1)}\\
    &\hat{a}\bigg(-\frac{1}{2}\ln\bigg[\frac{e^{2k_+}}{e^{2q_+}+e^{2k_+}-1}\bigg]\bigg)\hat{b}^{\dagger}\bigg(-\frac{1}{2}\ln\bigg[\frac{e^{2q_+}}{e^{2q_+}+e^{2k_+}-1}\bigg]\bigg){e_1^{\dagger}(k_+)e_2(q_+)}\\
    &-\int _0 ^{+\infty}dk_+ \int _0 ^{+\infty} dq_+ \frac{ e^{2q_+}}{(e^{2k_+}+1)(e^{2q_+}+1)}\\
    &\hat{a}\bigg(\frac{1}{2}\ln\bigg[\frac{e^{2q_+}+e^{2k_+}+1}{e^{2k_+}}\bigg]\bigg)\hat{b}^{\dagger}\bigg(\frac{1}{2}\ln\bigg[\frac{e^{2q_+}+e^{2k_+}+1}{e^{2q_+}}\bigg]\bigg){\epsilon_1^{\dagger}(k_+)\epsilon_2(q_+)}\\
    &-\int _0 ^{+\infty}dk_+ \int _0 ^{+\infty} dq_+ \frac{e^{2q_+}}{(e^{-2k_+}+1)(e^{2q_+}+1)}\\
    &\hat{a}\bigg(\frac{1}{2}\ln\bigg[\frac{e^{2q_+}+e^{-2k_+}+1}{e^{-2k_+}}\bigg]\bigg)\hat{b}^{\dagger}\bigg(\frac{1}{2}\ln\bigg[\frac{e^{2q_+}+e^{-2k_+}+1}{e^{2q_+}}\bigg]\bigg){\epsilon_1(k_+)\epsilon_2(q_+)}\\
    &-\int _0 ^{+\infty}dk_+ \int _0 ^{+\infty} dq_+ \frac{ e^{-2q_+}}{(e^{2k_+}+1)(e^{-2q_+}+1)}\\
    &\hat{a}\bigg(\frac{1}{2}\ln\bigg[\frac{e^{-2q_+}+e^{2k_+}+1}{e^{2k_+}}\bigg]\bigg)\hat{b}^{\dagger}\bigg(\frac{1}{2}\ln\bigg[\frac{e^{-2q_+}+e^{2k_+}+1}{e^{-2q_+}}\bigg]\bigg){\epsilon_1^{\dagger}(k_+)\epsilon_2^{\dagger}(q_+)}\\
    &-\int _0 ^{+\infty}dk_+ \int _0 ^{+\infty} dq_+ \frac{ e^{-2q_+}}{(e^{-2k_+}+1)(e^{-2q_+}+1)}\\
    &\hat{a}\bigg(\frac{1}{2}\ln\bigg[\frac{e^{-2q_+}+e^{-2k_+}+1}{e^{-2k_+}}\bigg]\bigg)\hat{b}^{\dagger}\bigg(\frac{1}{2}\ln\bigg[\frac{e^{-2q_+}+e^{-2k_+}+1}{e^{-2q_+}}\bigg]\bigg){\epsilon_1(k_+)\epsilon_2^{\dagger}(q_+)},\\
    J^{(2)}_1&=\int _0 ^{+\infty}dk_+ \int _0 ^{+\infty} dq_+ \frac{e^{2k_+}}{(e^{2k_+}+1)(e^{2q_+}+1)}\hat{\alpha}(k_+)\hat{\beta}^{\dagger}(q_+){\epsilon_1(k_+)\epsilon_2^{\dagger}(q_+)},   \\
    -J^{(2)}_2&=\int _0 ^{+\infty}dk_+ \int _0 ^{+\infty} dq_+ \frac{e^{2k_+}}{(e^{2k_+}+1)(e^{2q_+}+1)}\hat{\beta}^{\dagger}(q_+)\hat{\alpha}(k_+)\epsilon_2^{\dagger}(q_+)\epsilon_1(k_+)=\\
    &-\int_0^{\frac{1}{2}\ln(3)}dk_+ \int_{-\frac{1}{2}\ln(1-e^{-2k_+})}^{-\frac{1}{2}\ln(\frac{1-e^{-2k_+}}{2})}dq_+\frac{e^{2k_+}}{(e^{2k_+}-1)(e^{2q_+}-1)}\\
    &\hat{\beta}^{\dagger}\bigg(-\frac{1}{2}\ln\bigg[e^{2k_+}(1-e^{-2q_+})-1\bigg]\bigg)\hat{\alpha}\bigg(\frac{1}{2}\ln\bigg[\frac{e^{2(k_+-q_+)}}{e^{2k_+}(1-e^{-2q_+})-1}\bigg]\bigg){e_1(k_+)e_2^{\dagger}(q_+)}\\
    &-\int_{\frac{1}{2}\ln(3)}^{+\infty}dk_+ \int_{-\frac{1}{2}\ln(1-e^{-2k_+})}^{-\frac{1}{2}\ln(1-2e^{-2k_+})}dq_+\frac{e^{2k_+}}{(e^{2k_+}-1)(e^{2q_+}-1)}\\
    &\hat{\beta}^{\dagger}\bigg(-\frac{1}{2}\ln\bigg[e^{2k_+}(1-e^{-2q_+})-1\bigg]\bigg)\hat{\alpha}\bigg(\frac{1}{2}\ln\bigg[\frac{e^{2(k_+-q_+)}}{e^{2k_+}(1-e^{-2q_+})-1}\bigg]\bigg){e_1(k_+)e_2^{\dagger}(q_+)},\\
    J^{(2)}_3&=\int _0 ^{+\infty}dk_+ \int _0 ^{+\infty} dq_+ \frac{e^{2(k_++q_+)}}{(e^{2k_+}+1)(e^{2q_+}+1)}\hat{\alpha}(k_+)\hat{\alpha}(q_+){\epsilon_1(k_+)\epsilon_2(q_+)},\\
    -J^{(2)}_4&=\int _0 ^{+\infty}dk_+ \int _0 ^{+\infty} dq_+ \frac{e^{2(k_++q_+)}}{(e^{2k_+}+1)(e^{2q_+}+1)}\hat{\alpha}(q_+)\hat{\alpha}(k_+)\textcolor{black}{\epsilon_2(q_+)\epsilon_1(k_+)}=\\
    &-\int_{\frac{1}{2}\ln(3)}^{+\infty}dk_+\int_{-\frac{1}{2}\ln(1-2e^{-2k_+})}^{\frac{1}{2}\ln(\frac{1-e^{2k_+}}{2})}dq_+\frac{e^{2k_+}}{(e^{2k_+}-1)(e^{2q_+}-1)}\\
    &\hat{\alpha}\bigg(\frac{1}{2}\ln\bigg[e^{-2q_+}(e^{2(q_++k_+)}-e^{2k_+}-e^{2q_+})\bigg]\bigg)\hat{\alpha}\bigg(\frac{1}{2}\ln\bigg[\frac{e^{2k_+}}{e^{2(q_++k_+)}-e^{2k_+}-e^{2q_+}}\bigg]\bigg){e_1(k_+)e_2^{\dagger}(q_+)},\\
        J^{(2)}_5&=\int _0 ^{+\infty}dk_+ \int _0 ^{+\infty} dq_+ \frac{1}{(e^{2k_+}+1)(e^{2q_+}+1)}\hat{\beta}^{\dagger}(k_+)\hat{\beta}^{\dagger}(q_+){\epsilon_1^{\dagger}(k_+)\epsilon_2^{\dagger}(q_+)},
    \end{aligned}
\end{equation*}
\begin{equation*}
    \begin{aligned}
    -J^{(2)}_6&=\int _0 ^{+\infty}dk_+ \int _0 ^{+\infty} dq_+ \frac{1}{(e^{2k_+}+1)(e^{2q_+}+1)}\hat{\beta}^{\dagger}(q_+)\hat{\beta}^{\dagger}(k_+)\textcolor{black}{\epsilon_2^{\dagger}(q_+)\epsilon_1^{\dagger}(k_+)}=\\
    &-\int_0^{\frac{1}{2}\ln(2)}dk_+\int_{-\frac{1}{2}\ln(\frac{1-e^{-2k_+}}{2})}^{+\infty}dq_+\frac{e^{2k_+}}{(e^{2k_+}-1)(e^{2q_+}-1)}\\
    &\hat{\beta}^{\dagger}\bigg(\frac{1}{2}\ln\bigg[\frac{e^{2q_+}}{e^{2(k_++q_+)}-e^{2k_+}-e^{2q_+}}\bigg]\bigg)\hat{\beta}^{\dagger}\bigg(\frac{1}{2}\ln\bigg[e^{-2k_+}(e^{2(k_++q_+)}-e^{2k_+}-e^{2q_+})\bigg]\bigg){e_1(k_+)e_2^{\dagger}(q_+)}\\
    &-\int_{\frac{1}{2}\ln(2)}^{\frac{1}{2}\ln(3)}dk_+\int_{-\frac{1}{2}\ln(\frac{1-e^{-2k_+}}{2})}^{-\frac{1}{2}\ln(1-2e^{-2k_+})}dq_+\frac{e^{2k_+}}{(e^{2k_+}-1)(e^{2q_+}-1)}\\
    &\hat{\beta}^{\dagger}\bigg(\frac{1}{2}\ln\bigg[\frac{e^{2q_+}}{e^{2(k_++q_+)}-e^{2k_+}-e^{2q_+}}\bigg]\bigg)\hat{\beta}^{\dagger}\bigg(\frac{1}{2}\ln\bigg[e^{-2k_+}(e^{2(k_++q_+)}-e^{2k_+}-e^{2q_+})\bigg]\bigg){e_1(k_+)e_2^{\dagger}(q_+)},\\
   J^{(2)}_7&=\int _0 ^{+\infty}dk_+ \int _0 ^{+\infty} dq_+ \frac{e^{2q_+}}{(e^{2k_+}+1)(e^{2q_+}+1)}\hat{\beta}^{\dagger}(k_+)\hat{\alpha}(q_+){\epsilon_1^{\dagger}(k_+)\epsilon_2(q_+)} , \\
   -J^{(2)}_8&=\int _0 ^{+\infty}dk_+ \int _0 ^{+\infty} dq_+ \frac{e^{2q_+}}{(e^{2k_+}+1)(e^{2q_+}+1)}\hat{\alpha}(q_+)\hat{\beta}^{\dagger}(k_+)\textcolor{black}{\epsilon_2(q_+)}\epsilon_1^{\dagger}(k_+)=\\
   &-\int_{\frac{1}{2}\ln(2)}^{\frac{1}{2}\ln(3)}dk_+\int_{-\frac{1}{2}\ln(1-2e^{-2k_+})}^{+\infty}dq_+\frac{e^{2k_+}}{(e^{2k_+}-1)(e^{2q_+}-1)}\\
   &\hat{\alpha}\bigg(-\frac{1}{2}\ln\bigg[\frac{e^{2q_+}}{e^{2(k_++q_+)}-e^{2q_+}-e^{2k_+}}\bigg]\bigg)\hat{\beta}^{\dagger}\bigg(-\frac{1}{2}\ln\bigg[\frac{e^{2k_+}}{e^{2(k_++q_+)}-e^{2q_+}-e^{2k_+}}\bigg]\bigg){e_1(k_+)e_2^{\dagger}(q_+)}\\
   &-\int_{\frac{1}{2}\ln(3)}^{+\infty}dk_+\int_{-\frac{1}{2}\ln(\frac{1-e^{-2k_+}}{2})}^{+\infty}dq_+\frac{e^{2k_+}}{(e^{2k_+}-1)(e^{2q_+}-1)}\\
   &\hat{\alpha}\bigg(-\frac{1}{2}\ln\bigg[\frac{e^{2q_+}}{e^{2(k_++q_+)}-e^{2q_+}-e^{2k_+}}\bigg]\bigg)\hat{\beta}^{\dagger}\bigg(-\frac{1}{2}\ln\bigg[\frac{e^{2k_+}}{e^{2(k_++q_+)}-e^{2q_+}-e^{2k_+}}\bigg]\bigg){e_1(k_+)e_2^{\dagger}(q_+)},\\
      H^{(2)}_1&=\int _0 ^{+\infty}dk_+ \int _0 ^{+\infty} dq_+ \frac{e^{2k_+}}{(e^{2k_+}-1)(e^{2q_+}+1)}\hat{a}(k_+)\hat{\beta}^{\dagger}(q_+){e_1(k_+)\epsilon_2^{\dagger}(q_+)},     \\
   -H^{(2)}_2&=\int _0 ^{+\infty}dk_+ \int _0 ^{+\infty} dq_+ \frac{e^{2k_+}}{(e^{2k_+}-1)(e^{2q_+}+1)}\hat{\beta}^{\dagger}(q_+)\hat{a}(k_+)\epsilon_2^{\dagger}(q_+)e_1(k_+)=\\
   &\int_{\frac{1}{2}\ln(3/2)}^{+\infty}dk_+\int_{\frac{1}{2}\ln(2(e^{2k_+}-1))}^{\frac{1}{2}\ln(2e^{2k_+}-1)}dq_+\frac{e^{2(q_+-k_+)}}{(1-e^{-2k_+})(e^{2q_+}+1)}\\
   &\hat{\beta}^{\dagger}\bigg(-\frac{1}{2}\ln\bigg[e^{2(q_+-k_+)}+e^{-2k_+}-1\bigg]\bigg)\hat{a}\bigg(\frac{1}{2}\ln\bigg[\frac{e^{2(q_+-k_+)}}{e^{2(q_+-k_+)}+e^{-2k_+}-1}\bigg]\bigg){e_1^{\dagger}(k_+)\epsilon_2(q_+)}\\
   &+\int_0^{\frac{1}{2}\ln(3/2)}dk_+\int_0^{\frac{1}{2}\ln(2e^{2k_+}-1)}dq_+\frac{e^{2(q_+-k_+)}}{(1-e^{-2k_+})(e^{2q_+}+1)}\\
   &\hat{\beta}^{\dagger}\bigg(-\frac{1}{2}\ln\bigg[e^{2(q_+-k_+)}+e^{-2k_+}-1\bigg]\bigg)\hat{a}(\frac{1}{2}\ln\bigg[\frac{e^{2(q_+-k_+)}}{e^{2(q_+-k_+)}+e^{-2k_+}-1}\bigg]\bigg){e_1^{\dagger}(k_+)\epsilon_2(q_+)}\\
   &+\int_{0}^{\frac{1}{2}\ln(2)}dk_+\int_{0}^{-\frac{1}{2}\ln(e^{2k_+}-1)}dq_+\frac{e^{-2(q_++k_+)}}{(1-e^{-2k_+})(e^{-2q_+}+1)}\\
   &\hat{\beta}^{\dagger}\bigg(-\frac{1}{2}\ln\bigg[e^{-2(q_++k_+)}+e^{-2k_+}-1\bigg]\bigg)\hat{a}\bigg(\frac{1}{2}\ln\bigg[\frac{e^{-2(q_++k_+)}}{e^{-2(q_++k_+)}+e^{-2k_+}-1}\bigg]\bigg){e_1^{\dagger}(k_+)\epsilon_2^{\dagger}(q_+)}\\
   &+\int_{\frac{1}{2}\ln(3/2)}^{\frac{1}{2}\ln(2)}dk_+\int_{0}^{\frac{1}{2}\ln(2(e^{2k_+}-1))}dq_+\frac{e^{2(q_+-k_+)}}{(1-e^{-2k_+})(e^{2q_+}+1)}\\
   &\hat{\beta}^{\dagger}\bigg(-\frac{1}{2}\ln\bigg[e^{2(q_+-k_+)}+e^{-2k_+}-1\bigg]\bigg)\hat{a}\bigg(\frac{1}{2}\ln\bigg[\frac{e^{2(q_+-k_+)}}{e^{2(q_+-k_+)}+e^{-2k_+}-1}\bigg]\bigg){e_1^{\dagger}(k_+)\epsilon_2(q_+)}\\
   &+\int_{\frac{1}{2}\ln(2)}^{+\infty}dk_+\int_{\frac{1}{2}\ln(e^{2k_+}-1)}^{\frac{1}{2}\ln(2(e^{2k_+}-1))}dq_+\frac{e^{2(q_+-k_+)}}{(1-e^{-2k_+})(e^{2q_+}+1)}\\
   &\hat{\beta}^{\dagger}\bigg(-\frac{1}{2}\ln\bigg[e^{2(q_+-k_+)}+e^{-2k_+}-1\bigg])\hat{a}\bigg(\frac{1}{2}\ln\bigg[\frac{e^{2(q_+-k_+)}}{e^{2(q_+-k_+)}+e^{-2k_+}-1}\bigg]\bigg){e_1^{\dagger}(k_+)\epsilon_2(q_+)}
    \end{aligned}
\end{equation*}
\begin{equation*}
    \begin{aligned}
&+\int_0^{+\infty}dk_+\int_{\frac{1}{2}\ln(1+e^{-2k_+})}^{\frac{1}{2}\ln(1+2e^{-2k_+})}dq_+\frac{e^{2(q_++k_+)}}{(e^{2k_+}+1)(e^{2q_+}-1)}\\
&\hat{\beta}^{\dagger}\bigg(-\frac{1}{2}\ln\bigg[e^{2(k_++q_+)}-e^{2k_+}-1\bigg]\bigg)\hat{a}\bigg(\frac{1}{2}\ln\bigg[\frac{e^{2(k_++q_+)}}{e^{2(k_++q_+)}-e^{2k_+}-1}\bigg]\bigg){\epsilon_1(k_+)e_2(q_+)}\\
    &+\int_0^{+\infty}dk_+\int_{\frac{1}{2}\ln(1+e^{2k_+})}^{\frac{1}{2}\ln(1+2e^{2k_+})}dq_+\frac{e^{2(q_+-k_+)}}{(e^{-2k_+}+1)(e^{2q_+}-1)}\\
&\hat{\beta}^{\dagger}\bigg(-\frac{1}{2}\ln\bigg[e^{2(-k_++q_+)}-e^{-2k_+}-1\bigg]\bigg)\hat{a}\bigg(\frac{1}{2}\ln\bigg[\frac{e^{2(-k_++q_+)}}{e^{2(-k_++q_+)}-e^{-2k_+}-1}\bigg]\bigg){\epsilon_1^{\dagger}(k_+)e_2(q_+)},\\
  H^{(2)}_3&=\int _0 ^{+\infty}dk_+ \int _0 ^{+\infty} dq_+ \frac{e^{2(k_++q_+)}}{(e^{2k_+}-1)(e^{2q_+}+1)}\hat{a}(k_+)\hat{\alpha}(q_+){e_1(k_+)\epsilon_2(q_+)},   \\
  -H^{(2)}_4&=\int _0 ^{+\infty}dk_+ \int _0 ^{+\infty} dq_+ \frac{e^{2(k_++q_+)}}{(e^{2k_+}-1)(e^{2q_+}+1)}\hat{\alpha}(q_+)\hat{a}(k_+)\textcolor{black}{\epsilon_2(q_+)e_1(k_+)}=\\
  &\int_0^{+\infty}dk_+\int_{\frac{1}{2}\ln(2e^{2k_+}-1)}^{+\infty}dq_+\frac{e^{2(q_+-k_+)}}{(1-e^{-2k_+})(e^{2q_+}+1)}\\
  &\hat{\alpha}\bigg(\frac{1}{2}\ln\bigg[e^{2(q_+-k_+)}+e^{-2k_+}-1\bigg]\bigg)\hat{a}\bigg(\frac{1}{2}\ln\bigg[\frac{e^{2(q_+-k_+)}}{e^{2(q_+-k_+)}+e^{-2k_+}-1}\bigg]\bigg){e_1^{\dagger}(k_+)\epsilon_2(q_+)}\\
  &+\int_0^{+\infty}dk_+\int_{\frac{1}{2}\ln(2(e^{-2k_+}+1))}^{+\infty}dq_+\frac{e^{2(q_++k_+)}}{(1+e^{2k_+})(e^{2q_+}-1)}\\
  &\hat{\alpha}\bigg(\frac{1}{2}\ln\bigg[e^{2(q_++k_+)}-e^{2k_+}-1\bigg]\bigg)\hat{a}\bigg(\frac{1}{2}\ln\bigg[\frac{e^{2(q_++k_+)}}{e^{2(q_++k_+)}-e^{2k_+}-1}\bigg]\bigg){\epsilon_1(k_+)e_2(q_+)}\\
  &+\int_0^{+\infty}dk_+\int_{\frac{1}{2}\ln(2(e^{2k_+}+1))}^{+\infty}dq_+\frac{e^{2(q_+-k_+)}}{(1+e^{-2k_+})(e^{2q_+}-1)}\\
  &\hat{\alpha}\bigg(\frac{1}{2}\ln\bigg[e^{2(q_+-k_+)}-e^{-2k_+}-1\bigg]\bigg)\hat{a}\bigg(\frac{1}{2}\ln\bigg[\frac{e^{2(q_+-k_+)}}{e^{2(q_+-k_+)}-e^{-2k_+}-1}\bigg]\bigg){\epsilon^{\dagger}_1(k_+)e_2(q_+)}\\
  &+\int_0^{+\infty}dk_+\int_{\frac{1}{2}\ln(2e^{-2k_+}+1)}^{\frac{1}{2}\ln(2(e^{-2k_+}+1))}dq_+\frac{e^{2(q_++k_+)}}{(1+e^{2k_+})(e^{2q_+}-1)}\\
  &\hat{\alpha}\bigg(\frac{1}{2}\ln\bigg[e^{2(q_++k_+)}-e^{2k_+}-1\bigg]\bigg)\hat{a}\bigg(\frac{1}{2}\ln\bigg[\frac{e^{2(q_++k_+)}}{e^{2(q_++k_+)}-e^{2k_+}-1}\bigg]\bigg){\epsilon_1(k_+)e_2(q_+)}\\
  &+\int_0^{+\infty}dk_+\int_{\frac{1}{2}\ln(2e^{2k_+}+1)}^{\frac{1}{2}\ln(2(e^{2k_+}+1))}dq_+\frac{e^{2(q_+-k_+)}}{(1+e^{-2k_+})(e^{2q_+}-1)}\\
  &\hat{\alpha}\bigg(\frac{1}{2}\ln\bigg[e^{2(q_+-k_+)}-e^{-2k_+}-1\bigg]\bigg)\hat{a}\bigg(\frac{1}{2}\ln\bigg[\frac{e^{2(q_+-k_+)}}{e^{2(q_+-k_+)}-e^{-2k_+}-1}\bigg]\bigg){\epsilon^{\dagger}_1(k_+)e_2(q_+)},\\
  H^{(2)}_5&=\int _0 ^{+\infty}dk_+ \int _0 ^{+\infty} dq_+ \frac{1}{(e^{2k_+}-1)(e^{2q_+}+1)}\hat{b}^{\dagger}(k_+)\hat{\beta}^{\dagger}(q_+){e_1^{\dagger}(k_+)\epsilon_2^{\dagger}(q_+)}\\
  -H^{(2)}_6&=\int _0 ^{+\infty}dk_+ \int _0 ^{+\infty} dq_+ \frac{1}{(e^{2k_+}-1)(e^{2q_+}+1)}\hat{\beta}^{\dagger}(q_+)\hat{b}^{\dagger}(k_+)\textcolor{black}{\epsilon_2^{\dagger}(q_+)e_1^{\dagger}(k_+)}=\\
  &=\int_0^{\frac{1}{2}\ln(2)}dk_+\int_{-\frac{1}{2}\ln(2e^{-2k_+}-1)}^{+\infty}dq_+\frac{e^{2k_+}}{(e^{2k_+}-1)(e^{2q_+}+1)}\\
  &\hat{\beta}^{\dagger}\bigg(\frac{1}{2}\ln\bigg[\frac{e^{2q_+}}{e^{2(k_++q_+)}+e^{2k_+}-e^{2q_+}}\bigg]\bigg)\hat{b}^{\dagger}\bigg(\frac{1}{2}\ln\bigg[e^{-2k_+}(e^{2(k_++q_+)}+e^{2k_+}-e^{2q_+})\bigg]\bigg){e_1(k_+)\epsilon_2^{\dagger}(q_+)},\\
  H^{(2)}_7&=\int _0 ^{+\infty}dk_+ \int _0 ^{+\infty} dq_+ \frac{e^{2q_+}}{(e^{2k_+}-1)(e^{2q_+}+1)}\hat{b}^{\dagger}(k_+)\hat{\alpha}(q_+){e_1^{\dagger}(k_+)\epsilon_2(q_+)},\\
     -H^{(2)}_8&=\int _0 ^{+\infty}dk_+ \int _0 ^{+\infty} dq_+ \frac{e^{2q_+}}{(e^{2k_+}-1)(e^{2q_+}+1)}\hat{\alpha}(q_+)\hat{b}^{\dagger}(k_+)\textcolor{black}{\epsilon_2(q_+)}e_1^{\dagger}(k_+)= \\
   &\int_{\frac{1}{2}\ln(2)}^{+\infty}dk_+\int_0^{+\infty}dq_+\frac{e^{2(k_++q_+)}}{(e^{2k_+}-1)(e^{2q_+}+1)}\\
   &\hat{\alpha}\bigg(\frac{1}{2}\ln\bigg[e^{2(k_++q_+)}+e^{2k_+}-1\bigg]\bigg)\hat{b}^{\dagger}\bigg(-\frac{1}{2}\ln\bigg[\frac{e^{2(q_++k_+)}}{e^{2(k_++q_+)}+e^{2k_+}-1}\bigg]\bigg){e_1(k_+)\epsilon_2(q_+)}\\
    \end{aligned}
\end{equation*}
\begin{equation*} \begin{aligned}
   &+\int_{\frac{1}{2}\ln(2)}^{+\infty}dk_+\int_0^{+\infty}dq_+\frac{e^{2(k_+-q_+)}}{(e^{2k_+}-1)(e^{-2q_+}+1)}\\
   &\hat{\alpha}\bigg(\frac{1}{2}\ln\bigg[e^{2(k_+-q_+)}+e^{2k_+}-1\bigg]\bigg)\hat{b}^{\dagger}\bigg(-\frac{1}{2}\ln\bigg[\frac{e^{2(-q_++k_+)}}{e^{2(k_+-q_+)}+e^{2k_+}-1}\bigg]\bigg){e_1(k_+)\epsilon_2^{\dagger}(q_+)}\\
   &+\int_0^{\frac{1}{2}\ln(2)}dk_+\int_0^{+\infty}dq_+\frac{e^{2(k_++q_+)}}{(e^{2k_+}-1)(e^{2q_+}+1)}\\
   &\hat{\alpha}\bigg(\frac{1}{2}\ln\bigg[e^{2(k_++q_+)}+e^{2k_+}-1\bigg]\bigg)\hat{b}^{\dagger}\bigg(-\frac{1}{2}\ln\bigg[\frac{e^{2(q_++k_+)}}{e^{2(k_++q_+)}+e^{2k_+}-1}\bigg]\bigg){e_1(k_+)\epsilon_2(q_+)}\\
   &+\int_0^{\frac{1}{2}\ln(2)}dk_+\int_0^{-\frac{1}{2}\ln(2e^{-2k_+}-1)}dq_+\frac{e^{2(k_+-q_+)}}{(e^{2k_+}-1)(e^{-2q_+}+1)}\\
   &\hat{\alpha}\bigg(\frac{1}{2}\ln\bigg[e^{2(k_+-q_+)}+e^{2k_+}-1\bigg]\bigg)\hat{b}^{\dagger}\bigg(-\frac{1}{2}\ln\bigg[\frac{e^{2(-q_++k_+)}}{e^{2(k_+-q_+)}+e^{2k_+}-1}\bigg]\bigg){e_1(k_+)\epsilon_2^{\dagger}(q_+)},\\
   H^{(2)}_9&=\int _0 ^{+\infty}dk_+ \int _0 ^{+\infty} dq_+ \frac{e^{2k_+}}{(e^{2k_+}+1)(e^{2q_+}-1)}\hat{\alpha}(k_+)\hat{b}^{\dagger}(q_+){\epsilon_1(k_+)e_2^{\dagger}(q_+)},\\
       -H^{(2)}_{10}&=\int _0 ^{+\infty}dk_+ \int _0 ^{+\infty} dq_+ \frac{e^{2k_+}}{(e^{2k_+}+1)(e^{2q_+}-1)}\hat{b}^{\dagger}(q_+)\hat{\alpha}(k_+)e_2^{\dagger}(q_+)\epsilon_1(k_+)=\\
    &\int_0^{\frac{1}{2}\ln(2)}dk_+\int_{-\frac{1}{2}\ln(e^{2k_+}-1)}^{-\frac{1}{2}\ln(\frac{e^{2k_+}-1}{2})}dq_+\frac{e^{-2(k_++q_+)}}{(1-e^{-2k_+})(1+e^{-2q_+})}\\
    &\hat{b}^{\dagger}\bigg(-\frac{1}{2}\ln\bigg[1-e^{-2k_+}-e^{-2(k_++q_+)}\bigg]\bigg)\hat{\alpha}\bigg(\frac{1}{2}\ln\bigg[\frac{e^{-2(k_++q_+)}}{1-e^{-2k_+}-e^{-2(k_++q_+)}}\bigg]\bigg){e_1^{\dagger}(k_+)\epsilon_2^{\dagger}(q_+)}\\
    &+\int_{\frac{1}{2}\ln(2)}^{\frac{1}{2}\ln(3)}dk_+\int_0^{-\frac{1}{2}\ln(\frac{e^{2k_+}-1}{2})}dq_+\frac{e^{-2(k_++q_+)}}{(1-e^{-2k_+})(1+e^{-2q_+})}\\
    &\hat{b}^{\dagger}\bigg(-\frac{1}{2}\ln\bigg[1-e^{-2k_+}-e^{-2(k_++q_+)}\bigg]\bigg)\hat{\alpha}\bigg(\frac{1}{2}\ln\bigg[\frac{e^{-2(k_++q_+)}}{1-e^{-2k_+}-e^{-2(k_++q_+)}}\bigg]\bigg){e_1^{\dagger}(k_+)\epsilon_2^{\dagger}(q_+)}\\
     &+\int_{\frac{1}{2}\ln(2)}^{\frac{1}{2}\ln(3)}dk_+\int_0^{\frac{1}{2}\ln(e^{2k_+}-1)}dq_+\frac{e^{-2(k_+-q_+)}}{(1-e^{-2k_+})(1+e^{2q_+})}\\
    &\hat{b}^{\dagger}\bigg(-\frac{1}{2}\ln\bigg[1-e^{-2k_+}-e^{-2(k_+-q_+)}\bigg]\bigg)\hat{\alpha}\bigg(\frac{1}{2}\ln\bigg[\frac{e^{-2(k_+-q_+)}}{1-e^{-2k_+}-e^{-2(k_+-q_+)}}\bigg]\bigg){e_1^{\dagger}(k_+)\epsilon_2(q_+)}\\
    &+\int_{\frac{1}{2}\ln(3)}^{+\infty}dk_+\int_{\frac{1}{2}\ln(\frac{e^{2k_+}-1}{2})}^{\frac{1}{2}\ln(e^{2k_+}-1)}dq_+\frac{e^{-2(k_+-q_+)}}{(1-e^{-2k_+})(1+e^{2q_+})}\\
    &\hat{b}^{\dagger}\bigg(-\frac{1}{2}\ln\bigg[1-e^{-2k_+}-e^{-2(k_+-q_+)}\bigg]\bigg)\hat{\alpha}\bigg(\frac{1}{2}\ln\bigg[\frac{e^{-2(k_+-q_+)}}{1-e^{-2k_+}-e^{-2(k_+-q_+)}}\bigg]\bigg){e_1^{\dagger}(k_+)\epsilon_2(q_+)}\\
    &+\int_0^{+\infty}dk_+\int_0^{\frac{1}{2}\ln(1+e^{-2k_+})}dq_+\frac{e^{2(k_++q_+)}}{(e^{2k_+}+1)(e^{2q_+}-1)}\\
    &\hat{b}^{\dagger}\bigg(\frac{1}{2}\ln\bigg[\frac{1}{1+e^{2k_+}-e^{2(k_++q_+)}}\bigg]\bigg)\hat{\alpha}\bigg(\frac{1}{2}\ln\bigg[\frac{e^{2(k_++q_+)}}{1+e^{2k_+}-e^{2(k_++q_+)}}\bigg]\bigg){\epsilon_1(k_+)e_2(q_+)}\\
    &+\int_0^{+\infty}dk_+\int_{\frac{1}{2}\ln(\frac{1+e^{2k_+}}{2})}^{\frac{1}{2}\ln(1+e^{2k_+})}dq_+\frac{e^{2(-k_++q_+)}}{(e^{-2k_+}+1)(e^{2q_+}-1)}\\
    &\hat{b}^{\dagger}\bigg(\frac{1}{2}\ln\bigg[\frac{1}{1+e^{-2k_+}-e^{2(-k_++q_+)}}\bigg]\bigg)\hat{\alpha}\bigg(\frac{1}{2}\ln\bigg[\frac{e^{2(-k_++q_+)}}{1+e^{-2k_+}-e^{2(-k_++q_+)}}\bigg]\bigg){\epsilon_1^{\dagger}(k_+)e_2(q_+)},\\
        H^{(2)}_{11}&=\int _0 ^{+\infty}dk_+ \int _0 ^{+\infty} dq_+ \frac{e^{2(k_++q_+)}}{(e^{2k_+}+1)(e^{2q_+}-1)}\hat{\alpha}(k_+)\hat{a}(q_+){\epsilon_1(k_+)e_2(q_+)},
\end{aligned}\end{equation*}
\begin{equation*} \begin{aligned}
   -H^{(2)}_{12}&=\int _0 ^{+\infty}dk_+ \int _0 ^{+\infty} dq_+ \frac{e^{2(k_++q_+)}}{(e^{2k_+}+1)(e^{2q_+}-1)}\hat{a}(q_+)\hat{\alpha}(k_+)\textcolor{black}{e_2(q_+)\epsilon_1(k_+)}=\\
    &\int_0^{+\infty}dk_+\int_0^{-\frac{1}{2}\ln(\frac{1+e^{-2k_+}}{2})}dq_+\frac{e^{2(k_+-q_+)}}{(e^{2k_+}+1)(1-e^{-2q_+})}\\
    &\hat{a}\bigg(\frac{1}{2}\ln\bigg[1+e^{2k_+}-e^{2(k_+-q_+)}\bigg]\bigg)\hat{\alpha}\bigg(\frac{1}{2}\ln\bigg[\frac{e^{2(k_+-q_+)}}{1+e^{2k_+}-e^{2(k_+-q_+)}}\bigg]\bigg){\epsilon_1(k_+)e_2^{\dagger}(q_+)},\\
    H^{(2)}_{13}&=\int _0 ^{+\infty}dk_+ \int _0 ^{+\infty} dq_+ \frac{1}{(e^{2k_+}+1)(e^{2q_+}-1)}\hat{\beta}^{\dagger}(k_+)\hat{b}^{\dagger}(q_+){\epsilon_1^{\dagger}(k_+)e_2^{\dagger}(q_+)},\\
  -H^{(2)}_{14}&=\int _0 ^{+\infty}dk_+ \int _0 ^{+\infty} dq_+ \frac{1}{(e^{2k_+}+1)(e^{2q_+}-1)}\hat{b}^{\dagger}(q_+)\hat{\beta}^{\dagger}(k_+)\textcolor{black}{e_2^{\dagger}(q_+)\epsilon_1^{\dagger}(k_+)}=  \\
  &\int_{\frac{1}{2}\ln(3)}^{+\infty}dk_+\int_0^{\frac{1}{2}\ln(\frac{e^{2k_+}-1}{2})}dq_+\frac{e^{2(q_+-k_+)}}{(e^{2q_+}+1)(1-e^{-2k_+})}\\
  &\hat{b}^{\dagger}\bigg(\frac{1}{2}\ln\bigg[\frac{1}{1-e^{-2k_+}-e^{2(q_+-k_+)}}\bigg]\bigg)\hat{\beta}^{\dagger}\bigg(\frac{1}{2}\ln\bigg[\frac{1-e^{-2k_+}-e^{2(q_+-k_+)}}{e^{2(q_+-k_+)}}\bigg]\bigg){e_1^{\dagger}(k_+)\epsilon_2(q_+)}\\
  &+\int_{\frac{1}{2}\ln(3)}^{+\infty}dk_+\int_0^{+\infty}dq_+\frac{e^{-2(q_++k_+)}}{(e^{-2q_+}+1)(1-e^{-2k_+})}\\
  &\hat{b}^{\dagger}\bigg(\frac{1}{2}\ln\bigg[\frac{1}{1-e^{-2k_+}-e^{-2(q_++k_+)}}\bigg]\bigg)\hat{\beta}^{\dagger}\bigg(\frac{1}{2}\ln\bigg[\frac{1-e^{-2k_+}-e^{-2(q_++k_+)}}{e^{-2(q_++k_+)}}\bigg]\bigg){e_1^{\dagger}(k_+)\epsilon_2^{\dagger}(q_+)}\\
  &+\int_0^{\frac{1}{2}\ln(3)}dk_+\int_{-\frac{1}{2}\ln(\frac{e^{2k_+}-1}{2})}^{+\infty}dq_+\frac{e^{-2(q_++k_+)}}{(e^{-2q_+}+1)(1-e^{-2k_+})}\\
  &\hat{b}^{\dagger}\bigg(\frac{1}{2}\ln\bigg[\frac{1}{1-e^{-2k_+}-e^{-2(q_++k_+)}}\bigg]\bigg)\hat{\beta}^{\dagger}\bigg(\frac{1}{2}\ln\bigg[\frac{1-e^{-2k_+}-e^{-2(q_++k_+)}}{e^{-2(q_++k_+)}}\bigg]\bigg){e_1^{\dagger}(k_+)\epsilon_2^{\dagger}(q_+)}\\
  &+\int_0^{+\infty}dk_+\int_0^{\frac{1}{2}\ln(\frac{1+e^{2k_+}}{2})}dq_+\frac{e^{2(q_+-k_+)}}{(e^{-2k_+}+1)(e^{2q_+}-1)}\\
  &\hat{b}^{\dagger}\bigg(\frac{1}{2}\ln\bigg[\frac{1}{1+e^{-2k_+}-e^{2(q_+-k_+)}}\bigg]\bigg)\hat{\beta}^{\dagger}\bigg(\frac{1}{2}\ln\bigg[\frac{1+e^{-2k_+}-e^{2(q_+-k_+)}}{e^{2(q_+-k_+)}}\bigg]\bigg){\epsilon_1^{\dagger}(k_+)e_2(q_+)},\\
  H^{(2)}_{15}&=\int _0 ^{+\infty}dk_+ \int _0 ^{+\infty} dq_+ \frac{e^{2q_+}}{(e^{2k_+}+1)(e^{2q_+}-1)}\hat{\beta}^{\dagger}(k_+)\hat{a}(q_+){\epsilon_1^{\dagger}(k_+)e_2(q_+)}\\
  -H^{(2)}_{16}&=\int _0 ^{+\infty}dk_+ \int _0 ^{+\infty} dq_+ \frac{e^{2q_+}}{(e^{2k_+}+1)(e^{2q_+}-1)}\hat{a}(q_+)\hat{\beta}^{\dagger}(k_+)\textcolor{black}{e_2(q_+)}\epsilon_1^{\dagger}(k_+)=\\
&\int_0^{+\infty}dk_+\int_0^{+\infty}dq_+\frac{e^{-2(k_++q_+)}}{(e^{-2k_+}+1)(1-e^{-2q_+})}\\
&\hat{a}\bigg(\frac{1}{2}\ln\bigg[1+e^{-2k_+}-e^{-2(k_++q_+)}\bigg]\bigg)\hat{\beta}^{\dagger}\bigg(-\frac{1}{2}\ln\bigg[\frac{e^{-2(k_++q_+)}}{1+e^{-2k_+}-e^{-2(k_++q_+)}}\bigg]\bigg){\epsilon_1^{\dagger}(k_+)e_2^{\dagger}(q_+)}\\
&+\int_0^{+\infty}dk_+\int_{-\frac{1}{2}\ln(\frac{1+e^{-2k_+}}{2})}^{+\infty}dq_+\frac{e^{2(k_+-q_+)}}{(e^{2k_+}+1)(1-e^{-2q_+})}\\
&\hat{a}\bigg(\frac{1}{2}\ln\bigg[1+e^{2k_+}-e^{2(k_+-q_+)}\bigg]\bigg)\hat{\beta}^{\dagger}\bigg(-\frac{1}{2}\ln\bigg[\frac{e^{2(k_+-q_+)}}{1+e^{2k_+}-e^{2(k_+-q_+)}}\bigg]\bigg){\epsilon_1(k_+)e_2^{\dagger}(q_+)}.
\end{aligned}\end{equation*}
Also in this case, we re-wrote the even-numbered integrals by moving $x_2$ to the right.

Lastly, the left hand-side of the third equation in~(\ref{eq:commrel}),  $[\hat{\phi}^{\dagger}(x_1),\hat{\phi}^{\dagger}(x_2)]=0$, can be written as
\begin{equation}
    [\hat{\phi}^{\dagger}(x_1),\hat{\phi}^{\dagger}(x_2)]=\sum_{i=1}^{8}I^{(3)}_i+\sum_{i=1}^{8}J^{(3)}_i+\sum_{i=1}^{16}H^{(3)}_i \,,
\end{equation}
where
\begin{equation}
    \begin{aligned}
   I^{(3)}_1&=\int _0 ^{+\infty}dk_+ \int _0 ^{+\infty} dq_+ \frac{e^{2q_+}}{(e^{2k_+}-1)(e^{2q_+}-1)}\hat{b}(k_+)\hat{a}^{\dagger}(q_+){e_1(k_+)e_2^{\dagger}(q_+)}    ,\\
   -I^{(3)}_2&=\int _0 ^{+\infty}dk_+ \int _0 ^{+\infty} dq_+ \frac{e^{2q_+}}{(e^{2k_+}-1)(e^{2q_+}-1)}\hat{a}^{\dagger}(q_+)\hat{b}(k_+)e_2^{\dagger}(q_+)e_1(k_+)=\\
   &\int _0 ^{+\infty}dk_+ \int _0 ^{-\frac{1}{2}\ln(1-e^{-2{k_{+}}})} dq_+ \frac{e^{2q_+}}{(e^{2k_+}-1)(e^{2q_+}-1)}\\
   &\hat{a}^{\dagger}\bigg(\frac{1}{2}\ln\bigg[\frac{e^{2q_+}}{e^{2k_+}+e^{2q_+}-e^{2(q_++k_+)}}\bigg]\bigg)\hat{b}\bigg(\frac{1}{2}\ln\bigg[\frac{e^{2k_+}}{e^{2k_+}+e^{2q_+}-e^{2(q_++k_+)}}\bigg]\bigg){e_1(k_+)e_2^{\dagger}(q_+)},\\
   I^{(3)}_3&=\int _0 ^{+\infty}dk_+ \int _0 ^{+\infty} dq_+ \frac{1}{(e^{2k_+}-1)(e^{2q_+}-1)}\hat{b}(k_+)\hat{b}(q_+){e_1(k_+)e_2(q_+)},\\
   -I^{(3)}_4&=\int _0 ^{+\infty}dk_+ \int _0 ^{+\infty} dq_+ \frac{1}{(e^{2k_+}-1)(e^{2q_+}-1)}\hat{b}(q_+)\hat{b}(k_+)\textcolor{black}{e_2(q_+)e_1(k_+)}=\\
   &\int _0 ^{+\infty}dk_+ \int _0 ^{+\infty} dq_+ \frac{1}{(e^{2k_+}-1)(e^{2q_+}-1)}\\
   &\hat{b}\bigg(\frac{1}{2}\ln\bigg[e^{2(q_++k_+)}-e^{2k_+}+1\bigg]\bigg)\hat{b}\bigg(\frac{1}{2}\ln\bigg[\frac{e^{2(q_++k_+)}}{e^{2(q_++k_+)}-e^{2k_+}+1}\bigg]\bigg){e_1(k_+)e_2(q_+)},\\
   I^{(3)}_5&=\int _0 ^{+\infty}dk_+ \int _0 ^{+\infty} dq_+ \frac{e^{2(k_++q_+)}}{(e^{2k_+}-1)(e^{2q_+}-1)}\hat{a}^{\dagger}(k_+)\hat{a}^{\dagger}(q_+){e_1^{\dagger}(k_+)e_2^{\dagger}(q_+)},\\
   -I^{(3)}_6&=\int _0 ^{+\infty}dk_+ \int _0 ^{+\infty} dq_+ \frac{e^{2(k_++q_+)}}{(e^{2k_+}-1)(e^{2q_+}-1)}\hat{a}^{\dagger}(q_+)\hat{a}^{\dagger}(k_+)\textcolor{black}{e_2^{\dagger}(q_+)e_1^{\dagger}(k_+)}=\\
   &\int _0 ^{+\infty}dk_+ \int _0 ^{+\infty} dq_+ \frac{e^{2(k_++q_+)}}{(e^{2k_+}-1)(e^{2q_+}-1)}\\
   &\hat{a}^{\dagger}\bigg(\frac{1}{2}\ln\bigg[\frac{e^{2(q_++k_+)}}{e^{2(q_++k_+)}-e^{2q_+}+1}\bigg]\bigg)\hat{a}^{\dagger}\bigg(\frac{1}{2}\ln\bigg[e^{2(q_++k_+)}-e^{2q_+}+1\bigg]\bigg){e_1^{\dagger}(k_+)e_2^{\dagger}(q_+)},\\
   I^{(3)}_7&=\int _0 ^{+\infty}dk_+ \int _0 ^{+\infty} dq_+ \frac{e^{2k_+}}{(e^{2k_+}-1)(e^{2q_+}-1)}\hat{a}^{\dagger}(k_+)\hat{b}(q_+){e_1^{\dagger}(k_+)e_2(q_+)},\\
   -I^{(3)}_8&=\int _0 ^{+\infty}dk_+ \int _0 ^{+\infty} dq_+ \frac{e^{2k_+}}{(e^{2k_+}-1)(e^{2q_+}-1)}\hat{b}(q_+)\hat{a}^{\dagger}(k_+)\textcolor{black}{e_2(q_+)}e_1^{\dagger}(k_+)=\\
   &\int _0 ^{+\infty}dk_+ \int _0 ^{+\infty} dq_+ \frac{e^{2k_+}}{(e^{2k_+}-1)(e^{2q_+}-1)}\\
   &\hat{b}\bigg(-\frac{1}{2}\ln\bigg[\frac{e^{2k_+}}{e^{2q_+}+e^{2k_+}-1}\bigg]\bigg)\hat{a}^{\dagger}\bigg(-\frac{1}{2}\ln\bigg[\frac{e^{2q_+}}{e^{2q_+}+e^{2k_+}-1}\bigg]\bigg){e_1^{\dagger}(k_+)e_2(q_+)}\\
      &-\int _0 ^{+\infty}dk_+ \int _0 ^{+\infty} dq_+ \frac{ e^{2k_+}}{(e^{2k_+}+1)(e^{2q_+}+1)}\\
        &\hat{b}\bigg(\frac{1}{2}\ln\bigg[\frac{e^{2q_+}+e^{2k_+}+1}{e^{2k_+}}\bigg]\bigg)\hat{a}^{\dagger}\bigg(\frac{1}{2}\ln\bigg[\frac{e^{2q_+}+e^{2k_+}+1}{e^{2q_+}}\bigg]\bigg){\epsilon_1^{\dagger}(k_+)\epsilon_2(q_+)}\\
        &-\int _0 ^{+\infty}dk_+ \int _0 ^{+\infty} dq_+ \frac{e^{-2k_+}}{(e^{-2k_+}+1)(e^{2q_+}+1)}\\
        &\hat{b}\bigg(\frac{1}{2}\ln\bigg[\frac{e^{2q_+}+e^{-2k_+}+1}{e^{-2k_+}}\bigg]\bigg)\hat{a}^{\dagger}\bigg(\frac{1}{2}\ln\bigg[\frac{e^{2q_+}+e^{-2k_+}+1}{e^{2q_+}}\bigg]\bigg){\epsilon_1(k_+)\epsilon_2(q_+)}\\
        &-\int _0 ^{+\infty}dk_+ \int _0 ^{+\infty} dq_+ \frac{ e^{2k_+}}{(e^{2k_+}+1)(e^{-2q_+}+1)}\\
        &\hat{b}\bigg(\frac{1}{2}\ln\bigg[\frac{e^{-2q_+}+e^{2k_+}+1}{e^{2k_+}}\bigg]\bigg)\hat{a}^{\dagger}\bigg(\frac{1}{2}\ln\bigg[\frac{e^{-2q_+}+e^{2k_+}+1}{e^{-2q_+}}\bigg]\bigg){\epsilon_1^{\dagger}(k_+)\epsilon_2^{\dagger}(q_+)}\\
        &-\int _0 ^{+\infty}dk_+ \int _0 ^{+\infty} dq_+ \frac{ e^{-2k_+}}{(e^{-2k_+}+1)(e^{-2q_+}+1)}\\
        &\hat{b}\bigg(\frac{1}{2}\ln\bigg[\frac{e^{-2q_+}+e^{-2k_+}+1}{e^{-2k_+}}\bigg]\bigg)\hat{a}^{\dagger}\bigg(\frac{1}{2}\ln\bigg[\frac{e^{-2q_+}+e^{-2k_+}+1}{e^{-2q_+}}\bigg]\bigg){\epsilon_1(k_+)\epsilon_2^{\dagger}(q_+)},
    \end{aligned}
\end{equation}
\begin{equation*}
    \begin{aligned}
        J^{(3)}_1&=\int _0 ^{+\infty}dk_+ \int _0 ^{+\infty} dq_+ \frac{e^{2q_+}}{(e^{2k_+}+1)(e^{2q_+}+1)}\hat{\beta}(k_+)\hat{\alpha}^{\dagger}(q_+){\epsilon_1(k_+)\epsilon_2^{\dagger}(q_+)},\\
        -J^{(3)}_2&=\int _0 ^{+\infty}dk_+ \int _0 ^{+\infty} dq_+ \frac{e^{2q_+}}{(e^{2k_+}+1)(e^{2q_+}+1)}\hat{\alpha}^{\dagger}(q_+)\hat{\beta}(k_+)\epsilon_2^{\dagger}(q_+)\epsilon_1(k_+)=\\
        &-\int_0^{\frac{1}{2}\ln(3)}dk_+ \int_{-\frac{1}{2}\ln(1-e^{-2k_+})}^{-\frac{1}{2}\ln(\frac{1-e^{-2k_+}}{2})}dq_+\frac{e^{2q_+}}{(e^{2k_+}-1)(e^{2q_+}-1)}\\
        &\hat{\alpha}^{\dagger}\bigg(-\frac{1}{2}\ln\bigg[e^{2k_+}(1-e^{-2q_+})-1\bigg]\bigg)\hat{\beta}\bigg(\frac{1}{2}\ln\bigg[\frac{e^{2(k_+-q_+)}}{e^{2k_+}(1-e^{-2q_+})-1}\bigg]\bigg){e_1(k_+)e_2^{\dagger}(q_+)}\\
        &-\int_{\frac{1}{2}\ln(3)}^{+\infty}dk_+ \int_{-\frac{1}{2}\ln(1-e^{-2k_+})}^{-\frac{1}{2}\ln(1-2e^{-2k_+})}dq_+\frac{e^{2q_+}}{(e^{2k_+}-1)(e^{2q_+}-1)}\\
        &\hat{\alpha}^{\dagger}\bigg(-\frac{1}{2}\ln\bigg[e^{2k_+}(1-e^{-2q_+})-1\bigg]\bigg)\hat{\beta}\bigg(\frac{1}{2}\ln\bigg[\frac{e^{2(k_+-q_+)}}{e^{2k_+}(1-e^{-2q_+})-1}\bigg]\bigg){e_1(k_+)e_2^{\dagger}(q_+)},\\
        J^{(3)}_3&=\int _0 ^{+\infty}dk_+ \int _0 ^{+\infty} dq_+ \frac{1}{(e^{2k_+}+1)(e^{2q_+}+1)}\hat{\beta}(k_+)\hat{\beta}(q_+){\epsilon_1(k_+)\epsilon_2(q_+)},\\
        -J^{(3)}_4&=\int _0 ^{+\infty}dk_+ \int _0 ^{+\infty} dq_+ \frac{1}{(e^{2k_+}+1)(e^{2q_+}+1)}\hat{\beta}(q_+)\hat{\beta}(k_+)\textcolor{black}{\epsilon_2(q_+)\epsilon_1(k_+)}=\\
        &-\int_{\frac{1}{2}\ln(3)}^{+\infty}dk_+\int_{-\frac{1}{2}\ln(1-2e^{-2k_+})}^{\frac{1}{2}\ln(\frac{1-e^{2k_+}}{2})}dq_+\frac{e^{2q_+}}{(e^{2k_+}-1)(e^{2q_+}-1)}\\
        &\hat{\beta}\bigg(\frac{1}{2}\ln\bigg[\frac {e^{2(q_++k_+)}-e^{2k_+}-e^{2q_+}}{e^{2q_+}}\bigg]\bigg)\hat{\beta}\bigg(\frac{1}{2}\ln\bigg[\frac{e^{2k_+}}{e^{2(q_++k_+)}-e^{2k_+}-e^{2q_+}}\bigg]\bigg){e_1(k_+)e_2^{\dagger}(q_+)},\\
        J^{(3)}_5&=\int _0 ^{+\infty}dk_+ \int _0 ^{+\infty} dq_+ \frac{e^{2(k_++q_+)}}{(e^{2k_+}+1)(e^{2q_+}+1)}\hat{\alpha}^{\dagger}(k_+)\hat{\alpha}^{\dagger}(q_+){\epsilon_1^{\dagger}(k_+)\epsilon_2^{\dagger}(q_+)},\\
        -J^{(3)}_6&=\int _0 ^{+\infty}dk_+ \int _0 ^{+\infty} dq_+ \frac{e^{2(k_++q_+)}}{(e^{2k_+}+1)(e^{2q_+}+1)}\hat{\alpha}^{\dagger}(q_+)\hat{\alpha}^{\dagger}(k_+)\textcolor{black}{\epsilon_2^{\dagger}(q_+)\epsilon_1^{\dagger}(k_+)}=\\
        &-\int_0^{\frac{1}{2}\ln(2)}dk_+\int_{-\frac{1}{2}\ln(\frac{1-e^{-2k_+}}{2})}^{+\infty}dq_+\frac{e^{2q_+}}{(e^{2k_+}-1)(e^{2q_+}-1)}\\
        &\hat{\alpha}^{\dagger}\bigg(\frac{1}{2}\ln\bigg[\frac{e^{2q_+}}{e^{2(k_++q_+)}-e^{2k_+}-e^{2q_+}}\bigg]\bigg)\hat{\alpha}^{\dagger}\bigg(\frac{1}{2}\ln\bigg[\frac{e^{2(k_++q_+)}-e^{2k_+}-e^{2q_+}}{e^{2k_+}}\bigg]\bigg){e_1(k_+)e_2^{\dagger}(q_+)}\\
        &-\int_{\frac{1}{2}\ln(2)}^{\frac{1}{2}\ln(3)}dk_+\int_{-\frac{1}{2}\ln(\frac{1-e^{-2k_+}}{2})}^{-\frac{1}{2}\ln(1-2e^{-2k_+})}dq_+\frac{e^{2q_+}}{(e^{2k_+}-1)(e^{2q_+}-1)}\\
        &\hat{\alpha}^{\dagger}\bigg(\frac{1}{2}\ln\bigg[\frac{e^{2q_+}}{e^{2(k_++q_+)}-e^{2k_+}-e^{2q_+}}\bigg]\bigg)\hat{\alpha}^{\dagger}\bigg(\frac{1}{2}\ln\bigg[\frac{e^{2(k_++q_+)}-e^{2k_+}-e^{2q_+}}{e^{2k_+}}\bigg]\bigg){e_1(k_+)e_2^{\dagger}(q_+)},\\
        J^{(3)}_7&=\int _0 ^{+\infty}dk_+ \int _0 ^{+\infty} dq_+ \frac{e^{2k_+}}{(e^{2k_+}+1)(e^{2q_+}+1)}\hat{\alpha}^{\dagger}(k_+)\hat{\beta}(q_+){\epsilon_1^{\dagger}(k_+)\epsilon_2(q_+)},\\
        -J^{(3)}_8&=\int _0 ^{+\infty}dk_+ \int _0 ^{+\infty} dq_+ \frac{e^{2k_+}}{(e^{2k_+}+1)(e^{2q_+}+1)}\hat{\beta}(q_+)\hat{\alpha}^{\dagger}(k_+)\textcolor{black}{\epsilon_2(q_+)}\epsilon_1^{\dagger}(k_+)=     \\
   &-\int_{\frac{1}{2}\ln(2)}^{\frac{1}{2}\ln(3)}dk_+\int_{-\frac{1}{2}\ln(1-2e^{-2k_+})}^{+\infty}dq_+\frac{e^{2q_+}}{(e^{2k_+}-1)(e^{2q_+}-1)}\\
   &\hat{\beta}\bigg(\frac{1}{2}\ln\bigg[\frac{e^{2(k_++q_+)}-e^{2q_+}-e^{2k_+}}{e^{2q_+}}\bigg]\bigg)\hat{\alpha}^{\dagger}\bigg(\frac{1}{2}\ln\bigg[\frac{e^{2(k_++q_+)}-e^{2q_+}-e^{2k_+}}{e^{2k_+}}\bigg]\bigg){e_1(k_+)e_2^{\dagger}(q_+)}\\
   &-\int_{\frac{1}{2}\ln(3)}^{+\infty}dk_+\int_{-\frac{1}{2}\ln(\frac{1-e^{-2k_+}}{2})}^{+\infty}dq_+\frac{e^{2q_+}}{(e^{2k_+}-1)(e^{2q_+}-1)}\\
   &\hat{\beta}\bigg(\frac{1}{2}\ln\bigg[\frac{e^{2(k_++q_+)}-e^{2q_+}-e^{2k_+}}{e^{2q_+}}\bigg]\bigg)\hat{\alpha}^{\dagger}\bigg(\frac{1}{2}\ln\bigg[\frac{e^{2(k_++q_+)}-e^{2q_+}-e^{2k_+}}{e^{2k_+}}\bigg]\bigg){e_1(k_+)e_2^{\dagger}(q_+)},
    \end{aligned}
\end{equation*}
\begin{equation*}
     \begin{aligned}
   H^{(3)}_1&=\int _0 ^{+\infty}dk_+ \int _0 ^{+\infty} dq_+ \frac{e^{2q_+}}{(e^{2k_+}-1)(e^{2q_+}+1)}\hat{b}(k_+)\hat{\alpha}^{\dagger}(q_+){e_1(k_+)\epsilon_2^{\dagger}(q_+)} ,     \\
   -H^{(3)}_2&=\int _0 ^{+\infty}dk_+ \int _0 ^{+\infty} dq_+ \frac{e^{2q_+}}{(e^{2k_+}-1)(e^{2q_+}+1)}\hat{\alpha}^{\dagger}(q_+)\hat{}b(k_+)\epsilon_2^{\dagger}(q_+)e_1(k_+)=\\
   &\int_{\frac{1}{2}\ln(3/2)}^{+\infty}dk_+\int_{\frac{1}{2}\ln(2(e^{2k_+}-1))}^{\frac{1}{2}\ln(2e^{2k_+}-1)}dq_+\frac{1}{(1-e^{-2k_+})(e^{2q_+}+1)}\\
   &\hat{\alpha}^{\dagger}\bigg(-\frac{1}{2}\ln\bigg[e^{2(q_+-k_+)}+e^{-2k_+}-1\bigg]\bigg)\hat{b}\bigg(\frac{1}{2}\ln\bigg[\frac{e^{2(q_+-k_+)}}{e^{2(q_+-k_+)}+e^{-2k_+}-1}\bigg]\bigg){e_1^{\dagger}(k_+)\epsilon_2(q_+)}\\
   &+\int_0^{\frac{1}{2}\ln(3/2)}dk_+\int_0^{\frac{1}{2}\ln(2e^{2k_+}-1)}dq_+\frac{1}{(1-e^{-2k_+})(e^{2q_+}+1)}\\
   &\hat{\alpha}^{\dagger}\bigg(-\frac{1}{2}\ln\bigg[e^{2(q_+-k_+)}+e^{-2k_+}-1\bigg]\bigg)\hat{b}\bigg(\frac{1}{2}\ln\bigg[\frac{e^{2(q_+-k_+)}}{e^{2(q_+-k_+)}+e^{-2k_+}-1}\bigg]\bigg){e_1^{\dagger}(k_+)\epsilon_2(q_+)}\\
   &+\int_{0}^{\frac{1}{2}\ln(2)}dk_+\int_{0}^{-\frac{1}{2}\ln(e^{2k_+}-1)}dq_+\frac{1}{(1-e^{-2k_+})(e^{-2q_+}+1)}\\
   &\hat{\alpha}^{\dagger}\bigg(-\frac{1}{2}\ln\bigg[e^{-2(q_++k_+)}+e^{-2k_+}-1\bigg]\bigg)\hat{b}\bigg(\frac{1}{2}\ln\bigg[\frac{e^{-2(q_++k_+)}}{e^{-2(q_++k_+)}+e^{-2k_+}-1}\bigg]\bigg){e_1^{\dagger}(k_+)\epsilon_2^{\dagger}(q_+)}\\
   &+\int_{\frac{1}{2}\ln(3/2)}^{\frac{1}{2}\ln(2)}dk_+\int_{0}^{\frac{1}{2}\ln(2(e^{2k_+}-1))}dq_+\frac{1}{(1-e^{-2k_+})(e^{2q_+}+1)}\\
   &\hat{\alpha}^{\dagger}\bigg(-\frac{1}{2}\ln\bigg[e^{2(q_+-k_+)}+e^{-2k_+}-1\bigg]\bigg)\hat{b}\bigg(\frac{1}{2}\ln\bigg[\frac{e^{2(q_+-k_+)}}{e^{2(q_+-k_+)}+e^{-2k_+}-1}\bigg]\bigg){e_1^{\dagger}(k_+)\epsilon_2(q_+)}\\
&+\int_{\frac{1}{2}\ln(2)}^{+\infty}dk_+\int_{\frac{1}{2}\ln(e^{2k_+}-1)}^{\frac{1}{2}\ln(2(e^{2k_+}-1))}dq_+\frac{1}{(1-e^{-2k_+})(e^{2q_+}+1)}\\
&\hat{\alpha}^{\dagger}\bigg(-\frac{1}{2}\ln\bigg[e^{2(q_+-k_+)}+e^{-2k_+}-1\bigg]\bigg)\hat{b}\bigg(\frac{1}{2}\ln\bigg[\frac{e^{2(q_+-k_+)}}{e^{2(q_+-k_+)}+e^{-2k_+}-1}\bigg]\bigg){e_1^{\dagger}(k_+)\epsilon_2(q_+)}\\
&+\int_0^{+\infty}dk_+\int_{\frac{1}{2}\ln(1+e^{-2k_+})}^{\frac{1}{2}\ln(1+2e^{-2k_+})}dq_+\frac{1}{(e^{2k_+}+1)(e^{2q_+}-1)}\\
&\hat{\alpha}^{\dagger}\bigg(-\frac{1}{2}\ln\bigg[e^{2(k_++q_+)}-e^{2k_+}-1\bigg]\bigg)\hat{b}\bigg(\frac{1}{2}\ln\bigg[\frac{e^{2(k_++q_+)}}{e^{2(k_++q_+)}-e^{2k_+}-1}\bigg]\bigg){\epsilon_1(k_+)e_2(q_+)}\\
&+\int_0^{+\infty}dk_+\int_{\frac{1}{2}\ln(1+e^{2k_+})}^{\frac{1}{2}\ln(1+2e^{2k_+})}dq_+\frac{1}{(e^{-2k_+}+1)(e^{2q_+}-1)}\\
&\hat{\alpha}^{\dagger}\bigg(-\frac{1}{2}\ln\bigg[e^{2(-k_++q_+)}-e^{-2k_+}-1\bigg]\bigg)\hat{b}\bigg(\frac{1}{2}\ln\bigg[\frac{e^{2(-k_++q_+)}}{e^{2(-k_++q_+)}-e^{-2k_+}-1}\bigg]\bigg){\epsilon_1^{\dagger}(k_+)e_2(q_+)},\\
H^{(3)}_3&=\int _0 ^{+\infty}dk_+ \int _0 ^{+\infty} dq_+ \frac{1}{(e^{2k_+}-1)(e^{2q_+}+1)}\hat{b}(k_+)\hat{\beta}(q_+){e_1(k_+)\epsilon_2(q_+)}\\
     -H^{(3)}_4&=\int _0 ^{+\infty}dk_+ \int _0 ^{+\infty} dq_+ \frac{1}{(e^{2k_+}-1)(e^{2q_+}+1)}\hat{\beta}(q_+)\hat{b}(k_+)\textcolor{black}{\epsilon_2(q_+)e_1(k_+)}=\\
&  \int_0^{+\infty}dk_+\int_{\frac{1}{2}\ln(2e^{2k_+}-1)}^{+\infty}dq_+\frac{1}{(1-e^{-2k_+})(e^{2q_+}+1)}\\
&\hat{\beta}\bigg(\frac{1}{2}\ln\bigg[e^{2(q_+-k_+)}+e^{-2k_+}-1\bigg]\bigg)\hat{b}\bigg(\frac{1}{2}\ln\bigg[\frac{e^{2(q_+-k_+)}}{e^{2(q_+-k_+)}+e^{-2k_+}-1}\bigg]\bigg){e_1^{\dagger}(k_+)\epsilon_2(q_+)} \\
&+\int_0^{+\infty}dk_+\int_{\frac{1}{2}\ln(2(e^{-2k_+}+1))}^{+\infty}dq_+\frac{1}{(1+e^{2k_+})(e^{2q_+}-1)}\\
&\hat{\beta}\bigg(\frac{1}{2}\ln\bigg[e^{2(q_++k_+)}-e^{2k_+}-1\bigg]\bigg)\hat{b}\bigg(\frac{1}{2}\ln\bigg[\frac{e^{2(q_++k_+)}}{e^{2(q_++k_+)}-e^{2k_+}-1}\bigg]\bigg){\epsilon_1(k_+)e_2(q_+)}\\
&\int_0^{+\infty}dk_+\int_{\frac{1}{2}\ln(2(e^{2k_+}+1))}^{+\infty}dq_+\frac{1}{(1+e^{-2k_+})(e^{2q_+}-1)}\\
&\hat{\beta}\bigg(\frac{1}{2}\ln\bigg[e^{2(q_+-k_+)}-e^{-2k_+}-1\bigg]\bigg)\hat{b}\bigg(\frac{1}{2}\ln\bigg[\frac{e^{2(q_+-k_+)}}{e^{2(q_+-k_+)}-e^{-2k_+}-1}\bigg]\bigg){\epsilon^{\dagger}_1(k_+)e_2(q_+)}
     \end{aligned}
 \end{equation*}
\begin{equation*}
    \begin{aligned}
    &+\int_0^{+\infty}dk_+\int_{\frac{1}{2}\ln(2e^{-2k_+}+1)}^{\frac{1}{2}\ln(2(e^{-2k_+}+1))}dq_+\frac{1}{(1+e^{2k_+})(e^{2q_+}-1)}\\
&\hat{\beta}\bigg(\frac{1}{2}\ln\bigg[e^{2(q_++k_+)}-e^{2k_+}-1\bigg]\bigg)\hat{b}\bigg(\frac{1}{2}\ln\bigg[\frac{e^{2(q_++k_+)}}{e^{2(q_++k_+)}-e^{2k_+}-1}\bigg]\bigg){\epsilon_1(k_+)e_2(q_+)},\\
&+\int_0^{+\infty}dk_+\int_{\frac{1}{2}\ln(2e^{2k_+}+1)}^{\frac{1}{2}\ln(2(e^{2k_+}+1))}dq_+\frac{1}{(1+e^{-2k_+})(e^{2q_+}-1)}\\
&\hat{\beta}\bigg(\frac{1}{2}\ln\bigg[e^{2(q_+-k_+)}-e^{-2k_+}-1\bigg]\bigg)\hat{b}\bigg(\frac{1}{2}\ln\bigg[\frac{e^{2(q_+-k_+)}}{e^{2(q_+-k_+)}-e^{-2k_+}-1}\bigg]\bigg){\epsilon^{\dagger}_1(k_+)e_2(q_+)},\\
H^{(3)}_5&=\int _0 ^{+\infty}dk_+ \int _0 ^{+\infty} dq_+ \frac{e^{2(k_++q_+)}}{(e^{2k_+}-1)(e^{2q_+}+1)}\hat{a}^{\dagger}(k_+)\hat{\alpha}^{\dagger}(q_+){e_1^{\dagger}(k_+)\epsilon_2^{\dagger}(q_+)}\\
-H^{(3)}_6&=\int _0 ^{+\infty}dk_+ \int _0 ^{+\infty} dq_+ \frac{e^{2(k_++q_+)}}{(e^{2k_+}-1)(e^{2q_+}+1)}\hat{\alpha}^{\dagger}(q_+)\hat{a}^{\dagger}(k_+)\textcolor{black}{\epsilon_2^{\dagger}(q_+)e_1^{\dagger}(k_+)}=\\
&\int_0^{\frac{1}{2}\ln(2)}dk_+\int_{-\frac{1}{2}\ln(2e^{-2k_+}-1)}^{+\infty}dq_+\frac{e^{2q_+}}{(e^{2k_+}-1)(e^{2q_+}+1)}\\
&\hat{\alpha}^{\dagger}\bigg(\frac{1}{2}\ln\bigg[\frac{e^{2q_+}}{e^{2(k_++q_+)}+e^{2k_+}-e^{2q_+}}\bigg]\bigg)\hat{a}^{\dagger}\bigg(\frac{1}{2}\ln\bigg[e^{-2k_+}(e^{2(k_++q_+)}+e^{2k_+}-e^{2q_+})\bigg]\bigg){e_1(k_+)\epsilon_2^{\dagger}(q_+)},\\
H^{(3)}_7&=\int _0 ^{+\infty}dk_+ \int _0 ^{+\infty} dq_+ \frac{e^{2k_+}}{(e^{2k_+}-1)(e^{2q_+}+1)}\hat{a}^{\dagger}(k_+)\hat{\beta}(q_+){e_1^{\dagger}(k_+)\epsilon_2(q_+)}\\
  -H^{(3)}_8&=\int _0 ^{+\infty}dk_+ \int _0 ^{+\infty} dq_+ \frac{e^{2k_+}}{(e^{2k_+}-1)(e^{2q_+}+1)}\hat{\beta}(q_+)\hat{a}^{\dagger}(k_+)\textcolor{black}{\epsilon_2(q_+)}e_1^{\dagger}(k_+)=  \\
  &\int_{\frac{1}{2}\ln(2)}^{+\infty}dk_+\int_0^{+\infty}dq_+\frac{1}{(e^{2k_+}-1)(e^{2q_+}+1)}\\
  &\hat{\beta}\bigg(\frac{1}{2}\ln\bigg[e^{2(k_++q_+)}+e^{2k_+}-1\bigg]\bigg)\hat{a}^{\dagger}\bigg(-\frac{1}{2}\ln\bigg[\frac{e^{2(q_++k_+)}}{e^{2(k_++q_+)}+e^{2k_+}-1}\bigg]\bigg){e_1(k_+)\epsilon_2(q_+)}\\
  &+\int_{\frac{1}{2}\ln(2)}^{+\infty}dk_+\int_0^{+\infty}dq_+\frac{1}{(e^{2k_+}-1)(e^{-2q_+}+1)}\\
  &\hat{\beta}\bigg(\frac{1}{2}\ln\bigg[e^{2(k_+-q_+)}+e^{2k_+}-1\bigg]\bigg)\hat{a}^{\dagger}\bigg(-\frac{1}{2}\ln\bigg[\frac{e^{2(-q_++k_+)}}{e^{2(k_+-q_+)}+e^{2k_+}-1}\bigg]\bigg){e_1(k_+)\epsilon_2^{\dagger}(q_+)}\\
  &+\int_0^{\frac{1}{2}\ln(2)}dk_+\int_0^{+\infty}dq_+\frac{1}{(e^{2k_+}-1)(e^{2q_+}+1)}\\
  &\hat{\beta}\bigg(\frac{1}{2}\ln\bigg[e^{2(k_++q_+)}+e^{2k_+}-1\bigg]\bigg)\hat{a}^{\dagger}\bigg(-\frac{1}{2}\ln\bigg[\frac{e^{2(q_++k_+)}}{e^{2(k_++q_+)}+e^{2k_+}-1}\bigg]\bigg){e_1(k_+)\epsilon_2(q_+)}\\
  &+\int_0^{\frac{1}{2}\ln(2)}dk_+\int_0^{-\frac{1}{2}\ln(2e^{-2k_+}-1)}dq_+\frac{1}{(e^{2k_+}-1)(e^{-2q_+}+1)}\\
  &\hat{\beta}\bigg(\frac{1}{2}\ln\bigg[e^{2(k_+-q_+)}+e^{2k_+}-1\bigg]\bigg)\hat{a}^{\dagger}\bigg(-\frac{1}{2}\ln\bigg[\frac{e^{2(-q_++k_+)}}{e^{2(k_+-q_+)}+e^{2k_+}-1}\bigg]\bigg){e_1(k_+)\epsilon_2^{\dagger}(q_+)},\\
   H^{(3)}_9&=\int _0 ^{+\infty}dk_+ \int _0 ^{+\infty} dq_+ \frac{e^{2q_+}}{(e^{2k_+}+1)(e^{2q_+}-1)}\hat{\beta}(k_+)\hat{a}^{\dagger}(q_+){\epsilon_1(k_+)e_2^{\dagger}(q_+)}\\
    -H^{(3)}_{10}&=\int _0 ^{+\infty}dk_+ \int _0 ^{+\infty} dq_+ \frac{e^{2q_+}}{(e^{2k_+}+1)(e^{2q_+}-1)}\hat{a}^{\dagger}(q_+)\hat{\beta}(k_+)e_2^{\dagger}(q_+)\epsilon_1(k_+)=\\
  &\int_0^{\frac{1}{2}\ln(2)}dk_+\int_{-\frac{1}{2}\ln(e^{2k_+}-1)}^{-\frac{1}{2}\ln(\frac{e^{2k_+}-1}{2})}dq_+\frac{1}{(1-e^{-2k_+})(1+e^{-2q_+})}\\
  &\hat{a}^{\dagger}\bigg(-\frac{1}{2}\ln\bigg[1-e^{-2k_+}-e^{-2(k_++q_+)}\bigg]\bigg)\hat{\beta}\bigg(\frac{1}{2}\ln\bigg[\frac{e^{-2(k_++q_+)}}{1-e^{-2k_+}-e^{-2(k_++q_+)}}\bigg]\bigg){e_1^{\dagger}(k_+)\epsilon_2^{\dagger}(q_+)}\\
  &+\int_{\frac{1}{2}\ln(2)}^{\frac{1}{2}\ln(3)}dk_+\int_0^{-\frac{1}{2}\ln(\frac{e^{2k_+}-1}{2})}dq_+\frac{1}{(1-e^{-2k_+})(1+e^{-2q_+})}\\
  &\hat{a}^{\dagger}\bigg(-\frac{1}{2}\ln\bigg[1-e^{-2k_+}-e^{-2(k_++q_+)}\bigg]\bigg)\hat{\beta}\bigg(\frac{1}{2}\ln\bigg[\frac{e^{-2(k_++q_+)}}{1-e^{-2k_+}-e^{-2(k_++q_+)}}\bigg]\bigg){e_1^{\dagger}(k_+)\epsilon_2^{\dagger}(q_+)}
    \end{aligned}
\end{equation*}
\begin{equation*}
    \begin{aligned}
    &+\int_{\frac{1}{2}\ln(2)}^{\frac{1}{2}\ln(3)}dk_+\int_0^{\frac{1}{2}\ln(e^{2k_+}-1)}dq_+\frac{1}{(1-e^{-2k_+})(1+e^{2q_+})}\\
  &\hat{a}^{\dagger}\bigg(-\frac{1}{2}\ln\bigg[1-e^{-2k_+}-e^{-2(k_+-q_+)}\bigg]\bigg)\hat{\beta}\bigg(\frac{1}{2}\ln\bigg[\frac{e^{-2(k_+-q_+)}}{1-e^{-2k_+}-e^{-2(k_+-q_+)}}\bigg]\bigg){e_1^{\dagger}(k_+)\epsilon_2(q_+)}\\
  &+\int_{\frac{1}{2}\ln(3)}^{+\infty}dk_+\int_{\frac{1}{2}\ln(\frac{e^{2k_+}-1}{2})}^{\frac{1}{2}\ln(e^{2k_+}-1)}dq_+\frac{1}{(1-e^{-2k_+})(1+e^{2q_+})}\\
  &\hat{a}^{\dagger}\bigg(-\frac{1}{2}\ln\bigg[1-e^{-2k_+}-e^{-2(k_+-q_+)}\bigg]\bigg)\hat{\beta}\bigg(\frac{1}{2}\ln\bigg[\frac{e^{-2(k_+-q_+)}}{1-e^{-2k_+}-e^{-2(k_+-q_+)}}\bigg]\bigg){e_1^{\dagger}(k_+)\epsilon_2(q_+)}\\
  &+\int_0^{+\infty}dk_+\int_0^{\frac{1}{2}\ln(1+e^{-2k_+})}dq_+\frac{1}{(e^{2k_+}+1)(e^{2q_+}-1)}\\
  &\hat{a}^{\dagger}\bigg(\frac{1}{2}\ln\bigg[\frac{1}{1+e^{2k_+}-e^{2(k_++q_+)}}\bigg]\bigg)\hat{\beta}\bigg(\frac{1}{2}\ln\bigg[\frac{e^{2(k_++q_+)}}{1+e^{2k_+}-e^{2(k_++q_+)}}\bigg]\bigg){\epsilon_1(k_+)e_2(q_+)}\\
  &+\int_0^{+\infty}dk_+\int_{\frac{1}{2}\ln(\frac{1+e^{2k_+}}{2})}^{\frac{1}{2}\ln(1+e^{2k_+})}dq_+\frac{1}{(e^{-2k_+}+1)(e^{2q_+}-1)}\\
  &\hat{a}^{\dagger}\bigg(\frac{1}{2}\ln\bigg[\frac{1}{1+e^{-2k_+}-e^{2(-k_++q_+)}}\bigg]\bigg)\hat{\beta}\bigg(\frac{1}{2}\ln\bigg[\frac{e^{2(-k_++q_+)}}{1+e^{-2k_+}-e^{2(-k_++q_+)}}\bigg]\bigg){\epsilon_1^{\dagger}(k_+)e_2(q_+)},\\
     H^{(3)}_{11}&=\int _0 ^{+\infty}dk_+ \int _0 ^{+\infty} dq_+ \frac{1}{(e^{2k_+}+1)(e^{2q_+}-1)}\hat{\beta}(k_+)\hat{b}(q_+){\epsilon_1(k_+)e_2(q_+)},\\
   -H^{(3)}_{12}&=\int _0 ^{+\infty}dk_+ \int _0 ^{+\infty} dq_+ \frac{1}{(e^{2k_+}+1)(e^{2q_+}-1)}\hat{b}(q_+)\hat{\beta}(k_+)\textcolor{black}{e_2(q_+)\epsilon_1(k_+)}=\\
   &\int_0^{+\infty}dk_+\int_0^{-\frac{1}{2}\ln(\frac{1+e^{-2k_+}}{2})}dq_+\frac{1}{(e^{2k_+}+1)(1-e^{-2q_+})}\\
   &\hat{b}\bigg(\frac{1}{2}\ln\bigg[1+e^{2k_+}-e^{2(k_+-q_+)}\bigg]\bigg)\hat{\beta}\bigg(\frac{1}{2}\ln\bigg[\frac{e^{2(k_+-q_+)}}{1+e^{2k_+}-e^{2(k_+-q_+)}}\bigg]\bigg){\epsilon_1(k_+)e_2^{\dagger}(q_+)},\\
   H^{(3)}_{13}&=\int _0 ^{+\infty}dk_+ \int _0 ^{+\infty} dq_+ \frac{e^{2(k_++q_+)}}{(e^{2k_+}+1)(e^{2q_+}-1)}\hat{\alpha}^{\dagger}(k_+)\hat{a}^{\dagger}(q_+){\epsilon_1^{\dagger}(k_+)e_2^{\dagger}(q_+)},\\
      -H^{(3)}_{14}&=\int _0 ^{+\infty}dk_+ \int _0 ^{+\infty} dq_+ \frac{e^{2(k_++q_+)}}{(e^{2k_+}+1)(e^{2q_+}-1)}\hat{a}^{\dagger}(q_+)\hat{\alpha}^{\dagger}(k_+)\textcolor{black}{e_2^{\dagger}(q_+)\epsilon_1^{\dagger}(k_+)}=\\
   &\int_{\frac{1}{2}\ln(3)}^{+\infty}dk_+\int_0^{\frac{1}{2}\ln(\frac{e^{2k_+}-1}{2})}dq_+\frac{1}{(e^{2q_+}+1)(1-e^{-2k_+})}\\
   &\hat{a}^{\dagger}\bigg(\frac{1}{2}\ln\bigg[\frac{1}{1-e^{-2k_+}-e^{2(q_+-k_+)}}\bigg]\bigg)\hat{\alpha}^{\dagger}\bigg(\frac{1}{2}\ln\bigg[\frac{1-e^{-2k_+}-e^{2(q_+-k_+)}}{e^{2(q_+-k_+)}}\bigg]\bigg){e_1^{\dagger}(k_+)\epsilon_2(q_+)}\\
   &+\int_{\frac{1}{2}\ln(3)}^{+\infty}dk_+\int_0^{+\infty}dq_+\frac{1}{(e^{-2q_+}+1)(1-e^{-2k_+})}\\
   &\hat{a}^{\dagger}\bigg(\frac{1}{2}\ln\bigg[\frac{1}{1-e^{-2k_+}-e^{-2(q_++k_+)}}\bigg]\bigg)\hat{\alpha}^{\dagger}\bigg(\frac{1}{2}\ln\bigg[\frac{1-e^{-2k_+}-e^{-2(q_++k_+)}}{e^{-2(q_++k_+)}}\bigg]\bigg){e_1^{\dagger}(k_+)\epsilon_2^{\dagger}(q_+)}\\
   &+\int_0^{\frac{1}{2}\ln(3)}dk_+\int_{-\frac{1}{2}\ln(\frac{e^{2k_+}-1}{2})}^{+\infty}dq_+\frac{1}{(e^{-2q_+}+1)(1-e^{-2k_+})}\\
   &\hat{a}^{\dagger}\bigg(\frac{1}{2}\ln\bigg[\frac{1}{1-e^{-2k_+}-e^{-2(q_++k_+)}}\bigg]\bigg)\hat{\alpha}^{\dagger}\bigg(\frac{1}{2}\ln\bigg[\frac{1-e^{-2k_+}-e^{-2(q_++k_+)}}{e^{-2(q_++k_+)}}\bigg]\bigg){e_1^{\dagger}(k_+)\epsilon_2^{\dagger}(q_+)}\\
   &+\int_0^{+\infty}dk_+\int_0^{\frac{1}{2}\ln(\frac{1+e^{2k_+}}{2})}dq_+\frac{1}{(e^{-2k_+}+1)(e^{2q_+}-1)}\\
   &\hat{a}^{\dagger}\bigg(\frac{1}{2}\ln\bigg[\frac{1}{1+e^{-2k_+}-e^{2(q_+-k_+)}}\bigg]\bigg)\hat{\alpha}^{\dagger}\bigg(\frac{1}{2}\ln\bigg[\frac{1+e^{-2k_+}-e^{2(q_+-k_+)}}{e^{2(q_+-k_+)}}\bigg]\bigg){\epsilon_1^{\dagger}(k_+)e_2(q_+)},
    \end{aligned}
\end{equation*}
\begin{equation*}
    \begin{aligned}
   H^{(3)}_{15}&=\int _0 ^{+\infty}dk_+ \int _0 ^{+\infty} dq_+ \frac{e^{2k_+}}{(e^{2k_+}+1)(e^{2q_+}-1)}\hat{\alpha}^{\dagger}(k_+)\hat{b}(q_+){\epsilon_1^{\dagger}(k_+)e_2(q_+)},\\
   -H^{(3)}_{16}&=\int _0 ^{+\infty}dk_+ \int _0 ^{+\infty} dq_+ \frac{e^{2k_+}}{(e^{2k_+}+1)(e^{2q_+}-1)}\hat{b}(q_+)\hat{\alpha}^{\dagger}(k_+)\textcolor{black}{e_2(q_+)}\epsilon_1^{\dagger}(k_+)=\\
   &\int_0^{+\infty}dk_+\int_0^{+\infty}dq_+\frac{1}{(e^{-2k_+}+1)(1-e^{-2q_+})}\\
   &\hat{b}\bigg(\frac{1}{2}\ln\bigg[1+e^{-2k_+}-e^{-2(k_++q_+)}\bigg]\bigg)\hat{\alpha}^{\dagger}\bigg(-\frac{1}{2}\ln\bigg[\frac{e^{-2(k_++q_+)}}{1+e^{-2k_+}-e^{-2(k_++q_+)}}\bigg]\bigg){\epsilon_1^{\dagger}(k_+)e_2^{\dagger}(q_+)}\\
   &+\int_0^{+\infty}dk_+\int_{-\frac{1}{2}\ln(\frac{1+e^{-2k_+}}{2})}^{+\infty}dq_+\frac{1}{(e^{2k_+}+1)(1-e^{-2q_+})}\\
   &\hat{b}\bigg(\frac{1}{2}\ln\bigg[1+e^{2k_+}-e^{2(k_+-q_+)}\bigg]\bigg)\hat{\alpha}^{\dagger}\bigg(-\frac{1}{2}\ln\bigg[\frac{e^{2(k_+-q_+)}}{1+e^{2k_+}-e^{2(k_+-q_+)}}\bigg]\bigg){\epsilon_1(k_+)e_2^{\dagger}(q_+)}.
    \end{aligned}
\end{equation*}
 The resulting algebra is deformed and $\kappa$-Poincaré invariant and it is reported in Section \ref{sec:algebra}.

\end{document}